\documentclass[twocolumn,pra,superscriptaddress]{revtex4}
\usepackage{amssymb}
\usepackage{amsmath}
\usepackage{graphicx,bm}
\usepackage{epstopdf}
\usepackage{subfigure}
\usepackage{natbib}
\usepackage{epsfig}
\usepackage{amsfonts}
\usepackage{mathrsfs}
\usepackage{CJK}
\usepackage[toc,page,title,titletoc,header]{appendix}
\usepackage{array,longtable}

\usepackage[usenames, dvipsnames, x11names]{xcolor}

\usepackage[colorlinks]{hyperref}
\hypersetup{
 colorlinks=true,
 citecolor=red,
 linkcolor=blue,
 urlcolor=cyan}

\begin{document}
\title{Intrinsic and induced quantum quenches for enhancing qubit-based quantum noise spectroscopy}

\author{Yu-Xin Wang}
\affiliation{Pritzker School of Molecular Engineering,  University  of  Chicago, 5640  South  Ellis  Avenue,  Chicago,  Illinois  60637,  U.S.A.}

\author{Aashish A. Clerk}
\affiliation{Pritzker School of Molecular Engineering,  University  of  Chicago, 5640  South  Ellis  Avenue,  Chicago,  Illinois  60637,  U.S.A.}

\date{\today}

\begin{abstract}
We discuss how standard $T_2$-based quantum sensing and noise spectroscopy protocols often give rise to an inadvertent quench of the system or environment being probed:  there is an effective sudden change in the environmental Hamiltonian at the start of the sensing protocol.  These quenches are extremely sensitive to the initial environmental state, and lead to observable changes in the sensor qubit evolution.  We show how these new features can be used to directly access environmental response properties.  This enables methods for direct measurement of bath temperature, and methods to diagnose non-thermal equilibrium states.  We also discuss techniques that allow one to deliberately control and modulate this quench physics, which enables reconstruction of the bath spectral function.  Extensions to non-Gaussian quantum baths are also discussed, as is the direct applicability of our ideas to standard diamond NV-center based quantum sensing platforms. 
\end{abstract}

\maketitle

\section{Introduction}

A key technique in quantum sensing is to use a suitably driven sensor qubit to characterize a noisy, dissipative environment.  Commonly referred to as quantum noise spectroscopy (QNS)~\cite{Cappellaro2017}, this modality allows one to understand and possibly mitigate sources of decoherence that degrade a quantum processor~\cite{Kurizki2004,deSousa2009,Hanson2010,Suter2011,Oliver2011,Yacoby2013,Jelezko2015,Biercuk2017,Viola2018,Dzurak2018,Wiseman2018,Pershin2016,Viola2016nG,Cywinski2017,Oliver2018,Oliver2019,Oliver2020,Oliver2021}. It also serves as a powerful means to probe a complicated many-body target system via its fluctuation properties (see, e.g.,~\cite{Walsworth2012,Liu2015,Yacoby2018,Yao2020,Yao2021}).  While many QNS protocols focus on the more specific problem of characterizing classical Gaussian noise~\cite{Hanson2010,Suter2011,Oliver2011,Yacoby2013,Jelezko2015,Biercuk2017,Viola2018,Dzurak2018,Wiseman2018}, recent work has explored methods that go beyond these assumptions  
\cite{RBLiu2011,Du2011,Yang2016,Viola2016nG,Pershin2016,Cywinski2017,Oliver2018,RBLiu2019,Oliver2019,Oliver2020,Oliver2021,Clerk2020,Cywinski2020noise}.

\begin{figure}[t]
  \includegraphics[width=85mm]{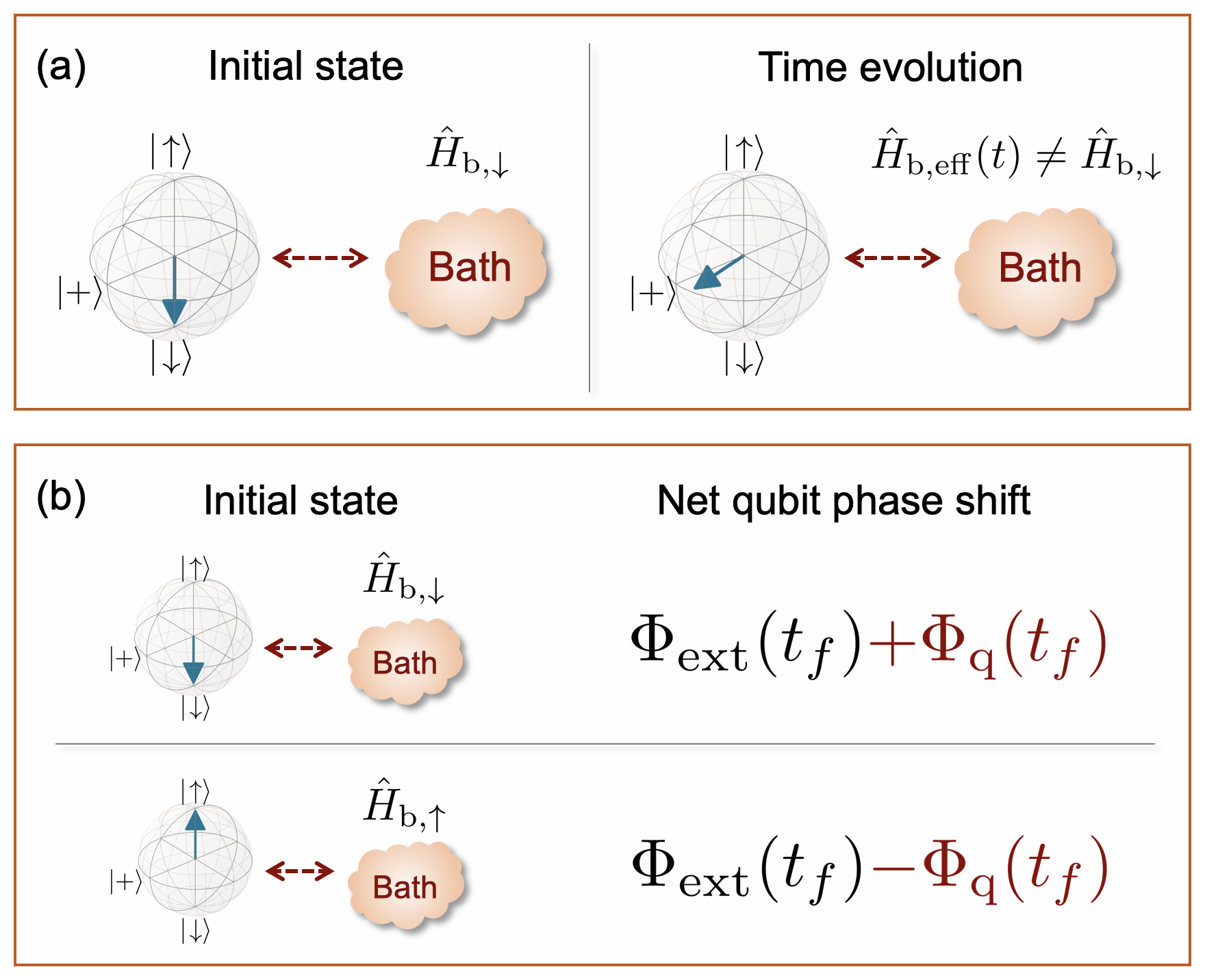}
      \caption{
      (a) Schematic illustrating an ``intrinsic'' quantum quench arising in standard $ T_2 $-type experiments, where the effective bath Hamiltonian undergoes a sudden change at the start of the protocol, c.f.~Eq.~\eqref{eq:H.beff.sum}.
      (b)  The quench manifests itself as an additional quench phase shift (QPS) $ \Phi _{\mathrm{q}} (t_f) $ of the sensor qubit. 
      The QPS can be distinguished from a phase $ \Phi _{\mathrm{ext}} (t_f) $ resulting from an external field: in the simplest case, it is crucially sensitive to the initial qubit state before the start of the sensing protocol. 
    }
    \label{fig:schematic}
\end{figure}

One crucial difference between a true quantum environment and a simple classical noise source is that the former is {\it dynamical}:  its properties can change in response to an external perturbation.  At a simple linear response level, this is encoded in the environment's susceptibility functions, or equivalently, asymmetric-in-frequency quantum noise spectral densities
\cite{Devoret2003,QN.RMP2010,Viola2016,Viola2017,Cywinski2020}. 
The most direct (and perhaps extreme) way to probe these properties is to induce a quantum quench, where the environment experiences a sudden change in its Hamiltonian.  Studying the consequences of deliberate quenches
has been an extremely useful tool for probing a variety of phenomena in correlated systems~\cite{Vengalattore2011}. 

In this paper, we show that the basic physics of a quantum quench is relevant to a wide variety of commonly employed QNS schemes and systems; crucially, this is the case {\it even if the protocol does not involve a deliberate quenching of the environment}.  We show how these quenches (whether intrinsic or deliberate) can be harnessed as a powerful new sensing modality: they reveal environmental response properties in previously unexplored ways.  By analyzing standard $ T_2 $-type qubit-based QNS protocols, we identify generic conditions under which an inadvertent quench of the environment influences the sensor qubit's evolution.
Surprisingly, the existence and properties of this quench effect are not simply a function of the initial environmental Hamiltonian, but instead depend on the initial environmental state.  The dominant effect of the quench is an unexpected phase shift of the sensor qubit coherence.  
For common cases where the environment is either a Gaussian quantum bath or the sensor-environment coupling is weak, we derive a simple, analytical expression connecting this quench phase shift (QPS) to a dissipative susceptibility of the environment (i.e.~an effective density of states).  We then use this to address a number of phenomena.  In particular, using the extra information provided by the QPS, a standard $T_2$-based QNS protocol can be enhanced to  independently characterize both fluctuation and response properties.  For the paradigmatic case of an environment with an Ohmic spectral density, we show that one can use the QPS (along with standard decoherence measurements) in a simple Hahn-echo protocol to directly extract the environmental temperature (something that cannot be done from decoherence measurements alone).  In Sec.~\ref{sec:V.gen.op}, we also show that the quench mechanism is relevant to generic initial bath states beyond equilibrium, and can be used to probe response properties in nonequilibrium systems.

We note that our work generalizes recent studies that analyzed anomalous phase shifts in $T_2$-type QNS protocols involving dynamical baths \cite{Viola2017,Cywinski2020}.  While those studies attributed these phases to an unusual ``biased'' qubit-environment coupling, we show that they can in fact arise in a far wider set of systems.  We also show how it is the initial bath state (and not the qubit-bath coupling) that plays a key role.  This realization will provide an important new control knob, as one can controllably change the properties of the quench via seemingly subtle changes in the initial bath state; this provides a powerful tool for reconstructing environmental spectral functions
(c.f.~Sec.~\ref{sec:gen.quench.spec}).  We further discuss extensions of this
physics in regimes beyond the validity of linear response
(see Sec.~\ref{sec:NonGaussian}).

\section{Intrinsic quantum quenches in standard $ T_2 $-type sensing protocols}
\label{sec:QuenchIntro}

While our ideas apply to a wide variety of settings, we focus throughout this paper on a standard QNS experiment where the sensor qubit is coupled to an environment via a pure-dephasing interaction. In the rotating frame with respect to free qubit Hamiltonian 
$ \Omega \hat \sigma _z /2  
= \left ( \Omega/2 \right )
\left (| \!  \! \uparrow \rangle
\langle \uparrow \!  \! |  \right.
- \left. | \!  \! \downarrow \rangle
\langle \downarrow \!  \! | \right )$, the qubit-bath Hamiltonian is given by 
\begin{align}
\label{eq:Htot.deph.gen}
 \hat H_{\mathrm{tot}} 
=  \left |  \uparrow \rangle \langle \uparrow \right | 
	 \otimes
	 \hat H _{\mathrm{b},\uparrow }
	 + \left |  \downarrow \rangle 
	 \langle \downarrow \right | 
	 \otimes
	 \hat H _{\mathrm{b},\downarrow }
	 ,
\end{align}
where $ \hat H _{\mathrm{b},\uparrow } $ 
($\hat H _{\mathrm{b}, \downarrow  } $) describes the bath Hamiltonian conditioned on qubit being in the state
$ \left |   \uparrow \right \rangle $
($ \left |  \downarrow  \right  \rangle $).
We set $\hbar = 1$ throughout.
As in standard  $ T_2 $-type measurements, the probe qubit is initialized in $ \left |  \downarrow  \right  \rangle$ and is initially unentangled with the bath. 
The quench physics we describe is crucially sensitive to the initial state of the bath. For illustrative purposes, we first focus on a simple but generic situation where the qubit  $ \left |  \downarrow  \right  \rangle $ state lifetime can be viewed as infinite, and the bath has relaxed to a thermal equilibrium state with respect to 
$\hat H _{\mathrm{b},\downarrow }$.  The initial density matrix of the qubit-bath system is thus 
\begin{subequations}
\label{eq:ini.down.sum}
\begin{align}
\label{eq:ini.down.tot}
& \hat{\rho}_{\mathrm{tot}} (t = 0^-)
= \left   |   \downarrow \rangle 
\langle \downarrow   \right | 
\otimes
\hat{\rho}_{\mathrm{b,i}}
    ,   \\
& \hat{\rho}_{\mathrm{b,i}}
= 
e^{-  \hat H_{\mathrm{b},\downarrow} / k_B T }
/ Z_T
 ,
\label{eq:ini.bath.down}
\end{align}
\end{subequations}
where $ Z _T $ is a normalization factor and $T$ is the initial bath temperature.

We consider a standard $T_2$-based sensing protocol.  At the start of the protocol ($ t = 0 $), an instantaneous $ \pi /{2} $-pulse is applied to prepare the qubit in an equal superposition state
$ | + \rangle \equiv
\left(  | \! \!   \uparrow \rangle 
+ | \!  \!  \downarrow \rangle \right ) 
/ \sqrt{2} $;
the system then evolves under  
$ \hat H_{\mathrm{tot}} $
for time $ t_f $, while the qubit is subject to
a sequence of instantaneous control $ \pi $-pulses.  At the end of the protocol, one measures qubit Pauli operator $\hat \sigma_x$ or $\hat \sigma_y$.  By repeating the measurements and varying $t_f$, one can obtain the qubit coherence 
$\langle 
\hat \sigma _- ( t_f  ) 
\rangle$ 
as a function of $t_f$.

Surprisingly, in many cases an intrinsic effective bath quench occurs as part of this standard sensing protocol.  To see this, we first rewrite 
$ \hat H_{\mathrm{tot}}$ as 
\begin{align}
    \hat H_{\mathrm{tot}} 
    = \frac{1 }{2} \hat \sigma _z \otimes \hat{\xi}
    + \hat \sigma _0   \otimes 
    \frac{1 }{2}( \hat H_{\mathrm{b},\uparrow} 
    + \hat H_{\mathrm{b},\downarrow} ) ,
    \label{eq:HTotIntro}
\end{align}
with $\hat{\xi}
\equiv \hat H_{\mathrm{b},\uparrow} - \hat H_{\mathrm{b},\downarrow} $, 
$ \hat \sigma _0   \equiv 
\left |   \uparrow \rangle
\langle \uparrow \! \!  | 
+  | \! \! \downarrow \rangle 
\langle \downarrow \right  |  $.
This suggests a simple picture for the evolution during the protocol:  the qubit dephases due to coupling to the bath noise operator 
$\hat{\xi}$, while the bath evolves under an effective averaged bath Hamiltonian.

Notably, the averaged bath Hamiltonian in Eq.~\eqref{eq:HTotIntro} may or may not commute with the initial bath state, which in our example is determined by $\hat H_{\mathrm{b},\downarrow}$.  This motivates defining a time-dependent effective bath Hamiltonian $ \hat H _{\mathrm{b,eff} }  ( t ) $ whose form reflects the change of the qubit state at $t=0$:
\begin{subequations}
\label{eq:H.beff.sum}
\begin{align}
\label{eq:H.beff.def}
& \hat H _{\mathrm{b,eff} }  ( t )
\equiv \text{Tr}_\mathrm{qb}
[ \hat{\rho}_{\mathrm{tot}} ( t )
    \hat H_{\mathrm{tot}} ( t ) ]
   , \\
   = & \begin{cases}
\hat H_{\mathrm{b},\downarrow} , 
    & t \le 0 , \\
    ( \hat H_{\mathrm{b},\uparrow} + 
\hat H_{\mathrm{b},\downarrow} ) /2
     ,
    &  0 < t < t_f  .
\end{cases}
\end{align}
\end{subequations}
We see that except for the trivial case 
$\hat H_{\mathrm{b},\uparrow} = 
\hat H_{\mathrm{b},\downarrow}  $, the bath Hamiltonian $ \hat H _{\mathrm{b,eff} }  ( t ) $ exhibits a sudden change (i.e., a quench) that is solely due to the sudden change in qubit state at $t=0$.  As we show below in Eq.~\eqref{eq:qb.dyn.gen}, this quench is physically meaningful:  it directly determines the evolution of the qubit coherence
$\langle 
\hat \sigma _- ( t_f  ) 
\rangle$, the very quantity that is measured in the protocol.

For our subsequent discussion, it is useful to rewrite 
$ \hat H _{\mathrm{b,eff} }  ( t )$ to make the quench more explicit:  
\begin{align}
\label{eq:Heff.bath.ini}
\hat H _{\mathrm{b,eff} }  ( t )
& = \hat H_{\mathrm{b,i}} 
+  \eta (t)  \hat V. 
\end{align} 
Here $\eta(t)$ is an effective quench control function, which encodes the temporal profile of the quench.  $\hat{V}$ represents the quench operator, which is defined as
\begin{align} 
\label{eq:V.bath.def}
\hat V \equiv
\hat H _{\mathrm{b,eff} }  ( t = 0 ^+ )  - 
 \hat H _{\mathrm{b,eff} }  ( t = 0 ^- ) 
 . 
\end{align} 
For the specific example considered here, we have $\hat H_{\mathrm{b,i}} =
    \hat H_{\mathrm{b},\downarrow} $
    and
\begin{subequations}
\begin{align}
    \eta (t)  & =  \Theta (t ) \Theta ( t_f - t ), 
        \label{eq:EtaSimple} \\ 
    \hat{V} & = \hat{\xi}/2,
        \label{eq:VSimple}
\end{align}
\end{subequations}
where $ \Theta (t)
$ is the Heaviside step function. As we will show in Sec.~\ref{sec:V.gen.op}, Eq.~\eqref{eq:Heff.bath.ini} describes quench physics corresponding to a much wider range of nonequilibrium initial bath states, even beyond the specific case in Eq.~\eqref{eq:ini.bath.down}.

We stress that in contrast to conventional quench experiments which require external temporal control of the bath, here the quench is intrinsic to the measurement protocol: it occurs unavoidably simply through the  ``back-action'' of the qubit on the bath associated with the start of the QNS protocol.  
While we have discussed a simple example here, the same physics also applies to more general initial bath states and more general quench functions $\eta(t)$.  
We show explicitly in Sec.~\ref{sec:V.gen.op} how in general, one can find the form of the effective quench operator $\hat{V}$ from the initial bath state (even if it is a non-thermal state unrelated to 
$H_{\mathrm{b},\uparrow/\downarrow}$).
Further, one can also generate more complicated quench functions $ \eta (t) $: as discussed in Sec.~\ref{sec:NV.impls}, one approach to achieve this is to use a qubit embedded in a multilevel physical system, e.g.~a nitrogen-vacancy center defect in diamond (see Fig.~\ref{fig:quench.filter}).

As our approach is more general than the specific example of Eq.~(\ref{eq:ini.bath.down}), in what follows, we will allow   
$ \eta (t) $ to have generic time dependence during the time evolution ($ 0 < t < t_f  $), and we will assume a general $ \hat V $ (unless specified otherwise).


\section{General sensor qubit evolution including effective 
quench}

We now rigorously show how the effective quench physics described in  Eqs.~(\ref{eq:Heff.bath.ini}) and (\ref{eq:V.bath.def}) 
manifests itself in our standard $T_2$-based sensing protocol. 
We first transform to an appropriate interaction picture, determined by the initial (static) bath Hamiltonian $ \hat H_{\mathrm{b,i}} $, and for the qubit, by the the standard toggling frame set by the choice of qubit control $ \pi $-pulses (see e.g.~\cite{Cappellaro2017}).  We thus have time-dependent interaction-picture bath operators  
$  \hat V  (t) $ and $ \hat{\xi} (t) $ whose time dependence is generated by $ \hat H_{\mathrm{b,i}} $.  Note that as the initial bath state is stationary in our interaction picture, $\hat{\xi}(t)$ will describe stationary quantum noise:  all its correlation functions will respect time-translational invariance.

Working in the above interaction picture, and letting $ F (t) $ denote the usual filter function that encodes the timing of qubit control $ \pi $-pulses, the time-dependent qubit coherence is given by (see Appendix~\ref{appsec:qubit.dyn.gen}):
\begin{subequations}
\label{eq:qb.dyn.gen}
\begin{align}
\label{eq:qb.dyn.gen.a}
&  \langle 
	\hat \sigma _- ( t_f  ) 
	\rangle
	= \frac{1}{2}
	\textrm{Tr} (
   \hat U _{ \uparrow }
    \hat{\rho}_{\mathrm{b,i}} 
    \hat U _{ \downarrow } ^\dag ) 
	, \\
&  \hat U _{ \uparrow (\downarrow) }
    =  \mathcal{T} 
    \exp \left\{
    -  i \int _{-\infty }^{ + \infty } \! 
    \left [ \eta (t')  \hat V  (t')   
    \pm \frac{F (t' )}{2} \hat{\xi} (t')
    \right ]
    dt'
    \right\}
    .
\end{align}
\end{subequations}
We stress that Eq.~\eqref{eq:qb.dyn.gen} is valid for a generic form of quench function $ \eta (t) $ and operator $\hat{V}(t)$, and not limited to the specific case described by Eq.~\eqref{eq:H.beff.sum}.  Note crucially that we {\it do not} include the quench operator $\hat{V}$ in the definition of our interaction picture.  While one could work in this alternate frame, it would obscure the fact that in general, $\hat{V}$ does not commute with the initial bath state.  It would also lead to a time-dependent bath noise operator $\hat{\xi}'(t)$ that is nonstationary.

To discuss the sensor qubit evolution, it is convenient to
separately parametrize the magnitude and phase of the qubit coherence function in Eq.~\eqref{eq:qb.dyn.gen}:
\begin{align}
\label{eq:qb.coh.gen}
\langle 
	\hat \sigma _- ( t_f  ) 
	\rangle
	&  = \frac{1}{2} \,
	e^{- \zeta (t_f)} \,
	e ^{- i \Phi(t_f)}.
\end{align}
The effects of the environment are now fully described by the (real, nonnegative) dephasing function
$ \zeta (t_f) $ (which controls the magnitude of the coherence) and the 
real bath-induced phase shift function $ \Phi(t_f) $.  Standard QNS protocols use information in $\zeta(t_f)$ to probe properties of the environment \cite{Cappellaro2017}.  As we will now see, due to our effective quench physics, key new features of the environment will also reveal themselves through the unexpected phase shift. 


\section{Quench-induced sensor-qubit phase shift}

The general goal of our QNS protocol is 
to measure properties of the environment.  
$T_2$-based QNS protocols typically have a sole focus on the {\it fluctuation} properties of the bath, specifically fluctuations of the bath noise operator $\hat\xi$.  The simplest quantity characterizing these is the symmetrized noise spectral density $\bar{S}[\omega]$, given by:
\begin{align}
    \bar{S} [ \omega ]
	 & \equiv \frac{1}{2}  
	\int _{-\infty }^{ + \infty } \!  d t   e^{ i \omega t}
	\langle
	\{ \hat{\xi} ( t )
	,
	\hat{\xi} ( 0 )
	\}
	\rangle
\end{align}
The average value here is with respect to the initial bath density matrix $ \hat{\rho}_{\mathrm{b,i}} $.  As discussed in many places (see e.g.~\cite{QN.RMP2010}), this quantity is symmetric in frequency, and plays the role of a classical noise spectral density.  

Another generic environmental property that is not typically probed in standard $T_2$-based QNS schemes is the dynamical response properties of the bath:  how does it change in response to a time-dependent external perturbation?  At the simplest linear-response level, this is described by conventional linear response susceptibilities (or equivalently, retarded Green's functions).  We will be interested in a particular susceptibility, describing how the average value of the noise operator $\hat\xi$ changes in response to a perturbation coupling to the quench operator $\hat{V}$.  This is described in linear response by  
 \begin{align}
    G ^R _{\xi V } [ \omega ] 
    & \equiv -i  \int _{-\infty }^{ + \infty } 
    \!  d t   e^{ i \omega t}
    \Theta (t)
    \langle
    [\hat{\xi} ( t )
	    ,
    \hat{ V } ( 0 )]
    \rangle.
\end{align}
We stress that in general, this susceptibility is {\it distinct} from the noise spectral density $\bar{S}[\omega]$.  Hence, being able to measure it would provide new information on the properties of our environment.

\subsection{Sensor qubit evolution in the weak coupling or Gaussian limit}

We return now to the evolution of our sensor qubit during the QNS protocol.  For a generic environment, Eqs.~\eqref{eq:qb.dyn.gen} and \eqref{eq:qb.coh.gen} (which describe the sensor qubit coherence) can be analyzed perturbatively in both $\hat\xi$ and $\hat{V}$; for the simple example of Eq.~\eqref{eq:H.beff.sum}, this amounts to perturbation theory in the qubit-bath coupling.  
The leading-order contributions to the dephasing and phase-shift functions in Eq.~(\ref{eq:qb.coh.gen}) can be succinctly written as (see App.~\ref{appsec:DepQPS.gen}) 
\begin{subequations}
\label{eq:qbdyn.Gau.gen}
\begin{align}
\label{eq:dep.Gau.gen}
& \zeta (t_f) \simeq  
\int _{-\infty }^{ + \infty } \! \frac{d  \omega }{4 \pi}
|F[\omega]|^2 \bar{S} [ \omega ] 
    , \\
\label{eq:Phi.Gau.gen}
&  \Phi  (t_f) =
\Phi _{\mathrm{q}}  (t_f) 
\simeq 
    \int _{-\infty }^{ + \infty } \! \frac{d  \omega }{2 \pi}
	F^*[\omega]
	\eta [\omega]
	G ^R _{\xi V }[ \omega ] .
\end{align}
\end{subequations}
Here we use the notation
$ z [\omega] 
\equiv 	
\int _{-\infty }^{ + \infty } \!  z (t) 
  e^{ i \omega t} d t
$ to denote the Fourier transform of a temporal function.  We stress that these expressions are valid for a general quench, and not just the specific example described by Eq.~\eqref{eq:H.beff.sum}.

Eqs.~\eqref{eq:qbdyn.Gau.gen} are generally valid for generic environments in the weak coupling limit; they also become exact for Gaussian quantum environments (i.e.~linear coupling to a bath of independent bosonic modes).  This covers many experimentally-relevant situations (e.g.~environments comprised of phononic or photonic modes)~\cite{Weiss2012book}.  Eq.~(\ref{eq:dep.Gau.gen}) is a standard textbook expression:  at the Gaussian level, the qubit dephasing is controlled by the environmental noise spectral density, weighted by the filter function.  In contrast, Eq.~(\ref{eq:Phi.Gau.gen}) is less appreciated:  because of the effective quench physics described above, and the dynamical nature of the bath, there is a bath-induced phase shift of the qubit sensor.  This phase shift $\Phi  _{\mathrm{q}} (t_f)$ depends both on the relevant bath susceptibility, the filter function $F[\omega]$ as well as the quench control function $\eta[\omega]$.  As we will see, this phase provides a new route to learning about the environment.
Note that the consequences of the quench can also be discussed beyond linear response, as would apply to more general environments and sensor-environment couplings; see Sec.~\ref{sec:NonGaussian}.

\subsection{Interpretation and measurement of the quench-induced phase shift}

It is worth unpacking Eq.~(\ref{eq:Phi.Gau.gen}) to provide a physical understanding of the quench-induced phase.  As discussed in Sec.~\ref{sec:QuenchIntro}, the effective quench at the start of our protocol at $t=0$ suddenly turns on a term $\eta(t) \hat{V}$ in the effective bath Hamiltonian (c.f. Eq.~(\ref{eq:Heff.bath.ini})).  At the linear response level, this perturbation causes a time-dependent shift in the average of the bath operator $\hat\xi$ that couples to the qubit.  
This shift is given from linear response by  
$	\langle {\delta \hat \xi (t )} \rangle_{V} 
=\int _{-\infty }^{ + \infty }  d t_2   \eta ( t_2 )
G ^R _{\xi V } (t - t_2) $.  Next, this induced average value of $\hat \xi$ has a direct consequence on the qubit:  it is equivalent to a time-dependent $z$ magnetic field on the sensor qubit.  This then leads to a net phase shift given by the integral of this effective field weighted by the filter function $F(t)$:
$ \Phi _{\mathrm{q}} (t_f) 
	=  \int _{-\infty }^{ + \infty }  d t_1 F(t_1) 
	\langle {\delta \hat \xi ( t_1 )} \rangle_{V}
$. Connections between sensor phase shifts and linear response were also discussed in Ref.~\cite{Cywinski2020}, though see \footnote{Our expression for the quench phase shift can thus be written as 
$ \Phi _{\mathrm{q}} (t_f) 
=  \int _{-\infty }^{ + \infty }  d t_1 F
\left (t_1 \right) 
\int _{-\infty }^{ + \infty }  d t_2  
\eta \left ( t_2 \right )
G ^R _{\xi V } \left (t_1 - t_2 \right)
$.  This is similar but not identical to Eq.~(15) in Ref.~\cite{Cywinski2020}, where the filter function $F (t_1 )$ erroneously occurs at a later time than the quench function $ \eta ( t_2 ) $}.

We stress that the quench-induced sensor phase shift
$ \Phi _{\mathrm{q}} (t_f) $ can be accessed in exactly the same type of experiments one would use to sense external DC or AC fields, making use of standard Ramsey, Hahn echo or more complex dynamical decoupling  sequences~\cite{Cappellaro2017}. 

A natural concern is whether this quench phase could be distinguished from more trivial phases resulting from external ambient magnetic fields.  In the presence of such fields, the net qubit phase shift in Eq.~\eqref{eq:qb.coh.gen} is now given by 
\begin{align}
\Phi(t_f)
	& = \Phi _\mathrm{ext} (t_f) 
	+  \Phi _{\mathrm{q}} (t_f) 
	,
	\\
\Phi _\mathrm{ext} (t_f) 
	&  = \int _{-\infty }^{ + \infty } \! 
	\frac{d  \omega }{2 \pi}
	F^*[\omega] B_{\mathrm{ext}}  [\omega] 
    ,
\end{align}
where $B_{\mathrm{ext}}(t)$ is the external ambient magnetic field, and the QPS 
$ \Phi _{\mathrm{q}} (t_f) $ is again given by  Eq.~\eqref{eq:Phi.Gau.gen}.  There is a critical difference between $\Phi_{\mathrm{ext}} $ and $ \Phi _{\mathrm{q}} $:  only the latter is sensitive to the initial state of the qubit.  One can thus easily exploit this feature to distinguish the QPS from other more trivial phase-shift mechanisms (see Fig.~\ref{fig:schematic}(b) for an example).

\subsection{Quench phase shift as a means to directly probe environmental density of states}

While the above discussion applies to the most general quench scenario, we will often be interested in cases where the quench Hamiltonian is static in the lab frame once the sensing protocol starts.  This corresponds to a quench control function $ \eta (t)  = \Theta (t ) \Theta ( t_f - t ) $.  This is the case for the specific example situation in Eq.~\eqref{eq:H.beff.sum}. As we will show in Eqs.~\eqref{eq:bath.ini.eq.gen} and \eqref{eq:quc.op.gen}, this also encompasses the case of more general forms of quench operator $\hat{V}$ beyond Eq.~\eqref{eq:H.beff.sum}, corresponding to a wide number of $T_2$-based sensing protocols with generic initial bath states. 

For the above cases, the QPS can be further recast in a form that only involves the imaginary part of response function $\text{Im}  G ^R _{\xi V } [ \omega ] $. For spin-echo control pulses satisfying $  F[ 0 ]  =  0 $, 
the expression for the QPS further simplifies (see App.~\ref{appsec:QPS.Imchi.gen}) 
\begin{align}
\label{eq:Phi.static.eta}
& \Phi _{\mathrm{q}} (t_f) = - \int _{-\infty }^{ + \infty } \!
    \frac{d  \omega }{  \pi \omega }
	 \text{Re} F [ \omega ] \,
	 \text{Im}  G ^R _{\xi V } [ \omega ].
\end{align}  
For a  general pulse sequences, $ F [ \omega ] $ above should be replaced by $ F [ \omega ] - F [ 0 ] $. 

Equation~(\ref{eq:Phi.static.eta}) becomes even more revealing in cases like our example of Eq.~\eqref{eq:H.beff.sum}, where the quench operator $\hat{V}$ is proportional to the noise operator $\hat \xi$, $\hat{V} = \beta _V \hat{\xi}$.  We can thus write:
\begin{equation}
    \Phi _{\mathrm{q}} (t_f) =  \beta _V \int _{-\infty }^{ + \infty } \!
    \frac{d  \omega }{  \omega }
	 \text{Re} F [ \omega ] \,
	 \mathcal{J}[\omega],
	 \label{eq:MainQPSResult}
\end{equation}
where we have introduced the environmental spectral function
\begin{equation}
\label{eq:spec.func.def}
\mathcal{J}[\omega] = - \frac{1 }{\pi} 
\text{Im} G^R_{\xi \xi}[\omega], 
\end{equation}
which determines the dissipative response of the environment, and also plays the role of an effective density of states (DOS).  Eq.~(\ref{eq:MainQPSResult})  shows that the quench phase shift provides a direct route to learning about properties of the environmental spectral function, a quantity that plays an important role both in quantum noise theory and in various areas of many body physics. Note that a related expression was derived in Ref.~\cite{Viola2017} (though this work did not consider the more general situations analyzed here, see e.g.~Eqs.~(\ref{eq:Phi.Gau.gen}) and (\ref{eq:Phi.static.eta})).

While the importance and utility of $\mathcal{J}[\omega]$ is clear in many contexts, it is useful to provide a simple but ubiquitous example.    Consider an environment comprised of independent bosonic modes $ b_k $ with 
$ H_{\mathrm{b},\downarrow} 
= \sum_k \Omega_k  
\hat b^\dag_k \hat b_k$ and a noise operator
$ \hat \xi (t) 
= \sum_k g_k e^{  i \Omega_k t }
\hat b^\dag_k + \text{H.c.}  $.
In this case, we have $\mathcal{J}[\omega] =   \sum_k g^2_k
\delta( \omega - \Omega_k ) $:  it is indeed a weighted DOS, with each mode's contribution weighted by its coupling constant.  Further, in this simple bosonic case $\mathcal{J}[\omega]$ is completely independent of the environmental state.


The above example highlights a general fact:  to understand whether a large bath noise spectral density $\bar{S}[\omega]$ (as revealed by a standard QNS measurement) is due to a large bath DOS or a large mode-occupancy (i.e.~temperature), one needs to also know the spectral function $\mathcal{J}[\omega]$.  As such, the information provided by the QPS provides crucial additional information which {\it complements} information provided by the dephasing factor.  To see this explicitly, consider the case where the initial bath state $\hat{\rho}_{b,i}$ is in thermal equilibrium at temperature $T$.  In this case, the quantum fluctuation-dissipation theorem (FDT) yields~\cite{Welton1951,Kubo1966}
\begin{align}
\label{eq:DFR.gen}
\bar{S} [ \omega  ]
	=  \pi \mathcal{J}[\omega] 
	\coth \frac{ \omega }{2 k_B T }. 
\end{align} 
This relation suggests something we will investigate in detail further:  if one knows {\it both} the noise spectrum and spectral function at a given frequency, one can extract (in a parameter free manner) the environmental temperature.  A standard dephasing-based QNS measurement does not provide sufficient information for such an extraction.  Only the extra information provided by the QPS makes this possible.  Note that our focus here is on dephasing-type couplings between a sensor qubit and the environment.  If one instead had a transverse coupling, then an extended version of $T_1$ relaxometry could also be used in principle to extract $\mathcal{J}[\omega]$, see App.~\ref{appsec:T1Relaxometry}.

Finally, we point out that the above characterization is useful even in more general situations where the initial bath state is not in thermal equilibrium.  In that case, the FDT relation in Eq.~(\ref{eq:DFR.gen}) can be used to define (at each frequency) an effective temperature $T_{\mathrm{eff} }[\omega]$ (see, e.g.,~\cite{Kurchan2005,QN.RMP2010,Cugliandolo2011}).  The fact that this quantity varies as a function of frequency would then be direct evidence of an initial nonequilibrium bath state.  This is also the kind of nontrivial information that can be addressed in a standard $T_2$-style QNS protocol using the extra information provided by the QPS.

\section{Using the quench phase shift to probe low-frequency environmental properties and non-thermal states}

\begin{figure}[t]
  \includegraphics[width=85mm]{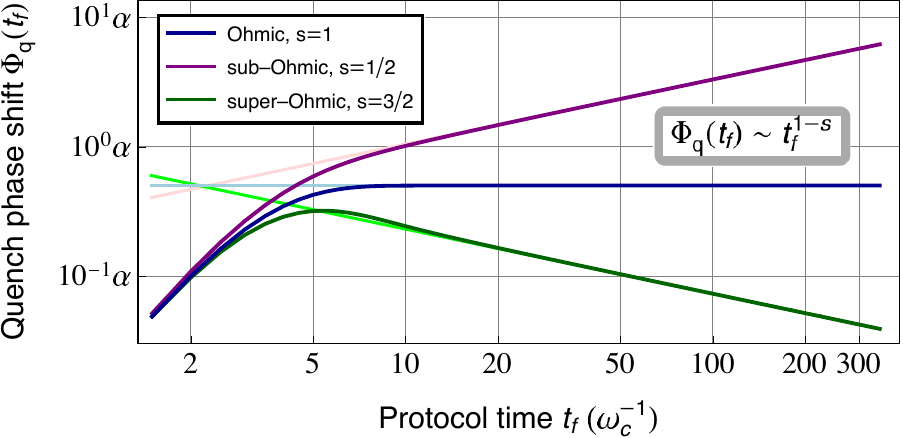}
      \caption{
      Quench phase shift (QPS) $ \Phi  _{\mathrm{q}} (t_f) $ of a sensor qubit coupled to a Gaussian quantum environment, 
     acquired during a Hahn echo sequence, as a function of protocol time $t_f$.
     We assume the environment has spectral function which behaves as a power law at low frequencies,  i.e. 
      $ \mathcal{J}[\omega] 
	= (\alpha/\pi)  \omega_\mathrm{c}
	\left (  \omega / \omega_\mathrm{c} \right ) ^{s}
	e^{ - \left (  \omega / \omega_\mathrm{c} \right ) ^2}
	$.  Curves correspond to different power laws:  Ohmic ($ s= 1 $, dark blue curve), sub-Ohmic  
	($ s= 1/2 $, purple curve), and super-Ohmic ($ s= 3/2 $, dark green curve).  We see that the QPS is extremely sensitive to the spectral function power law $s$. 
	Light-colored lines depict the asymptotic long-time dependence of the QPS 
	$ \Phi  _{\mathrm{q}} (t_f) \sim t_f ^{1-s} $ (see also Eq.~\eqref{eq:QPS.pl.asymp}), which shows excellent agreement with the exact results in the long-time regime 
	$ t_f \gg \omega _\mathrm{c}^{-1} $, as expected. 
	Note that for Gaussian, bosonic environments, the QPS is independent of temperature. 
    }
    \label{fig:Phi.powerlaw}
\end{figure}

\subsection{Estimation protocol}

\label{sec:QPS.est.powerlaw}

As an example of its utility, we show here how the quench phase shift can be used to extract low-frequency environmental properties, going beyond what could be done by studying the dephasing factor alone.  This information directly allows one to determine if the bath is in thermal equilibrium, i.e.~whether the FDT relation of Eq.~(\ref{eq:DFR.gen}) is violated.  In the case where the bath is in equilibrium, it provides a direct means to extract the environmental temperature.  While the estimation protocol we discuss here applies to general quench operators, for concreteness we focus on the specific quench configuration in Eq.~\eqref{eq:H.beff.sum}, where $ \hat{V} (t) = \hat{\xi} (t) /2 $.  The protocol applies essentially the same way for more general situations as long as the lab-frame quench operator is static during the protocol   
(i.e.~Eq.~\eqref{eq:Phi.static.eta} must hold). 

Our focus here is on a very generic scenario where both the environmental symmetrized noise spectrum and spectral function exhibit power-law behavior at low-frequency limit:
\begin{subequations}
\label{eq:sum.spec.powerlaw}
\begin{align}
\bar{S} [ \omega  ] 
& \sim  
S_0 \omega ^{ p } \quad
( \omega \to 0^+ ),
\\
\label{eq:ImG.powerlaw}
\mathcal{J}[\omega]
	& \sim  
	\frac{A_0}{\pi} 
 \omega ^s \quad 
( \omega \to 0^+ ) 
    .
\end{align} 
\end{subequations} 
Note this includes the case where these quantities tend to a constant at $\omega = 0$.
Note also that even if one or both of the exponents 
$ p $, $ s $ are negative, Eqs.~\eqref{eq:sum.spec.powerlaw} can still describe a physical bath, as long as one also introduces a low-frequency IR cutoff.   
We thus have four parameters characterizing the low-frequency features of the environment.  
Recall that a general, non-thermal environment can always be characterized by a  frequency-dependent effective temperature $T_{\mathrm{eff}} [\omega]$.  Using the asymptotic forms above, we have in the low-frequency limit:  
\begin{equation}
    T_{\mathrm{eff}} [\omega] \sim 
    \frac{S_0}{ 2 k_B A_0 }  \omega ^{ p +1 -s } 
    \label{eq:Teff}
\end{equation}
If the environment is in thermal equilibrium then $T_{\mathrm{eff}} [\omega] $ will be frequency independent and equal to the bath temperature.  We see this requires $p = s-1$.

Our goal is thus to estimate the power-law exponents 
$  p $, $ s $, and overall coefficients 
$ S_0 $, $ A_0 $ from the sensor qubit dynamics.  As we now show, this can be achieved by looking at both the phase and magnitude of the qubit coherence in the long-time limit.  As long as the asymptotic power-law dependence of bath NSD (response function) does not exhibit too strong a low-frequency divergence, the asymptotic long-time behavior of the dephasing function 
$ \zeta  (t_f) $ (QPS 
$ \Phi  _{\mathrm{q}} (t_f) $) under any specific spin-echo or dynamical-decoupling pulse becomes independent of details about the cutoff, and is solely determined by the low-frequency asymptotic behavior of NSD (spectral function) in Eq.~\eqref{eq:sum.spec.powerlaw}.  
The needed conditions are satisfied by most physical environments (including, e.g., Ohmic baths and baths producing $1/f$ noise)

Using Eqs.~\eqref{eq:dep.Gau.gen} and \eqref{eq:Phi.static.eta} we can rigorously show (see App.~\ref{appsec:QPS.asymp.deriv})
\begin{subequations}
\label{eq:tot.pl.asymp}
\begin{align}
\label{eq:dep.pl.asymp}
\zeta  (t_f) 
& \sim 
\mathcal{C} _{ \zeta } 
S_0  t_f ^{ 1- p } \quad 
(  t_f \to + \infty ,
-3 < p <1 )  ,
    \\
\label{eq:QPS.pl.asymp}
\Phi  _{\mathrm{q}} (t_f) 
& \sim 
  \mathcal{C} _{\Phi} 
 A_0 
 t_f ^{1-s } \quad 
( t_f \to + \infty ,
-2< s <2 )
, 
\end{align}
\end{subequations}
where $ \mathcal{C} _{ \zeta }  $ and 
$ \mathcal{C} _{\Phi}  $ are nonzero dimensionless coefficients determined by details of the qubit control pulse.
Eqs.~\eqref{eq:tot.pl.asymp} are valid for some of the most common types of physical environments, including Ohmic baths ($  p=0 $, $ s=1 $) and baths generating $1/f$ noise ($  p=-1 $, $ s=0 $). 
Comparing against Eq.~(\ref{eq:Teff}), we see that the combined information in the dephasing function and quench phase shift is exactly what is needed to characterize the effective temperature $T_{\mathrm{eff}} [\omega]$.  If this quantity is frequency-dependent, the bath is not in a thermal state.  Note that for exponents $  p $ and $ s $ falling out of the range of validity given in Eq.~\eqref{eq:tot.pl.asymp}, the long-time regime of qubit dynamics would also be sensitive to details of the cutoff, but it would still be possible to extract information about the bath NSD (response function) from the dephasing function 
$ \zeta  (t_f) $ (QPS 
$ \Phi  _{\mathrm{q}} (t_f) $) using parametric spectral estimation techniques.

The asymptotic result in Eq.~\eqref{eq:dep.pl.asymp} for dephasing is well established~\cite{Schoen2006,DasSarma2008b} and has been utilized for QNS in various experimental platforms~\cite{Hanson2010,Yacoby2013,Dzurak2018}. 
The corresponding result for the quench phase shift in Eq.~\eqref{eq:QPS.pl.asymp} provides complementary information, on the properties of the spectral function. 
We stress that to assess whether the bath is in equilibrium, and if so what the temperature is, both these quantities are needed. 
In Fig.~\ref{fig:Phi.powerlaw}, we show the evolution of the quench phase shift for a simple Hahn echo pulse sequence; curves correspond to Gaussian Ohmic, sub- and super-Ohmic baths with Gaussian cutoffs, where
$ s = 1$, $\frac{1}{2}$, $ \frac{3}{2}$ respectively. As expected, the exact QPS is accurately described by the asymptotic power-law function in the long-time regime.  For Hahn echo, the constants appearing in Eqs.~(\ref{eq:tot.pl.asymp}) are given by
$ \mathcal{C} _{ \zeta, \mathrm{H} } 
= \frac{ 1 - 2 ^{ p +1 }  }{\pi}
	\Gamma  \left ( { p -1} \right )
	\sin \frac{  p  \pi }{2}   $ and 
$ \mathcal{C} _{\Phi, \mathrm{H} }
=  \frac{ 1- 2 ^{ s } }{\pi}
	\Gamma  \left ( {s-1} \right )
	\cos \frac{  s \pi }{2}  $, where 
$ \Gamma (\cdot ) $ is the gamma function.

\begin{figure}[t]
  \includegraphics[width=85mm]{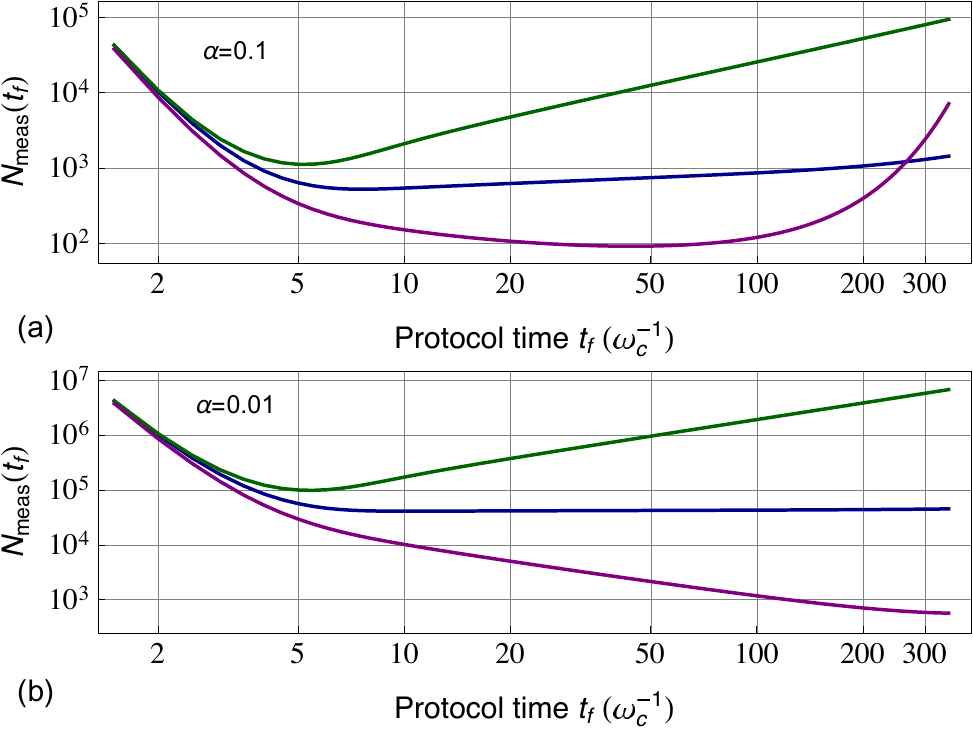}
      \caption{ 
     Because of bath induced dephasing, it will in general take many repeated measurements to resolve the quench phase shift (QPS).  Plotted here is the 
      minimum number of measurements $N_{\mathrm{meas}}$ needed to resolve the quench phase shift with a unit signal-to-noise ratio in a Hahn echo protocol of time $t_f$:  $ N _\mathrm{meas}  ( t_f  )  \equiv
    \left |\langle 
	\hat \sigma _y ( t_f  ) 
	\rangle  \right | ^{-2} $.  We use the same Gaussian baths (and labelling) as in Fig~\ref{fig:Phi.powerlaw}.
	Panel (a) corresponds to a dimensionless coupling parameter $ \alpha = 0.1 $, while (b) corresponds to $ \alpha = 0.01 $.  
	All plots correspond to a temperature $ k _B T  =  0.01 
	\omega _\mathrm{c} $.    }
    \label{fig:Nmeas.powerlaw}
\end{figure}

\subsection{Measuring the quench phase shift}

We have shown that the long-time properties of the sensor qubit coherence (both its magnitude and phase) reveal key features of our environment.  This sensing modality of course has a natural tension:  in the long-time limit, the loss of qubit coherence described by Eq.~(\ref{eq:dep.Gau.gen}) will make it difficult to resolve the quench phase shift (c.f.~Eq.~(\ref{eq:Phi.Gau.gen})).  This is not a fundamental problem, but necessitates sufficient averaging, i.e.~repeated evolutions and measurements of the sensor qubit under the chosen pulse protocol.  In what follows, we characterize the amount of averaging needed for given environmental parameters.

For convenience, in what follows we express the coefficient $A_0$ in Eq.~(\ref{eq:ImG.powerlaw}) as $ A_0 = \alpha  \omega_\mathrm{c} ^{1-s}$, 
i.e. the product of a dimensionless parameter 
$\alpha $ quantifying the qubit-bath coupling strength, and powers of a UV-cutoff frequency scale
$ \omega_\mathrm{c} $ characterizing the regime where Eq.~\eqref{eq:ImG.powerlaw} is valid. 
In the weak coupling limit $\alpha \ll 1$, we can calculate the number of repeated measurements required to achieve a unit signal-to-noise ratio (SNR)
for the measurement of the quench phase shift.
Focusing only on fundamental projection noise, this is given (as is standard) by the squared inverse norm of the qubit coherence signal~\cite{Cappellaro2017}, 
$ N _\mathrm{meas}  ( t_f  )  =
\left |\langle 
	\hat \sigma _y ( t_f  ) 
	\rangle  \right | ^{-2} $.  
This figure-of-merit is plotted in Fig.~\ref{fig:Nmeas.powerlaw} for weakly-coupled baths with different 
spectral functions $\mathcal{J}[\omega]$. 
Note that in many experimentally-relevant situations, the effective environment temperature scale is much lower than the UV energy scale, i.e. 
$ k _B T _\mathrm{eff} \ll 
\omega _\mathrm{c} $.  As a result, measuring the long-time quench phase shift is within reach of state-of-the-art systems realizing QNS.

\subsection{Direct thermometry for an Ohmic bath}
\label{subsec:OhmicThermometry}

In this subsection, we specialize to the case of an environment that is approximately Ohmic at low frequencies, i.e.~the low-frequency spectral function $\mathcal{J}[\omega]$ is proportional to frequency.  As we now show, this sole assumption allows one to directly extract the environmental temperature via simple measurements that require no curve fitting.    

When in thermal equilibrium, an Ohmic environment has a flat NSD at low frequencies, i.e.
$\bar{S}[\omega] \sim 2 A_0 k_B T$, 
c.f.~Eq.~\eqref{eq:sum.spec.powerlaw}.  A measurement of the low-frequency NSD alone only yields the product of $A_0$ and $T$, and hence does not permit direct thermometry.
Luckily, the missing information (i.e.~the value of the coupling constant $A_0$) is directly provided by the long-time limit quench phase shift.  
Defining 
$\Phi_{\mathrm{q}} (\infty)  \equiv 
\lim_{ t_f \to + \infty } 
\Phi  _{\mathrm{q}} (t_f) $, we find from Eq.~\eqref{eq:QPS.pl.asymp}: 
\begin{align}
    \Phi_{\mathrm{q}} (\infty) 
     & = 
    \frac{ \pi }{2} 
	\left . \frac{ d \mathcal{J}[\omega] }{d \omega }  
	\right|_{ \omega =0 }
	= \frac{1}{2} A_0
	.
	    \label{eq:QPS.Ohmic.asymp}
\end{align}
See App.~\ref{appsec:QPS.Ohmic.asymp} for an alternative, intuitive derivation of this expression.

Given this simple result, one can now {\it directly} extract the environment temperature.  Using the fact that for a Hahn echo sequence, the
$T_2$ decoherence time is given by $T_2 = 2/\bar{S} [  0 ] $ \cite{Cappellaro2017}, we obtain:
\begin{align}
\label{eq:thermo.Ohmic}
k_B T =  
    \frac{  1 }
    {2 T_2 \, \Phi_{\mathrm{q}} (\infty)}  
    . 
\end{align} 
The upshot is that for a thermal, Ohmic environment, simply measuring the Hahn-echo $T_2$ and the long-time quench phase shift directly yields the environmental temperature.  We stress that this does not require any curve fitting, nor further assumptions.
Note that our protocol is also applicable if in addition to low-frequency Ohmic noise, we also have large quasistatic noise; see App.~\ref{appsec:OhmThermo.NSD} for detail.  This is a common scenario in many systems. 

\begin{figure}[t]
  \includegraphics[width=85mm]{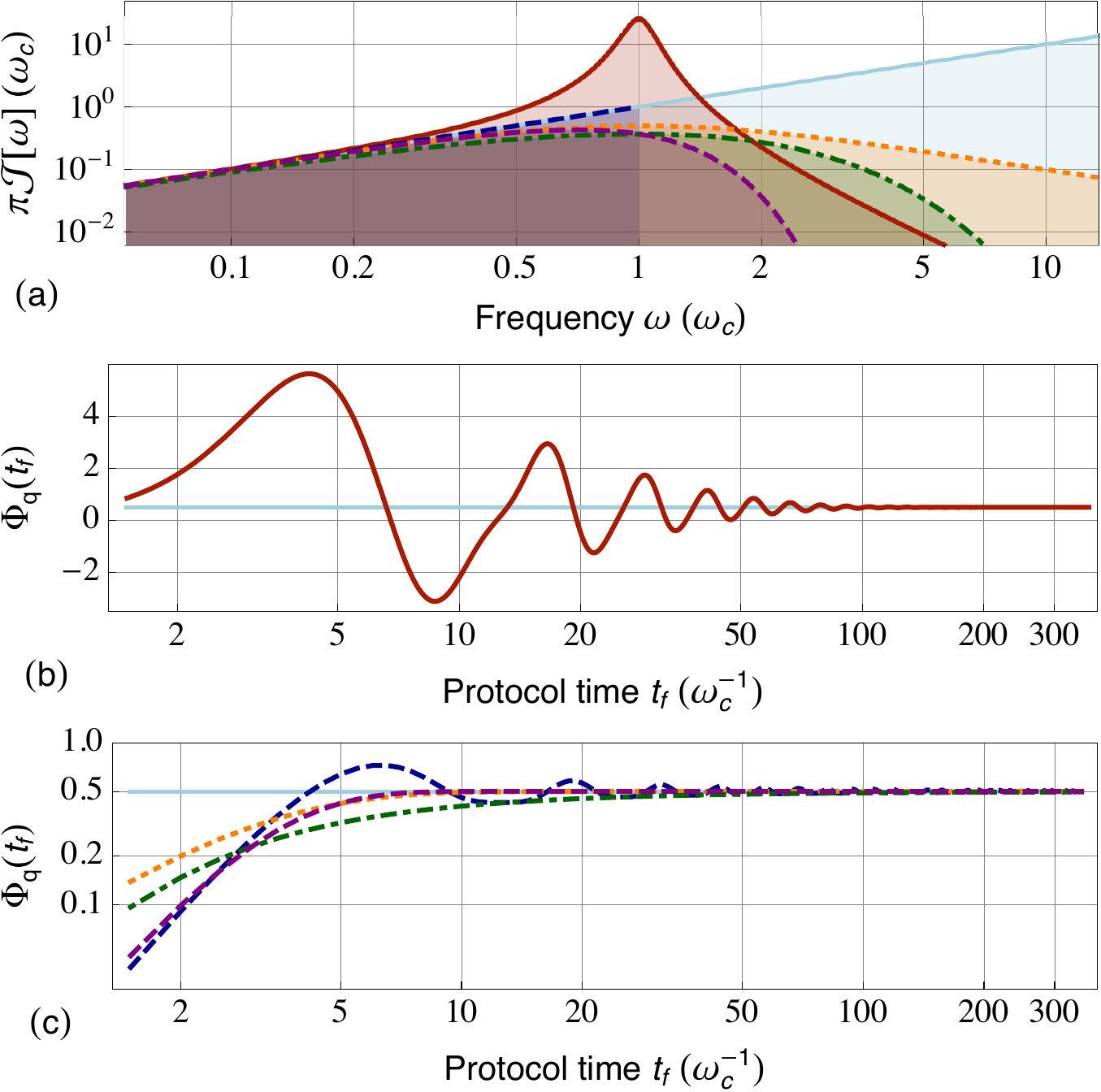}
      \caption{
      For practical applications, the crossover timescale at which the Hahn-echo quench phase shift (QPS) 
      $ \Phi  _{\mathrm{q}}  (t_f) $ approaches the asymptotic power law behavior (c.f.~Eq.~\eqref{eq:QPS.pl.asymp}) becomes important. This timescale depends on high-frequency deviations in the bath spectral function $\mathcal{J}[\omega]$ from power law; plotted here are two generic scenarios for this crossover dynamics.  
      As shown in (a), we consider bath spectral functions that are asymptotically Ohmic (i.e.,~proportional to frequency) at low frequencies. (b) Crossover dynamics for QPS with spectral function exhibiting a narrow Lorentzian peak in the range of high frequencies, i.e., 
      $\mathcal{J}[\omega] \sim 
      \alpha  \omega_\mathrm{c} ^3 / 4 \pi  
      [ \left (  \omega - \omega_\mathrm{c} \right )^2 + \Gamma _\mathrm{min} ^2 ] $ for $ \omega \simeq \omega_\mathrm{c} $ (red solid curve in (a); see Sec.~\ref{subsec:OhmicThermometry} for exact form of $\mathcal{J}[\omega] $). For narrow peak with linewidth $  \Gamma _\mathrm{min} / \omega_\mathrm{c} <  1 $, the crossover timescale is given by $ \Gamma _\mathrm{min} ^{-1} $. Parameters: $\alpha =1$,  $ \Gamma _\mathrm{min} / \omega_\mathrm{c}= 0.1 $. 
      (c) QPS for spectral functions that are Ohmic at low frequencies with a simple UV cutoff. In this case, the crossover time is given by inverse of the cutoff frequency 
      $  \omega_\mathrm{c}  ^{-1} $, and is independent of details about the cutoff. 
      Cutoffs used (see Sec.~\ref{subsec:OhmicThermometry} for specific forms of $\mathcal{J}[\omega] $): step-function (dashed blue), Lorentzian  
(dotted orange), exponential  
(dot-dashed green), 
and Gaussian cutoff  (dashed purple curve).
    }
    \label{fig:QPS.Ohmic}
\end{figure}

While Eq.~(\ref{eq:thermo.Ohmic}) is exact, it is also useful to understand how long one must wait to achieve the asymptotic long-time limit of the QPS.  
The answer to this question depends on features in the spectral function away from $\omega = 0$.  For convenience, in following discussion we rewrite the spectral function as $ \mathcal{J}[\omega]  
= ( \alpha  /\pi)  \omega \phi
\left (  \omega / \omega_\mathrm{c} \right ) $, where 
$ \phi \! \left ( \cdot  \right )$ 
encodes high-frequency dependence of the spectral function.   As illustrated in Fig.~\ref{fig:QPS.Ohmic}, there are two possible scenarios for the crossover dynamics of QPS. 
First, if the spectral function $\mathcal{J}[\omega]$ exhibits narrow peak(s) in the high-frequency regime, then the crossover timescale is given by 
$ \Gamma _\mathrm{min}  ^{-1} $, where $ \Gamma _\mathrm{min} < \omega_\mathrm{c}$ is the smallest linewidth of these peaked features.  This is shown in Fig.~\ref{fig:QPS.Ohmic}(b), where the corresponding spectral function exhibits a high-frequency narrow Lorentzian peak with linewidth 
$  \Gamma _\mathrm{min} = \epsilon \omega_\mathrm{c} = 0.1 \omega_\mathrm{c} $, as encoded by  
$ \phi \! \left ( x  \right )
      = (1 + \epsilon ^2 ) ^2 / 
      [ ( x -1 ) ^2 + \epsilon ^2]
      [ ( x +1 ) ^2 + \epsilon ^2]
      $.  
The second generic case is where there are no such sharp features at high frequencies, and only a smooth cutoff in $\mathcal{J}[\omega]$ characterized by the UV cutoff frequency $\omega _\mathrm{c} $.
In this case, the timescale for the QPS to saturate is $1/\omega _\mathrm{c} $ and independent of specific details of the form of the cutoff. This is confirmed in Fig.~\ref{fig:QPS.Ohmic}(c), where QPS crossover dynamics is plotted for step-function cutoff 
$ \phi \! \left ( x  \right )  = \Theta ( 1-x ) $ (dashed blue), Lorentzian cutoff 
$ \phi \! \left ( x  \right )  =  1 / ( 1 + x ^2 ) $ (dotted orange), exponential cutoff 
$ \phi \! \left ( x  \right )  = e ^{-x} $ (dot-dashed green), 
and Gaussian cutoff $ \phi \! \left ( x  \right )  = e ^{-x ^2 } $ (dashed purple curve), respectively.

\section{Generalized quenches based on arbitrary initial bath states}

\label{sec:V.gen.op}

Our discussion so far has focused on ``incidental'' environment quenches occurring during a generic QNS sensing protocol; for the most part, we considered a specific scenario where before the protocol starts, the environment is in the initial state described by Eq.~\eqref{eq:ini.bath.down}.  We now show that the basic quench physics we have described (and its impact on the sensor qubit) applied to a far wider set of circumstances, where the bath starts in an {\it arbitrary} initial state $ \hat{\rho}_{\mathrm{b,i}} $.  This provides an entire new modality for sensing:  one could deliberately prepare the environment in an interesting target state before the start of the sensing sequence, and then use the resulting quench physics (namely the influence on the sensor qubit's phase) to probe the environment.  

The simplest generalization is when the environment is initially in a thermal state corresponding to some arbitrary (bath-only) Hamiltonian $  \hat H_{\mathrm{b,i}} $:  
\begin{align}
\label{eq:bath.ini.eq.gen}
& \hat{\rho}_{\mathrm{b,i}}
= e^{-  \hat H_{\mathrm{b,i}}
   / k_B T }
/ { Z _T } ,
\end{align}
In this case, we can directly use Eq.~(\ref{eq:Heff.bath.ini}) to define our quench, and identify the quench operator $\hat{V}$ via Eq.~\eqref{eq:V.bath.def}.  
We stress that in this more general case, the initial bath Hamiltonian $ \hat H_{\mathrm{b,i}} $ need not have any simple relation to the qubit-conditioned bath Hamiltonians 
$ \hat H_{\mathrm{b},\uparrow(\downarrow) }$ appearing in Eq.~(\ref{eq:Htot.deph.gen}).
As a result, the quench operator $ \hat V $ will now be {\it independent} of the noise operator $\hat{\xi}
\equiv \hat H_{\mathrm{b},\uparrow} - \hat H_{\mathrm{b},\downarrow} $.
For systems where it is possible to initialize the environment in different initial equilibrium states, this provides a powerful new way to probe the environment:  different initial states yield different quenches, and hence different quench phase shifts via Eq.~(\ref{eq:Phi.static.eta}).

An even more general scenario is when the bath starts in an arbitrary {\it non-thermal equilibrium} initial state 
$ \hat{\rho}_{\mathrm{b,i}} $ that has no simple relation to a static Hamiltonian.  This could be achieved in numerous ways, e.g.~by explicitly driving the bath~\cite{Walsworth2018,Jayich2019}.  As we have stressed repeatedly, our general quench mechanism is ultimately controlled by the initial {\it state} $\hat{\rho}_{\mathrm{b,i}}$ of the environment.  A quench will occur as part of our $T_2$-style QNS protocol any time 
\begin{align}
    \label{eq:noncomm.ini.Heff}
    [ \hat{\rho}_{\mathrm{b,i}}
    ,
    \hat H _{\mathrm{b,eff} }  ( t  )  ]
    \ne 0, \quad 
    0 < t < t_f .
\end{align}
In cases where this state was thermal, this $\hat{\rho}_{\mathrm{b,i}}$ could easily be related to an initial bath Hamiltonian $\hat H_{\mathrm{b,i}}$, which we then used to identify the quench operator $\hat{V}$ in Eq.~\eqref{eq:V.bath.def}.  In contrast, for our more general case, there is no unique way to identify $\hat H_{\mathrm{b,i}}$.  In general, we may choose any bath Hamiltonian compatible with the initial bath state, i.e. satisfying
$  [\hat H_{\mathrm{b,i}}, \hat{\rho}_{\mathrm{b,i}} ] =0 $.  The choice of $\hat H_{\mathrm{b,i}}$ would then determine $\hat{V}$.  We stress that this seeming ambiguity is only a choice of bookkeeping:  the actual evolution of the sensor qubit (and the quench phase shift) is of course only determined by 
$\hat{\rho}_{\mathrm{b,i}}$ (see App.~\ref{appsec:qubit.dyn.gen}).

Given these caveats, we now present a simple (though non-unique) method to usefully parametrize the quench in the most general case.  We define the initial bath Hamiltonian 
$ \hat H' _{\mathrm{b,i}}   $ and quench operator 
$ \hat V' $  as the ``longitudinal'' (maximally commuting) and ``transverse'' (minimally non-commuting) components of the effective bath Hamiltonian with respect to 
$ \hat{\rho}_{\mathrm{b,i}} $. 
We can make this prescription explicit by first diagonalizing the initial bath state as  $  \hat{\rho}_{\mathrm{b,i}} 
= \sum _{n= 0 }^{N} p _n \hat P _n  $.  Here the eigenvalues $p_n$ are distinct with $ p_0 = 0$, and $\hat{P}_n$ is the projector onto the eigenspace corresponding to $p_n$~\cite{Sakurai2011book}.  
The initial bath Hamiltonian 
$ \hat H' _{\mathrm{b,i}}   $ and the quench 
$ \hat V' $ can now be defined as 
\begin{subequations}
\label{eq:Hquc.gen.ini.sum}
\begin{align}
& \hat H' _{\mathrm{b,i}}   
\equiv \sum _{ n=0 }^N  
\hat P _n 
( \hat H_{\mathrm{b},\uparrow} + 
\hat H_{\mathrm{b},\downarrow} ) 
\hat P _n  /2 
, \\ 
\label{eq:quc.op.gen}
& \hat V'  \equiv \sum_{m,n=0; \,
m \ne n }^N 
\hat P _m
( \hat H_{\mathrm{b},\uparrow} + 
\hat H_{\mathrm{b},\downarrow} ) 
\hat P _n /2.
\end{align}
\end{subequations}

For concreteness, we provide an example of this procedure for a bosonic bath that couples linearly to the sensor qubit:
\begin{subequations}
\begin{align}
& \hat H_{\mathrm{b},\downarrow} 
=  \sum_k \Omega_k  
\hat b^\dag_k \hat b_k 
, \\ 
& \hat H_{\mathrm{b},\uparrow}
=  \hat H_{\mathrm{b},\downarrow} 
+ \sum_k \left ( g_k  
\hat b^\dag_k + \text{H.c.} 
\right).
\end{align}
\end{subequations} 
We also assume that the initial bath state is not a thermal state, but a squeezed thermal state.  As a result, the initial state of the sensor and bath is given by
\begin{subequations}
\begin{align}
& \hat{\rho}_{\mathrm{tot}} (t = 0^-)
= | \! \downarrow \rangle \langle \downarrow \! | 
\otimes
\hat{\rho}_{\mathrm{b,i}}
    , \\
& \hat{\rho}_{\mathrm{b,i}}
= \hat{ S } ( \vec{r} ) \, 
 e ^{-  \hat H_{\mathrm{b,\downarrow}}
   / k_B T }
 \hat{ S } ^\dag ( \vec{r} ) 
/ { Z _T } 
, 
\end{align}
\end{subequations}
where 
$ \hat{ S } ( \vec{r} ) 
\equiv \exp ( \sum_k r_k 
\hat b_k ^2 /2 -  \text{H.c.} )$ denotes the squeezing operator, with real constants $r_k$ the corresponding mode squeezing parameters~\cite{Gardiner2004book}.

Using our above prescription, we find that initial bath Hamiltonian 
$ \hat H' _{\mathrm{b,i}} $ and the quench operator
$ \hat V' $ in Eq.~\eqref{eq:Hquc.gen.ini.sum} are given by  
\begin{subequations}
\begin{align}
\hat H' _{\mathrm{b,i}}   
    = & \sum_k \Omega_k  
\left [ 
\hat b^\dag_k \hat b_k 
 \cosh ^2 2 r_k  
 + \frac{ \sinh  4 r_k  }{4}   \left (  
    \hat b ^{\dag 2}_k 
    + \text{H.c.} 
    \right)
\right] 
    ,\\
    \label{eq:Vp.example.sq}
\hat V'  
    = & - \sum_k \Omega_k  
    \hat b^\dag_k \hat b_k 
    \sinh ^2 2 r_k  
     \nonumber \\
     & +\frac{ 1 }{4}  \sum_k \left ( 
    - \Omega_k  
    \hat b ^{\dag 2}_k  \sinh  4 r_k 
    + 2 g_k   \hat b^\dag_k 
    + \text{H.c.} 
\right)
    . 
\end{align}
\end{subequations} 
While the quench operator $ \hat V' $ in Eq.~\eqref{eq:Vp.example.sq} still contains an incidental contribution which depends on qubit-bath couplings $ g_k $, it also includes \textit{deliberate} quenches that can be tuned via the initial squeezing parameters $ r_k $. We again note that it is not the only way to introduce a quench operator 
$ \hat V $ satisfying Eq.~\eqref{eq:qb.dyn.gen}. However, this convention is useful for understanding effects on qubit dynamics due to the quench.


\section{Generalized quenches for frequency-space reconstruction of response functions}
\label{sec:gen.quench.spec}

In this section, we restrict attention to situations where the quench (whether intentional or accidental) yields a quench operator $\hat{V}$ (c.f.~Eq.(\ref{eq:Heff.bath.ini})) which commutes with the noise operator $\xi $ (Eq.~(\ref{eq:HTotIntro})).  In the simple and standard case where the quench temporal function
$\eta(t)$ is a step function (c.f.~Eq.~(\ref{eq:EtaSimple})), we showed in Sec.~\ref{sec:QPS.est.powerlaw}  that the quench phase shift can be used to extract the low-frequency properties of the environment's spectral function $\mathcal{J}[\omega]$ (i.e.~response function).  A natural question is to ask whether it is possible to perform a complete reconstruction of $\mathcal{J}[\omega]$ in some finite bandwidth window.  This would be then analogous to spectral reconstruction techniques used in conventional QNS measurements to reconstruct $\bar{S}[\omega]$.

It is worth noting that for the specific QPS given by Eq.~\eqref{eq:MainQPSResult}, Ref.~\cite{Viola2017} has proven a no-go theorem, which prevents systematical reconstructions of spectral function $\mathcal{J}[\omega]$ using the restricted form of quenches in Eq.~\eqref{eq:H.beff.sum}.  
Here we are interested in a more general question: can we utilize quenches with a more complex time-dependence, as encoded in quench function $ \eta (t) $, to overcome the limitation set by aforementioned no-go theorem? Indeed, as we show in App.~\ref{appsec:QPS.recon.specfunc}, the extra tunability in the quench function allows us to use the more general form of QPS in Eq.~\eqref{eq:Phi.Gau.gen} and reconstruct $\mathcal{J}[\omega]$ in a generic target frequency range.

The protocol we introduce below makes use of a generic structure, where the sensor qubit is controllably embedded in a multi-level system.  While this can be realized in many different experimental platforms (e.g.~a superconducting transmon qubit, as implemented in Ref.~\cite{Oliver2021}), we focus here on sensor based on a $S=1$ nitrogen-vacancy (NV) defect in diamond. For this system, we discuss a specific protocol to reconstruct finite-frequency spectral function $\mathcal{J}[\omega]$ by engineering time-dependent quenches.  We stress that our strategy can be used to implement generic forms of quench functions $ \eta (t) $.

\subsection{Realization of a time-dependent quench function using sensor qubits embedded in a spin-$1$ structure}

\label{sec:NV.impls}

We start by showing how to engineer time-dependent quenches using NV centers in diamond. 
NV-based qubits are an ideal candidate to implement $T_2$-style QNS: the spin relaxation timescale $ T_1 $ of NV centers is typically much longer than the dephasing timescale, so that $ \hat  S_z $ is conserved to a great approximation during $ T_2 $-type protocols. 
The dominating dephasing typically comes from coupling to environmental magnetic noise (due to surrounding nuclear spins, etc.); alternatively, this makes them a powerful magnetic sensor.  We can thus write NV-bath Hamiltonian as~\cite{Yang2016}
\begin{align}
\label{eq:H.NVbath}
& \hat H_{\mathrm{NV-bath}} 
= \sum _{ m_z = 0, \pm 1 }
|  m_z \rangle \langle m_z  | 
	 \otimes
	 \hat H _{\mathrm{b},m_z }. 
\end{align}
We will consider the NV-bath coupling to correspond to an effective bath-induced magnetic field $\hat{B}$, which then satisfies
$ \hat{B}  = \hat H _{\mathrm{b}, 0 } - \hat H _{\mathrm{b},-1 } 
= \hat H _{\mathrm{b}, +1  } - \hat H _{\mathrm{b}, 0 } $. 

The most common and straightforward way to experimentally initialize the NV center is via optical illumination, which  prepares it in the $ |  m_z =0 \rangle $ state~\cite{Walsworth2020}. 
Given this specific initial NV center state, the initial bath Hamiltonian 
$  \hat H_{\mathrm{b,i}}  $ in Eq.~\eqref{eq:Heff.bath.ini}  should be replaced by 
$ \hat H _{\mathrm{b}, 0 } $ (i.e., the bath Hamiltonian conditioned on qubit in $ |  m_z =0 \rangle $ state).
Turning to the sensing protocol, it is relatively easy and straightforward to rapidly produce a superposition state using any two of the three $ |  m_z \rangle$ states.  This
provides us then with three different choices for the specific form of the sensor qubit, each corresponding to different effective quench physics.  This is summarized in Table~\ref{tab:NV.quench}: the form of the quench physics given by Eqs.~\eqref{eq:Heff.bath.ini} and \eqref{eq:V.bath.def} can be  controlled or even turned off by choosing the sensor qubit subspace:  $\{ m_z = 0, m_z =  1 \} $, $\{ m_z = 0, m_z = -1 \}$ or $ \{ m_z = +1 , m_z = -1\}$. This extra knob in NV-based qubits can be used to distinguish the quench phase shift effect from other spurious phases due to the environment~\cite{Cywinski2020}. 

\begin{table} 
\caption{\label{tab:NV.quench}
Quench operator dependence on subspaces of NV center used to form the sensor qubit, assuming fixed initial NV state $ |  m_z =0 \rangle  $. The environment is coupled magnetically to the NV spin with    
$ \hat{B}  \equiv \hat H _{\mathrm{b}, 0 } - \hat H _{\mathrm{b},-1 } 
= \hat H _{\mathrm{b}, +1  } - \hat H _{\mathrm{b}, 0 } $ (see Eq.~\eqref{eq:H.NVbath}). } 
\begin{ruledtabular} 
\begin{tabular}{cccccc} 
$ | \! \!  \uparrow \rangle $
& 
$ | \! \!  \downarrow \rangle $ 
& Quench operator 
$ \hat V  $ 
& Noise operator 
$ \hat \xi  $ 
\\
\hline
$ |  m_z =0 \rangle $ 
& $ |  m_z =-1 \rangle $
& $ - \hat{B} /2 $ 
& $ \hat{B} $  
\\ 
$ |  m_z = +1 \rangle $ 
& $ |  m_z = 0 \rangle $
& $ + \hat{B} /2 $
& $ \hat{B} $ 
\\ 
$ |  m_z = +1  \rangle $ 
& $ |  m_z = -1 \rangle $
& 0 
& $ 2 \hat{B} $
\\ 
\end{tabular} 
\end{ruledtabular} 
\end{table}

We can now harness this freedom to generate a powerful new kind of quench protocol.  The basic idea is to engineer a nontrivial time-dependence of the quench function $ \eta (t) $ 
(c.f.~Eq.~(\ref{eq:Heff.bath.ini})) by deliberately switching the sensor spin between the different possible qubit subspaces at prescribed times during the protocol. As we show below, the time dependent $\eta(t)$ generated by this approach can be utilized to generate comb-based filter functions that enable spectral reconstruction of the response function. Figure~\ref{fig:quench.filter} illustrates a concrete example of control pulses that realize such time-dependent quenches: by periodically switching between the $\{ m_z = 0, m_z = -1 \} $ and 
$ \{ m_z = 0, m_z = +1 \}$ qubit subspaces (pulse period 
$ T = t_f/ 2M$) in addition to applying standard qubit-control 
$ \pi $-pulses at $ \ell T / 2 $ ($ \ell = 1,3, \ldots, 4M-1 $), we effectively realize the more general quench in Eq.~\eqref{eq:Heff.bath.ini} with $  \hat H_{\mathrm{b,i}} = \hat H _{\mathrm{b}, 0 }  $ and $ \hat V =  \hat{B} /2 $, whereas the corresponding quench function 
$ \eta (t)  = 
\sum _{ n = 0 } ^{2M-1 } 
(-) ^{n-1}
\Theta (t - nT ) 
\Theta ( nT + T - t ) $ is shown in Fig.~\ref{fig:quench.filter}.


\subsection{Reconstruction of spectral function $\mathcal{J}[\omega]$ via time-dependent quenches}

\label{sec:recon.ImGR}

\begin{figure}[t]
  \includegraphics[width=85mm]{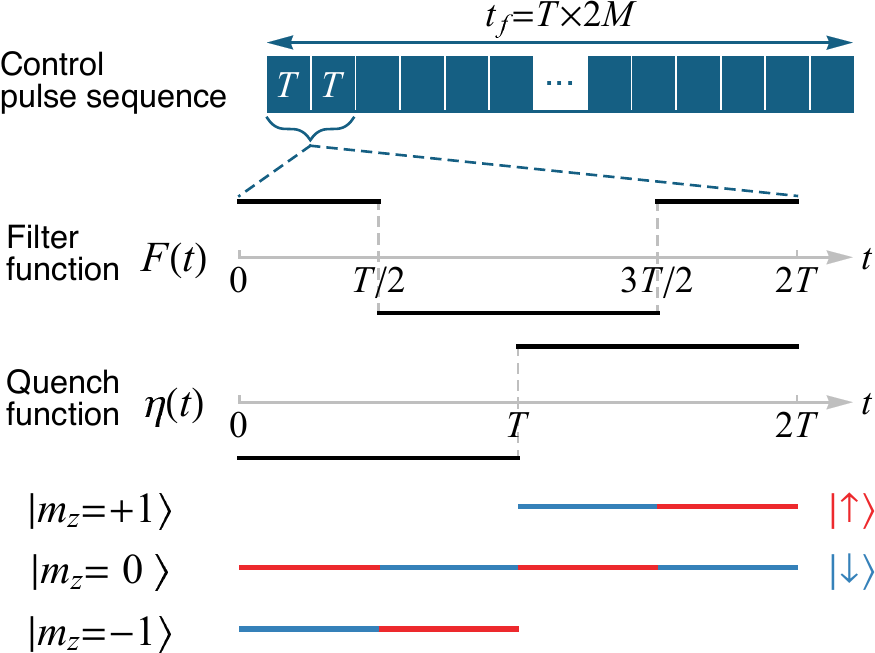}
      \caption{
      Schematic for example NV center control pulse sequence realizing time modulations in both the noise filter function $ F(t) $ and quench function $ \eta (t) $, which in turn enables reconstruction of spectral function $ \mathcal{J}[\omega]  $ (see Eq.~\eqref{eq:eta.comb.sum}).
      The control pulses are periodically structured as $ 2M $ repetitions of a base sequence (period 
$ T = t_f/ 2M$), so that $ F(t) $ and $ \eta (t) $ have period 
$ 2T = t_f/ M$.  The noise filter function $ F(t) $ encodes timings of the standard dynamical decoupling 
      $ \pi $-pulses within the qubit subspace.
      In addition, we can further engineer a modulating quench function 
      $ \eta (t) $ by periodically switching between qubit subspaces of $\{ m_z = 0, m_z = -1 \} $ and 
$ \{ m_z = 0, m_z = +1 \}$ (see also Table~\ref{tab:NV.quench}). The corresponding time-dependent NV levels used as sensor qubit states during the protocol are illustrated in the bottom part of the schematic. 
    }
    \label{fig:quench.filter}
\end{figure}

For the more general control sequence discussed above, the qubit dynamics is again given by Eq.~\eqref{eq:qb.dyn.gen}. The lab-frame noise operator is given by $ \hat \xi =  \hat{B} $; as before, we transform to the toggling frame defined by the standard qubit-control 
$ \pi $-pulses (the control pulses switching between qubit subspaces do not contribute here), with the resulting filter function shown in Fig.~\ref{fig:quench.filter}.

We thus obtain the filter function $ F [\omega] $ and quench function $ \eta [\omega] $ in frequency space as 
\begin{subequations}
\label{eq:eta.filter.sum}
\begin{align}
F [\omega]  &  
=  - \frac{4  }{\omega} 
e^{i \frac{\omega t_f }{2}  } 
\sin \frac{\omega t_f }{2}
\frac{ \sin^2  \frac{\omega   t_f }{ 8 M } }
{  \cos \frac{\omega t_f }{ 4 M }} 
    , \\
\eta [\omega]  &  = 
 \frac{ 2i }{\omega} 
e^{i \frac{\omega t_f }{2}  } 
\sin \frac{\omega t_f }{2}
\tan \frac{\omega   t_f }{ 4 M }
    . 
\end{align}
\end{subequations} 
Substituting above equations into Eq.~\eqref{eq:Phi.Gau.gen}, together they generate a frequency-comb filter function in the large pulse number $ M \gg 1 $ limit, which can be directly used to probe the spectral function $\mathcal{J}[\omega]$ (i.e.~response function). More specifically, noting that $ \hat V =  \hat{B} /2 = \hat{\xi} /2 $ we have  
\begin{align}
& \Phi _{\mathrm{q}} (t_f)  \simeq   
    \int _{ - \infty }^{ + \infty } \! 
    d  \omega
	\mathcal{F} _{ \mathcal{J} } [\omega ; t_f ] 
    \mathcal{J}[\omega] 
	 , \\
& \mathcal{F} _{ \mathcal{J} } [\omega ; t_f ] 
    =  \text{Im} 
    ( F^*[\omega]
	\eta [\omega] )
	/4 
	, 
\end{align}
and $ \mathcal{F} _{ \mathcal{J} } [\omega ; t_f ] $ forms a comb-like structure in frequency space if we fix pulse periodicity 
$ T = t_f/2M $ and take the asymptotic large pulse number limit, i.e.  
\begin{subequations}
\label{eq:eta.comb.sum}
\begin{align} 
\label{eq:eta.comb.asymp}
\mathcal{F} _{ \mathcal{J} } [\omega ; t_f ] 
& \sim   
 - \frac{ M }{ \omega _0  } 
    \sum _{\ell = -\infty } ^{ + \infty} 
    \mathcal{A} _\ell 
    \delta ( \omega 
    - \ell \omega _0 )
   \quad  ( M \gg 1)
   , \\
\omega _0 & =  \pi / T 
= 2 M \pi / t_f  
    ,  
\end{align}
\end{subequations}
where 
$  \mathcal{A} _\ell 
= (4 / \ell^2) 
\sin ( \ell \pi/2  )  $ are constant coefficients that depend on the control sequence and can be derived using Eq.~\eqref{eq:eta.filter.sum}.

Thus, by making use of all three levels of our spin-$1$ sensor, we can engineer a time-dependent quench function $ \eta (t) $ that enables the construction of a standard comb-based filter function.  This in turn allows spectral reconstruction of the imaginary bath response function (i.e.~the spectral function $\mathcal{J}[\omega]$) over a large frequency range.

Note that Ref.~\cite{Viola2017} have developed related techniques for reconstructing spatially correlated noise and response functions using multiple qubits; 
we stress that these are distinct from our \textit{multilevel} protocols. More specifically, the multiqubit protocols crucially require $2$-qubit SWAP gates in addition to standard dynamical-decoupling-type controls; further, the spectral function  $\mathcal{J}[\omega]$ is not directly accessible via those existing protocols. In contrast, the quench physics in Eq.~\eqref{eq:V.bath.def} provides a tool for directly probing bath response properties (i.e.~its density of states), and our quench-based protocols can be straightforwardly implemented using only local spin-echo-type, or dynamical-decoupling control sequences.

Reference~\cite{Cywinski2020} also discussed a surprising phase shift emerging from a dynamical environment coupled to an NV defect sensor, akin to the quench phase shift in Eq.~\eqref{eq:MainQPSResult}. In that work, the phase shift effect was again attributed to a specific class of biased coupling. Our work significantly extends this important study:  we show that the effective quench is directly controlled by the initial bath state, which can be used as a sensing tool to probe the spectral function, or general response functions. More significantly, as shown in this section, the quench physics can be actively modulated by embedding the qubit in a higher-dimensional Hilbert space of the physical system, and alternating the qubit subspace. This opens up the possibility of using the quench phase shift as a new tool for reconstructing environmental response properties in a variety of quantum platforms.

\section{Generalizations beyond Gaussian bath approximation}
\label{sec:NonGaussian}

Our discussion so far on effective quench physics in standard $T_2$-type quantum noise spectroscopy has focused on the common case where the environment is either a quantum Gaussian bath, or where the sensor is weakly coupled to a quantum environment.  While these cases make it convenient to describe the result quench physics (namely the emergence of a quench-induced sensor qubit phase shift), the physics we describe is far more general.  In particular, the quench has nontrivial consequences on the sensor qubit even beyond the weak-coupling or Gaussian regime.  We briefly discuss how this physics can be described below.  

For the case of a general environment, the qubit dynamics is still generally described by Eq.~\eqref{eq:qb.dyn.gen}.  For the general case, a powerful means of attack is provided by Keldysh field theory
techniques~\cite{Kubo1962,Diehl2016,Clerk2020}.  One finds that the quench does more that just induce a phase shift 
$ \Phi _{\mathrm{q}} (t_f) $ 
(c.f.~Eq.~\eqref{eq:Phi.Gau.gen}).  The quench physics can also modify the noise properties of the bath, changing the dephasing function in Eq.~(\ref{eq:qb.coh.gen}).

A general way to describe these effects is to use non-linear response theory, something that can be effectively calculated using the Keldysh approach.  For weakly coupled (or Gaussian) environments, the quench only induces a nonzero average of the noise operator $\xi(t)$ that is linear in the quench operator $\hat{V}$.  In the more general cases, this shift in mean will have higher-order terms in $\hat{V}$; in addition, the symmetrized noise correlator of $\xi(t)$ will also be modified.  Keeping terms to leading orders in $\hat{V}$, these effects can formally be written in terms of nonlinear response functions as  
\begin{align}
\langle {\delta \hat \xi (t )} \rangle_{V} 
= & \int _{-\infty }^{ + \infty }  \! d t_1   \eta ( t_1 )
G ^R _{\xi V } (t - t_1 )
\nonumber \\
+ \iint _{-\infty }^{ + \infty }  \! &  d t_1 d t_2 
  \eta ( t_1 )  \eta ( t_2 )
G ^R _{\xi, V V } (t - t_2, t_1 - t_2 )
\nonumber \\
& + \ldots , 
    \\
\langle {\delta \hat \xi (t )
\delta \hat \xi (t' )
} \rangle_{V} 
= & \int _{-\infty }^{ + \infty }  \! d t_1   \eta ( t_1 )
G ^R _{\xi \xi , V } (t - t_1, t' - t_1 )
\nonumber \\
& + \ldots , 
\end{align}
where the susceptibility functions  
$ G ^R _{\xi V } ( t ) $, 
$ G ^R _{\xi, V V } ( t_1 , t_2 ) $, and 
$ G ^R _{\xi \xi, V } ( t_1 , t_2 ) $
can be systematically computed using Keldysh field theory techniques~\cite{Diehl2016,Clerk2020}.
Note the appearance of a new function here, 
$ G ^R _{\xi \xi, V } ( t_1 , t_2 )$.  This is often referred to as a noise susceptibility, and describes how the environmental symmetrized noise spectral density is modified by the quench.  Noise susceptibilities often reveal subtle features of a physical system, and have been studied in a variety of contexts (e.g.~to uncover subtle features of coherent quantum electronic transport~\cite{Reulet2008}).

The upshot is that the quench physics described here is not limited to weakly-coupled or Gaussian environments.  For the general case, it can be described using both linear and nonlinear response functions~\cite{Stratonovich1992book,Reulet2008,Clerk2020}. This also highlights another utility of our quench approach to QNS:  it provides in principle access to higher-order nonlinear response properties of an unknown environment using $ T_2 $-type measurements.

\section{Conclusion and Outlook}

We have shown how intrinsic quenches arise in standard $ T_2 $-style noise spectroscopy experiments, and how quench-induced phase shift effects can be utilized to estimate or reconstruct the spectral function, or more general response functions of the environment.  
These response properties provide an independent environmental characterization from the standard noise spectral density, and encode useful information: in combination with standard NSD, we can use the estimated spectral function to extract effective temperature of a generic nonequilibrium bath. For environments in thermal equilibrium, the quench-enhanced QNS based on a single probe qubit also allows one to extract environmental temperature.

Our work highlights the critical role played by the initial state in controlling the quench physics associated with a generic $ T_2 $-style experiment. As such, our quench formalism greatly expands previous works that considered related examples of environment-induced phase shift effects: by analyzing quenches that arise in the most general settings of $ T_2 $-based QNS protocols, we show that one can engineer a generic quench operator, or a quench with complex time dependence. These generalizations allow us to further use the quench phase shift to probe general response functions, or reconstruct the spectral function in a generic frequency range, respectively.

We note that the quench formalism is also straightforwardly applicable to non-Gaussian environments beyond the Gaussian approximation. For these systems, the quench can lead to much richer dynamical effects than the Gaussian case. While we have shown concretely that these effects can be related to nonlinear response functions and noise susceptibilities of the environment, it remains an open question how to design sensing protocols to extract those higher order response functions.  Alternatively, quantum quenches have conventionally been used to explore correlated phenomena in many-body systems~\cite{Vengalattore2011}; our quench approach to QNS also opens up possibilities to explore new physics in these systems.  We leave these to future works.

\section*{Acknowledgements}
We thank M. Onizhuk, G. Galli, P. Jerger, J. Karsch, M. Fukami, N. Delegan, and  F. J. Heremans for useful discussions.  This work was supported as part of the Center for Novel Pathways to Quantum Coherence in Materials, an Energy Frontier Research Center funded by the U.S. Department of Energy, Office of Science, Basic Energy Sciences.

\onecolumngrid
\appendix

\section{Derivation of Eq.~\eqref{eq:qb.dyn.gen} on qubit dynamics during a standard $ T_2 $-type experiment}

\label{appsec:qubit.dyn.gen}

In this section, we provide a detailed derivation of Eq.~\eqref{eq:qb.dyn.gen} in the main text, which describes qubit dynamics due to pure-dephasing baths in a general $ T_2 $-type (e.g., spin echo, or general dynamical decoupling) experiment. For concreteness, we first reiterate the general setup of the $T_2 $-type experiment; more detail can be found in the main text. As standard, we assume that the qubit is initialized into a pure state with no qubit-bath entanglement, so that an instantaneous $ \pi / {2} $-pulse at the beginning of the protocol ($ t = 0 $) prepares the system in a product state given by 
\begin{subequations}
\label{appeq:tot.rho.ini}
\begin{align}
& \hat{\rho}_{\mathrm{tot}} (t = 0^+ )
= \left |   + \rangle \langle +  \right | 
\otimes
\hat{\rho}_{\mathrm{b,i}}
    , \\
& | + \rangle   \equiv
 \left ( | \!  \uparrow \rangle 
+ | \!   \downarrow \rangle \right ) 
/ \sqrt{2}
    .
\end{align}
\end{subequations}
The system then evolves under a pure-dephasing-type total Hamiltonian 
$ \hat H_{\mathrm{tot}} $ 
for time $ t_f $, while the qubit is subject to
a sequence of instantaneous control $ \pi $-pulses. At each instant $ t $ during the time evolution, the total qubit-bath system Hamiltonian can be rewritten as 
\begin{subequations}
\begin{align}
 \hat H_{\mathrm{tot}}  
& =  \left |   \uparrow \rangle 
    \langle \uparrow  \right| 
    \otimes
    \hat H _{\mathrm{b},\uparrow }   
    + \left |   \downarrow \rangle 
    \langle \downarrow \right | 
	 \otimes
	 \hat H _{\mathrm{b},\downarrow }  
	  \\
	  \label{appeq:Htot.noise}
& = \hat \sigma _0 
     \otimes  H_{\mathrm{b,avg}}
	 + \frac{ \hat \sigma _z }{2}
	 \otimes
	\hat{\xi} 
    , \quad \quad  
    0 < t < t_f ,
\end{align}
\end{subequations}
where we introduce 
\begin{subequations}
\begin{align}
& \hat \sigma _0  \equiv
    \left |   \uparrow \rangle 
    \langle \uparrow  \right| 
    +  \left |   \downarrow \rangle 
    \langle \downarrow \right |  
    , \quad 
    \hat \sigma _z  \equiv
     \left |   \uparrow \rangle 
    \langle \uparrow  \right| 
    -  \left |   \downarrow \rangle 
    \langle \downarrow \right |  
    , \\
& H_{\mathrm{b,avg}} \equiv 
    \frac{1 }{2}( \hat H_{\mathrm{b},\uparrow} + \hat H_{\mathrm{b},\downarrow} ) 
    , \quad 
\hat{\xi}   \equiv 
    \hat H _{\mathrm{b},\uparrow }
    - \hat H _{\mathrm{b},\downarrow }  
    .
\end{align}
\end{subequations}
To keep our discussion general, we will assume a generic initial bath state 
$ \hat{\rho}_{\mathrm{b,i}} $, and a corresponding initial bath Hamiltonian 
$ \hat H_{\mathrm{b,i}} $ satisfying   
$  [\hat H_{\mathrm{b,i}} 
, \hat{\rho}_{\mathrm{b,i}} ] =0 $ (see also Sec.~\ref{sec:V.gen.op} in the main text). It is thus convenient to introduce a time-dependent effective bath Hamiltonian $ \hat H _{\mathrm{b,eff} }  ( t ) $ as 
\begin{align}
\label{appeq:H.beff.def}
\hat H _{\mathrm{b,eff} }  ( t )
\equiv  & \begin{cases}
\hat H_{\mathrm{b,i}}  , 
    & t \le 0 , \\
    H_{\mathrm{b,avg}}  
     ,
    &  0 < t < t_f  .
\end{cases}
\end{align}
As shown in Eq.~\eqref{eq:noncomm.ini.Heff}, a nontrivial quench $ \hat V $ generally arises in this standard $ T_2 $-type protocol, if the initial bath state 
$ \hat{\rho}_{\mathrm{b,i}} $ does not commute with the effective bath Hamiltonian governing subsequent bath dynamics, i.e., 
$ [ \hat{\rho}_{\mathrm{b,i}}
,
\hat H_{\mathrm{b,avg}}  ]
\ne 0 $. We can rewrite $ \hat H_{\mathrm{b,eff}}  (t) $ in terms of the initial $ \hat H_{\mathrm{b,i}} $ and this quench operator $ \hat V $ as (see Sec.~\ref{sec:V.gen.op} and \ref{sec:NV.impls} in the main text)  
\begin{align}
\label{appeq:Hbath.avg.td}
\hat H_{\mathrm{b,eff}}  (t) 
&  = \hat H_{\mathrm{b,i}} 
    + \eta (t)  \hat V 
    ,
\end{align}
where the quench function $ \eta (t) $ vanishes, i.e. $ \eta (t) =0$, unless $ 0 < t<   t_f$.

Transforming to the standard toggling frame with respect to qubit control pulses, as well as the rotating frame defined by initial bath Hamiltonian 
$ \hat H_{\mathrm{b,i}} $, we obtain the rotating-frame Hamiltonian  
\begin{align}
 \hat H_{\mathrm{I}}  (t)
 	= \eta (t) \hat \sigma _0 
     \otimes \hat V  (t)
	 + F (t) \frac{ \hat \sigma _z }{2}
	 \otimes
	\hat{\xi} (t)
    , \quad  
    0 < t < t_f 
    , 
\end{align}
where $ F (t) $ denotes the usual noise filter function that encodes the timing of qubit control $ \pi $-pulses, and 
$ \hat V  (t) $ and $ \hat \xi (t) $ refer to the rotating-frame bath-only operators. Thus, we can compute the qubit coherence function $\langle 
\hat \sigma _- ( t_f  ) 
\rangle = \langle \uparrow \! | 
	\textrm{Tr} _{\mathrm{bath}} \left [
	\hat{\rho}_{\mathrm{tot}} ( t_f )
	\right ] 
| \!  \downarrow \rangle  $ in this toggle-rotating frame as 
\begin{align}
& \hat{\rho}_{\mathrm{tot}} ( t_f )
    = \mathcal{T} 
    e^{-  i \int_0^{t_f} dt' 
    \hat H_{\mathrm{I}}  (t') }
	\hat{\rho}_{\mathrm{tot}} ( t = 0^+ )
	\tilde{\mathcal{T}} 
    e^{   i \int_0^{t_f} dt' 
    \hat H_{\mathrm{I}}  (t') } 
	, \\
	\label{appeq:qb.coh.genini}
\Rightarrow 
&  \frac{\langle 
	\hat \sigma _- ( t_f  ) 
	\rangle}
	{\langle 
	\hat \sigma _- ( 0^+  ) 
	\rangle}
 = \textrm{Tr} \left\{
    \mathcal{T} 
    e^{-  i \int_0^{t_f} dt' 
    \hat H _{ \uparrow  }  (t')}
    \hat{\rho}_{\mathrm{b,i}} 
    \tilde{\mathcal{T}} 
    e^{   i \int_0^{t_f} dt' 
    \hat H _{ \downarrow  }  (t') }
    \right\} 
    , \\
& \hat H _{ \uparrow (\downarrow) } (t)
    \equiv \eta (t)  \hat V  (t) 
     \pm \frac{1}{2} F (t) \hat{\xi} (t)
     ,
\end{align}
where $ \mathcal{T} $ and $ \tilde{\mathcal{T}} $ denote time- and anti-time orderings, respectively. From Eq.~\eqref{appeq:tot.rho.ini} we have $ \langle 
\hat \sigma _- ( 0^+  ) 
\rangle = 1/2 $, so that the qubit coherence function 
$ \langle 
	\hat \sigma _- ( t_f  ) 
\rangle $ can be rewritten as 
\begin{subequations}
\label{appeq:qb.dyn.gen}
\begin{align}
&  \langle 
	\hat \sigma _- ( t_f  ) 
	\rangle
	= \frac{1}{2}
	\textrm{Tr} (
   \hat U _{ \uparrow }
    \hat{\rho}_{\mathrm{b,i}} 
    \hat U _{ \downarrow } ^\dag ) 
	, \\
&  \hat U _{ \uparrow (\downarrow) }
    =  \mathcal{T} 
    \exp \left\{
    -  i \int _{-\infty }^{ + \infty } \! 
    \left [ \eta (t')  \hat V  (t')   
    \pm \frac{F (t' )}{2} \hat{\xi} (t')
    \right ]
    dt'
    \right\} 
    ,
\end{align}
\end{subequations}
i.e., we obtain Eq.~\eqref{eq:qb.dyn.gen} in the main text.

From above derivation, it is straightforward to see that the exact qubit dynamics only depends on initial bath state $ \hat{\rho}_{\mathrm{b,i}} $, and Eq.~\eqref{appeq:qb.dyn.gen}  (i.e., Eq.~\eqref{eq:qb.dyn.gen} in the main text) holds for a generic quench operator 
$ \hat V $ associated with any 
$ \hat H_{\mathrm{b,i}} $ satisfying   
$  [\hat H_{\mathrm{b,i}} 
, \hat{\rho}_{\mathrm{b,i}} ] =0 $.  
For a generic bath initial state $ \hat{\rho}_{\mathrm{b,i}} $, as discussed in Sec.~\ref{sec:V.gen.op} in the main text, a useful way to resolve such ambiguity in $ \hat V $ is to choose the particular $ \hat V' \equiv
H_{\mathrm{b,avg}} ( t = 0 ^+ ) 
- \hat H' _{\mathrm{b,i}} $ as the minimally non-commuting component of $ H_{\mathrm{b,avg}} $ with respect to 
$  \hat{\rho}_{\mathrm{b,i}} $.  While this is not the only workable choice, it highlights the fact that the quench is more directly controlled by the initial bath state rather than the physical bath Hamiltonian at the beginning of the protocol. Identifying the quench in this way also allows us to compute qubit dynamics using a systematic perturbative expansion of Eq.~\eqref{appeq:qb.dyn.gen} in terms of the quench operator $ \hat V$.

\section{Derivation of Eqs.~\eqref{eq:qbdyn.Gau.gen} on leading order dephasing and phase shift effects to the sensor qubit due to a quenched environment}

\label{appsec:DepQPS.gen}

In Eqs.~\eqref{eq:qbdyn.Gau.gen} in the main text, we present general formulae relating leading order bath-induced dephasing (quench-induced phase shift) effects to the bath noise spectral density (linear response susceptibility function). While there are multiple ways to derive this result, in this section we provide a detailed derivation based on the Keldysh field theory technique and cumulant expansion~\cite{Clerk2020}. Our approach is directly applicable to non-Gaussian baths, and elucidates the regime where Eqs.~\eqref{eq:qbdyn.Gau.gen} becomes exact.

We start with Eq.~\eqref{appeq:qb.coh.genini} (of which Eq.~\eqref{eq:qb.dyn.gen} constitutes a special case), where the sensor-qubit coherence function by the end of a $ T_2 $-type experiment is given by 
\begin{subequations}
\begin{align}
\label{appeq:Keld.op.tr}
\frac{\langle 
	\hat \sigma _- ( t_f  ) 
	\rangle}
	{\langle 
	\hat \sigma _- ( 0^+  ) 
	\rangle}
 &  = \textrm{Tr} \left\{
    \mathcal{T} 
    e^{-  i \int_0^{t_f} dt' 
    \hat H _{ \uparrow  }  (t')}
    \hat{\rho}_{\mathrm{b,i}} 
    \tilde{\mathcal{T}} 
    e^{   i \int_0^{t_f} dt' 
    \hat H _{ \downarrow  }  (t') }
    \right\} 
    , \\
\hat H _{ \uparrow (\downarrow) } (t)
    & = \eta (t)  \hat V  (t) 
     \pm \frac{1}{2} F (t) \hat{\xi} (t)
     .
\end{align}
\end{subequations} 
For convenience, we introduce the Keldysh-ordered cumulant generating function (CGF) 
$ \chi[F(t), \eta (t) ;t_f]  $ of the bath noise and quench operators as 
\begin{align}
\label{appeq:CGF.def.gen}
\chi[F(t), \eta (t) ;t_f]  
    &  \equiv \ln
    \left[ 
    \frac{\langle 
	\hat \sigma _- ( t_f  ) 
	\rangle}
	{\langle 
	\hat \sigma _- ( 0^+  ) 
	\rangle} 
	\right ] 
	. 
\end{align}
See Ref.~\cite{Clerk2020} for discussions on the physical implications of this function. In terms of qubit dynamics, the real and imaginary parts of $ \chi[F(t), \eta (t) ;t_f]  $ correspond to qubit dephasing and bath-induced phase shift effects, respectively.

We can now compute qubit coherence $ \langle 
\hat \sigma _- ( t_f  ) 
\rangle $ by perturbatively expanding the CGF $ \chi[F(t), \eta (t) ;t_f]   $ in terms of $ \hat \xi (t) $ and $ \hat V (t) $. The Keldysh technique offers a systematic way to perform this expansion~\cite{Kubo1962,Diehl2016,Clerk2020}. In the Keldysh approach, for each bath operator $  \hat A $ there are a corresponding classical field $A _{ \mathrm{cl} }(t)$ and a quantum field $A _{\mathrm{ q} }(t)$. Stochastic averages between these fields can be computed in a well-defined way, with respect to the so-called Keldysh action. The very construction of the Keldysh action (see Ref.~\cite{Clerk2020}) ensures that the quantum operator expectation value in Eq.~\eqref{appeq:Keld.op.tr} can be directly related to averages that only involve $ \xi _{ \mathrm{cl} } $ and $ V _{\mathrm{ q} } $. Taking the logarithm of Eq.~\eqref{appeq:Keld.op.tr}, the CGF $ \chi[F(t), \eta (t) ;t_f]  $ is in turn given by 
\begin{align}
\label{appeq:qbath.cml.gen}
\chi[ F(t), \eta (t) ;t_f]  
    = &  \sum_{\ell=1}^\infty
     \sum_{m = 0 }^\infty
    \!\frac{(-i)^ {\ell+m} }{  \ell ! m! }
    \prod_{j=1}^\ell 
    \left[ 
    \int_0^{t_f}\!\! dt_j F(t_j) \right]\!
    \prod_{k=\ell+1 } ^{\ell+m }
    \left[ 
    \int_0^{t_f}\!\! dt_k \eta  (t_k) \right]\!
    C^{(\ell ,m  )}(\vec{t}_{\ell+m }) 
     ,
\end{align}
where we define $\vec{t}_n \equiv (t_1,\ldots,t_n)$. Here $ C^{(\ell ,m  )}(\vec{t}_{\ell+m}) $ denote Keldysh-ordered cumulants, which can be directly generated from Keldysh-ordered moments of the form 
$ \overline{ \xi _{ \mathrm{cl} }(t_1) 
\ldots
\xi _{ \mathrm{cl} }(t _{\ell }) 
V _{\mathrm{ q} }(t _{\ell+1 })
\ldots
V _{\mathrm{ q} }(t _{\ell+m })} $ and lower-order averages. Note that the terms in Eq.~\eqref{appeq:qbath.cml.gen} involving cumulants of the form $ C^{(\ell ,0  )}(\vec{t}_{\ell}) $ correspond to contributions solely from noise fluctuations, which would determine qubit dynamics in the absence of quenches. In contrast, the cumulants 
$ C^{(\ell ,m  )}(\vec{t}_{\ell+m }) $ for $ m>0$ encode bath (linear and nonlinear) response properties ($ \ell =1 $), as well as noise susceptibilities ($ \ell >1 $).

Without loss of generality, we assume zero-average noise operator, i.e. $ \langle 
\hat \xi  
\rangle = 0$. The first order contribution to the cumulant expansion in Eq.~\eqref{appeq:qbath.cml.gen} thus vanishes, and the leading-order nontrivial Keldysh-ordered cumulants can be written explicitly as 
\begin{subequations}
\begin{align}
 C^{(2 ,0  )}(\vec{t}_{ 2 }) 
    &=  \overline{ 
    \xi _{ \mathrm{cl} }(t_1) 
   \xi _{ \mathrm{cl} } (t _{ 2 }) }
    = \frac{1}{2}  
	\langle
	\{ \hat{\xi} ( t_1 )
	,
	\hat{\xi} ( t_2 )
	\}
	\rangle
    , \\
C^{( 1 , 1  )}(\vec{t}_{ 2 }) 
    & = \overline{ \xi _{ \mathrm{cl} }(t_1) 
V _{ \mathrm{q} }(t _{ 2 }) }
    = 
     \Theta (t_1 - t_2 )
	\langle
	[ \hat{\xi} ( t_1 )
	,
	\hat{V } ( t_2 )
    ]
	\rangle
     , 
\end{align}
\end{subequations}
where the bath operator average is defined with respect to 
$ \hat{\rho}_{\mathrm{b,i}}  $ as 
$ \langle 
\hat A  
\rangle \equiv 
\mathrm{Tr} ( \hat A 
\hat{\rho}_{\mathrm{b,i}} )$. These leading order cumulants can be directly related to the bath noise spectral density (NSD) 
$ \bar{S} [ \omega ] $ and response susceptibility function 
$ G ^R _{\xi V }[ \omega ] $, as  
\begin{subequations}
\begin{align}
    \bar{S} [ \omega ]
	 & \equiv \frac{1}{2}  
	\int _{-\infty }^{ + \infty } \!  
	d t   e^{ i \omega t}
	\langle
	\{ \hat{\xi} ( t )
	,
	\hat{\xi} ( 0 )
	\}
	\rangle
	= \int _{-\infty }^{ + \infty } \!  
	d t   e^{ i \omega t} 
	C^{(2 ,0  )}( t, 0 ) 
	,\\
    G ^R _{\xi V } [ \omega ] 
    & \equiv -i  \int _{-\infty }^{ + \infty } 
    \!  d t   e^{ i \omega t}
    \Theta (t)
    \langle
    [\hat{\xi} ( t )
	    ,
    \hat{ V } ( 0 )]
    \rangle
    = - i \int _{-\infty }^{ + \infty } \!  
	d t   e^{ i \omega t} 
	C^{( 1 ,1 )}( t, 0 ) .
\end{align}
\end{subequations}
Substituting above relations into Eqs.~\eqref{appeq:CGF.def.gen} and \eqref{appeq:qbath.cml.gen}, we can rewrite the leading order contributions in terms of bath NSD and linear response susceptibility, so that we obtain 
\begin{subequations}
\label{appeq:qbdyn.Gau.gen}
\begin{align}
& \frac{\langle 
	\hat \sigma _- ( t_f  ) 
	\rangle}
	{\langle 
	\hat \sigma _- ( 0^+  ) 
	\rangle} 
	= e^{- \zeta (t_f)
	- i \Phi(t_f) } 
    ,\\
\label{appeq:dep.Gau.gen}
& \zeta (t_f) \simeq  
    \frac{1}{2}  
    \int _{-\infty }^{ + \infty } 
    d t_1 F(t_1) 
    \int _{-\infty }^{ + \infty }  d t_2  
    F ( t_2 )
    C^{(2 ,0  )}(\vec{t}_{ 2 }) 
    = \int _{-\infty }^{ + \infty } \! 
    \frac{d  \omega }{4 \pi}
|F[\omega]|^2 \bar{S} [ \omega ] 
    , \\
\label{appeq:Phi.Gau.gen}
&  \Phi  (t_f)  
    \simeq 
    - i \int _{-\infty }^{ + \infty } 
    d t_1 F(t_1) 
    \int _{-\infty }^{ + \infty }  d t_2   \eta ( t_2 )
    C^{( 1 , 1  )}(\vec{t}_{ 2 })  
    = 
    \int _{-\infty }^{ + \infty } \! 
    \frac{d  \omega }{2 \pi}
	F^*[\omega]
	\eta [\omega]
	G ^R _{\xi V }[ \omega ] .
\end{align}
\end{subequations}
Above equations reproduce Eqs.~\eqref{eq:qbdyn.Gau.gen}.

The cumulant expansion in Eq.~\eqref{appeq:qbath.cml.gen} terminates at the second order if $ \xi _{ \mathrm{cl} }(t)$ and $ V _{\mathrm{ q} }(t)$ are Gaussian random variables, which is satisfied by linearly coupled harmonic oscillator bath models discussed in the main text. For this case, the dephasing and phase shift expressions in Eqs.~\eqref{appeq:dep.Gau.gen} and \eqref{appeq:Phi.Gau.gen} become exact. Thus, we conclude that Eqs.~\eqref{eq:qbdyn.Gau.gen}, or equivalently Eqs.~\eqref{appeq:dep.Gau.gen} and \eqref{appeq:Phi.Gau.gen}, hold exactly for Gaussian baths, where noise and quench operators can take arbitrarily strong coupling strengths. Alternatively for more general non-Gaussian baths, Eqs.~\eqref{appeq:qbdyn.Gau.gen} describe leading-order approximations for qubit dynamics in terms of bath noise and quench operators.

\section{Derivation of Eq.~\eqref{eq:Phi.static.eta} relating the quench phase shift to the imaginary part of the spectral bath response function}

\label{appsec:QPS.Imchi.gen}

In Eq.~\eqref{eq:Phi.Gau.gen} of the main text, we provide a general formula to compute the quench-induced phase shift using the Green-Kubo linear response theory, which relates the phase shift to the bath susceptibility function 
$ G ^R _{\xi V }[ \omega ] $. We then claim that for the specific quench function 
$ \eta (t)  =  \Theta (t ) \Theta ( t_f - t ) $ emerging in a typical $ T_2 $-type experiment, this phase shift can be rewritten as Eq.~\eqref{eq:Phi.static.eta}, which only involves the imaginary part of response function. In this section, we explicitly derive the latter equation using the general formula. Recall that for a generic Gaussian bath, the quench phase shift (QPS) is given by Eq.~\eqref{eq:Phi.Gau.gen} as
\begin{align}
\Phi _{\mathrm{q}} (t_f) 
   &  =   \int _{-\infty }^{ + \infty } \! 
    \frac{d  \omega }{2 \pi}
	F^*[\omega]
	\eta [\omega]
	G ^R _{\xi V }[ \omega ] 
	\\
\label{appeq:Phi.Gau.gen.temp}
	& =  \int _{-\infty }^{ + \infty }  d t_1 F(t_1) 
    \int _{-\infty }^{ + \infty }  d t_2   \eta ( t_2 )
    G ^R _{\xi V } (t_1 - t_2)
    ,
\end{align}
where $  G ^R _{\xi V } (t )
\equiv 
-i \Theta (t)  
\langle
[\hat{\xi} ( t ) ,
\hat{ V } ( 0 )]
\rangle$ is the standard  Green-Kubo linear response susceptibility function. Noting that the real and imaginary parts of the spectral response function 
$ G ^R _{\xi V }[ \omega ] $ are related to each other via the Kramers-Kronig relation, we can rewrite the QPS in terms of only 
$ \text{Im}  G ^R _{\xi V } [ \omega ] $. More specifically, given that the retarded Green's function $ G ^R _{\xi V } ( t ) $ is real by definition, the imaginary part of the response function 
$ \text{Im}  G ^R _{\xi V } [ \omega ] $ can be written as 
\begin{align} 
\text{Im}  G ^R _{\xi V } [ \omega ]
    =  \frac{ 1 }{2 i } 
    \int _{-\infty }^{ + \infty } 
    \!  d t   e^{ i \omega t} 
    \left [ G ^R _{\xi V } (t ) 
    - G ^R _{\xi V } ( -t ) 
    \right ]
    .
\end{align}
Thus, we can rewrite the QPS as 
\begin{align}
& \Phi _{\mathrm{q}} (t_f) 
   = 
    \int _{-\infty }^{ + \infty }  d t_1 F(t_1) 
    \int _{-\infty }^{ + \infty }  d t_2   \eta ( t_2 )
    G ^R _{\xi V } (t_1 - t_2)
   \nonumber   \\
= &  2 \int_0^{ t_f  }   d t_1 F(t_1) 
    \int_0^{ t_1  }   d t_2   \eta ( t_2 )
    \frac{  G ^R _{\xi V } (t_1 - t_2)
    -  G ^R _{\xi V } ( t_2 - t_1 ) }{2 }  
      \\
      \label{appeq:Phi.mid}
= &  2 \int _{-\infty }^{ + \infty } \frac{d  \omega }{2 \pi}
	 \text{Im}  G ^R _{\xi \xi } [ \omega ] 
	\int_0^{ t_f  }  d t_1 F(t_1) 
	\int_0^{ t_1  }  d t_2  \eta (t_2) 
	\sin\omega ( t_1 - t_2) 
    ,
\end{align}
where we have made use of the fact that the susceptibility $ G ^R _{\xi V } (t ) =0  $ for $ t<0 $. Substituting the specific quench function $ \eta (t) 
=\Theta (t ) \Theta ( t_f - t ) $ into above equation, we can explicitly compute the integral involving $ \eta (t_2) $ in Eq.~\eqref{appeq:Phi.mid} to obtain 
\begin{align}
\label{appeq:Phi.q.stat}
& \Phi _{\mathrm{q}} (t_f) 
    =  \int _{-\infty }^{ + \infty } \!
    \frac{d  \omega }{  \pi }
    \frac{ \text{Re} F [ 0 ]
    - \text{Re} F [ \omega ]}
    { \omega } \,
	 \text{Im}  G ^R _{\xi V } [ \omega ] 
    ,
\end{align}
so that for spin-echo or dynamical-decoupling control pulses with $ F [ 0 ]
\equiv	\int_0^{ t_f  } 
   \! F (t)  d t
= 0 
$, we recover Eq.~\eqref{eq:Phi.static.eta}.  For the specific form of the quench operator $ \hat{V} (t) = \hat{\xi} (t) /2 $ as given by Eq.~\eqref{eq:VSimple}, we can rewrite above expression for the QPS $ \Phi _{\mathrm{q}} (t_f)  $ in terms of imaginary part of the response function 
$ \text{Im}  G ^R _{\xi \xi } [ \omega ]  $ as 
\begin{align}
\label{appeq:Phi.q.DOS.stat}
& \Phi _{\mathrm{q}} (t_f) 
    =  \int _{-\infty }^{ + \infty } \!
    \frac{d  \omega }{ 2 \pi }
    \frac{ \text{Re} F [ 0 ]
    - \text{Re} F [ \omega ]}
    { \omega } \,
	 \text{Im}  G ^R _{\xi \xi  } [ \omega ] 
    .
\end{align}

Alternatively, for general forms of quench function $ \eta ( t ) $, it is still possible to represent the QPS using a frequency-space integral of imaginary part of the response function, as weighted by the control functions 
$ F [ \omega ]  $ and 
$  \eta [ \omega ]$. However, in general the integral would be nonlocal in frequency space. From the general expression in Eq.~\eqref{appeq:Phi.mid}, we obtain   
\begin{align}
& \Phi _{\mathrm{q}} (t_f) 
    =  \int _{-\infty }^{ + \infty } \frac{d  \omega }{2 \pi}
	 \text{Im} G ^R _{\xi V }  [ \omega ]
	 \left (  
	- \text{Re}   F [ \omega ]  \text{Im}  \eta [ \omega ]
	+  \mathcal{P} \int _{-\infty }^{ + \infty } \frac{d  \omega _1 }{ \pi}
	\frac{ \text{Re}   F  [ \omega _1 ] \text{Re}  \eta [ \omega _1 ]   }
	{\omega  -\omega  _1 }
	\right )
    .
\end{align}
As shown in Eq.~\eqref{appeq:Phi.q.stat}, this equation for QPS greatly simplifies when using the specific quench function $ \eta (t) 
=\Theta (t ) \Theta ( t_f - t ) $.

\section{QPS in comparison to relaxometry-based techniques to measure response}
\label{appsec:T1Relaxometry}

In the main text, we have focused on the specific type of quantum noise spectroscopy measurements where a sensor qubit is coupled to its environment via pure-dephasing-type interactions. Within this setting, we have shown the quench phase shift (QPS) lets one probe the response properties, or spectral function of the environment, which would be otherwise inaccessible using standard dephasing-based noise spectroscopy. 

Interestingly, in principle one can extract similar information about the imaginary part of response function $\text{Im}  G ^R _{\xi \xi } [ \omega ] $, or equivalently the spectral function, using an extended version of standard $T_1$ relaxometry experiments. Conventional $T_1$-type experiments specifically probes transversely coupled bath fields (e.g., via 
$ \hat H_{\mathrm{int}} 
=  \hat \sigma _x \otimes \hat{\xi} $), which induces transitions between the qubit levels. In this setting, typically one would measure the qubit population decay rate 
$ \Gamma _{\mathrm{tot} } \equiv 1 / T_1  $~\cite{Abragam1961book}, which corresponds to the sum of qubit relaxation and excitation rates. A straightforward calculation based on Fermi's Golden rule can then relate  
$ \Gamma _{\mathrm{tot} } $ to the symmetrized noise spectral density (NSD) via $ \Gamma _{\mathrm{tot} } 
= 2 \bar{S} [ \Omega ] $, whereas the difference between relaxation and excitation rates corresponds to the response function 
$\text{Im}  G ^R _{\xi \xi } [ \Omega ] $ (see the following paragraph for more detail). While the bath NSD can be directly inferred from $ T_1 $-decay rate 
$ \Gamma _{\mathrm{tot} } $, to further probe response function 
$\text{Im}  G ^R _{\xi \xi } [ \omega ] $ one would also need to measure the qubit steady state population 
$ \langle  \hat \sigma _z 
\rangle _\mathrm{ss} $: the latter measurement is not a part of standard $T_1$ relaxometry~\cite{Cappellaro2017,Veldhorst2018}.
In comparison, the QPS in Eq.~\eqref{eq:Phi.Gau.gen} is readily accessible in standard $T_2$-type measurements, and as we show in the main text, arguably offers a more direct knob to probe the response properties of longitudinal bath fields.

We now concretely show the relation between qubit decay rate 
$ \Gamma _{\mathrm{tot} } $ and the steady state population 
$ \langle  \hat \sigma _z 
\rangle _\mathrm{ss} $ to environmental properties. 
For the case of transverse coupling to the bath, it is more illuminating to represent bath properties in terms of the quantum noise spectra 
$ S [ \omega ]
	 \equiv  
	\int _{-\infty }^{ + \infty } \!  d t   e^{ i \omega t}
	\langle
	\hat{\xi} ( t )
	\hat{\xi} ( 0 )
	\rangle
$, so that the Fermi's Golden rule transition rates for qubit excitation and relaxation  
$ \Gamma _{\pm} $
are given by $ \Gamma _{\pm} = S [ \mp \Omega ] $ 
($ \Omega $ denotes qubit transition frequency; see Ref.~\cite{QN.RMP2010} for a pedagogical introduction). 
One can use a few lines of algebra to show that the symmetrized and anti-symmetrized components of 
$ S [ \omega ]$ are related to the bath NSD 
$ \bar{S} [ \omega ] = 
(S [ + \omega ] + S [ - \omega ] ) /2 $, and the imaginary part of response function 
$\text{Im}  G ^R _{\xi \xi } [ \omega ]= 
(S [ - \omega ] - S [ + \omega ] ) /2  $, respectively.  
The qubit population decay rate  
$ \Gamma _{\mathrm{tot} }  $ and the steady state population $ \langle  \hat \sigma _z 
\rangle _\mathrm{ss}$ can then be computed explicitly as 
$ \Gamma _{\mathrm{tot} } 
= \Gamma _{+ } + \Gamma _{- } =2 \bar{S} [ \Omega ] $, and 
$ \langle  \hat \sigma _z 
\rangle _\mathrm{ss} = 
\text{Im}
G ^R _{\xi \xi  } [ \Omega  ]
/ \bar{S} [ \Omega  ] $. Thus, given both the qubit decay rate and steady state population simultaneously, we can use them to infer the response function~\cite{Devoret2003,QN.RMP2010}.

\section{Asymptotic analysis on qubit dephasing and quench-induced phase shift for environments with power-law noise spectra and response functions}

\label{appsec:QPS.asymp.deriv}

In the main text, we present the asymptotic long-time behavior of qubit dephasing function and quench phase shift (QPS) in Eqs.~\eqref{eq:tot.pl.asymp}, assuming that the bath noise spectral density (NSD) and density of states functions exhibit power-law dependence in the asymptotic low-frequency limit. For clarity, in this Appendix we provide a detailed derivation of the asymptotic results. We start with the general expressions for the dephasing function 
$ \zeta  (t_f) $ and QPS 
$ \Phi  _{\mathrm{q}} (t_f) $ under any spin-echo qubit control pulse, given by Eqs.~\eqref{eq:dep.Gau.gen} and \eqref{eq:Phi.static.eta} in the main text as
\begin{subequations}
\label{appeq:qbdyn.Gau.gen.2}
\begin{align}
\label{appeq:dep.Gau.gen.2}
\zeta (t_f) &  =  
\int _{-\infty }^{ + \infty } \! \frac{d  \omega }{4 \pi}
|F[\omega]|^2 \bar{S} [ \omega ] 
    , \\
\label{appeq:Phi.static.eta.2}
\Phi _{\mathrm{q}} (t_f) 
& = - \int _{-\infty }^{ + \infty } \!
    \frac{d  \omega }{  \pi \omega }
	 \text{Re} F [ \omega ] \,
	 \text{Im}  G ^R _{\xi V } [ \omega ]
	 .
\end{align}
\end{subequations}
We also assume that the bath NSD $ \bar{S} [ \omega ] $ and response functions $ \text{Im}  G ^R _{\xi V } [ \omega ] $ exhibit power-law dependence in the asymptotic low-frequency regime (see Eqs.~\eqref{eq:sum.spec.powerlaw} in the main text)
\begin{subequations}
\label{appeq:sum.spec.powerlaw}
\begin{align}
\bar{S} [ \omega  ] 
& \sim  
S_0 \omega ^{ p } \quad
( \omega \to 0^+ ),
\\
\label{appeq:ImG.powerlaw}
\text{Im}  
G ^R _{ \xi V } [ \omega ]
	& \sim  
	- \frac{A_0}{2}  
 \omega ^s \quad 
( \omega \to 0^+ )
    .
\end{align} 
\end{subequations}
For convenience, we rewrite the bath NSD $ \bar{S} [ \omega ] $ and response functions $ \text{Im}  G ^R _{\xi V } [ \omega ] $ in the full frequency range in terms of cutoff functions 
$ \mu _{A} \left ( x \right ) $ 
$ \left ( A =S, G \right ) $ as 
\begin{subequations}
\begin{align}
\bar{S} [ \omega  ] 
& = 
S_0 \omega ^{ p } 
\mu _{S} \left ( 
\omega / \omega_\mathrm{c} \right ) 
,
\\
\label{appeq:ImG.powerlaw}
 \text{Im}  
G ^R _{ \xi V} [ \omega ]
	& = -   
	 \frac{A_0}{2}   
 \omega ^s 
 \mu _{G} \left (  
 \omega / \omega_\mathrm{c} \right ) 
    ,
\end{align} 
\end{subequations}
where we introduce a UV cutoff frequency $ \omega_\mathrm{c} $ below which the asymptotic power-law function provides a good approximation for the exact function. By definition, the cutoff functions 
$ \mu _{A} \left ( x \right ) $ 
satisfy following conditions 
\begin{align}
\mu _{A} \left ( 0 \right ) =1
    , \quad
\lim_{x \to \infty} \mu _{A} \left ( x \right ) = 0
    , \quad
    \left ( A =S, G \right )
    ,
\end{align}
and we further assume both cutoff functions 
$ \mu _{A} \left ( x \right ) $ 
$ \left ( A =S, G \right ) $ 
corresponding to any physical bath are smooth near $ x =0 $.

We now consider a generic qubit control pulse satisfying 
$ F [ 0 ] = 0 $, which consists of $ L $ instantaneous $ \pi $-pulses at times $ t = \alpha _\ell t_f $. Without loss of generality, we assume the coefficients 
$ \alpha _\ell \, (\ell = 1,2,\ldots, L) $ satisfy following conditions 
\begin{align}
\alpha _1 < \alpha _2 
    < \ldots < \alpha _L 
    , \quad
F [ 0 ] = 0
\Leftrightarrow
2 \sum_{ \ell =1} ^{L} 
    (-) ^{ \ell } \alpha _\ell 
    + (-) ^{ L+1 } = 0 
    ,
\end{align}
so that we can explicitly compute the filter function 
$ F [ \omega ] $ as 
\begin{align}
 F [ \omega ]  & = 
    \int_0^{ t_f  }  d t_1 F(t_1) 
    e^{ i \omega t_1 } 
    = \frac{ 2 \sum_{ \ell =1} ^{L} 
    (-) ^{ \ell -1 }  e^{ 
    i \alpha _\ell \omega t_f }
    - 1 
    + (-) ^{ L } 
   e^{ i \omega t_f } 
    }{ i \omega }
	.
\end{align}
Substituting above equation into Eq.~\eqref{appeq:qbdyn.Gau.gen} and noting that the integrands are even functions of frequency, we obtain
\begin{subequations}
\begin{align}
\zeta (t_f) &  = \frac{ S_0 }{ 2\pi }
\int_0^{ +\infty }  \!
    \omega ^{ p -2 }
    \left | 2 \sum_{ \ell =1} ^{L} 
    (-) ^{ \ell }  e^{ 
    i \alpha _\ell \omega t_f }
    + 1 
    + (-) ^{ L+1 } 
   e^{ i \omega t_f } \right | ^2
\mu _{S} \left (  
    \omega / \omega_\mathrm{c} \right ) 
    d \omega 
    \nonumber \\
    &  = \frac{ S_0 }{ 2\pi }
    t_f ^{ 1 -p }
\int_0^{ +\infty }  \!
    x ^{ p -2 }
    \left | 2 \sum_{ \ell =1} ^{L} 
    (-) ^{ \ell }  e^{ 
    i \alpha _\ell x }
    + 1 
    + (-) ^{ L+1 } 
   e^{ i x } \right | ^2
\mu _{S} \left ( x / 
    \omega_\mathrm{c} t_f \right ) 
    d x  
    , \\
\Phi _{\mathrm{q}} (t_f) & = \frac{ A_0 }{ \pi }
\int_0^{ +\infty }  \!
    \omega ^{ s -2 }
\left [ 2 \sum_{ \ell =1} ^{L} 
    (-) ^{ \ell-1 }  \sin 
    \alpha _\ell \omega t_f
    + (-) ^{ L } 
    \sin  \omega t_f \right ]
    \mu _{G} \left (  
    \omega / \omega_\mathrm{c} \right ) 
    d \omega 
	\nonumber \\
\label{appeq:phi.pl.gen.expr}
    &  = \frac{ A_0 }{ \pi }
    t_f ^{ 1 -s }
\int_0^{ +\infty }  \!
    x ^{ s -2 }
\left [ 2 \sum_{ \ell =1} ^{L} 
    (-) ^{ \ell-1 }  \sin 
    \alpha _\ell x
    + (-) ^{ L } 
    \sin  x \right ]
\mu _{G} \left ( x / 
    \omega_\mathrm{c} t_f \right ) 
    d x 
	.
\end{align}
\end{subequations}
In the long-time limit $ t_f \to + \infty $, the integrals above would tend asymptotically to universal limits that are independent of details about the physical cutoffs, if and only if the integrals when setting 
$ \mu _{A} \left ( x \right ) \equiv 1 $ 
$ \left ( A =S, G \right ) $ are well defined. For this scenario, the asymptotic limits of dephasing function and phase shift functions can be derived as 
\begin{subequations}
\label{appeq:qbdyn.pl}
\begin{align}
\label{appeq:dep.pl.asymp}
-3 < p <1 : \, 
\zeta  (t_f) 
& \sim 
\mathcal{C} _{ \zeta } ( p ) 
S_0  t_f ^{ 1- p } \quad 
(  t_f \to + \infty  )  ,
    \\
\label{appeq:phi.pl.asymp}
-2< s <2: 
\Phi  _{\mathrm{q}} (t_f) 
& \sim 
  \mathcal{C} _{\Phi} ( s ) 
 \frac{A_0}{2}    
 t_f ^{1-s } \quad 
( t_f \to + \infty  )
, 
\end{align}
\end{subequations}
where the dimensionless coefficients $ \mathcal{C} _{ \zeta } ( p ) $ and $  \mathcal{C} _{\Phi} ( s )$ are determined by the spin-echo pulse parameters as
\begin{subequations}
\label{appeq:coeff.asymp}
\begin{align}
\mathcal{C} _{ \zeta } ( p ) 
& = \frac{ \Gamma  \left ( { p -1} \right ) }{\pi}
	\left \{ 4 \sum_{ \ell>\ell' } ^{L} 
	(-) ^{ \ell + \ell' }  
	\left (  \alpha _\ell 
	- \alpha _{\ell'} \right )  ^{1-p } 
	+ 2 \sum_{ \ell =1} ^{L} 
    (-) ^{ \ell }  
   \left [ (-) ^{ L+1 }  
    \left ( 1-\alpha _\ell  \right )  ^{1-p } 
    + \alpha _\ell  ^{1-p }  \right ]
    + (-) ^{ L+1 }  \right \}
	\sin \frac{  p  \pi }{2}  
,
    \\
\mathcal{C} _{\Phi} ( s ) 
& = \frac{ \Gamma  \left ( {s-1} \right ) }{\pi}
	\left [ 2 \sum_{ \ell =1} ^{L} 
    (-) ^{ \ell }  
    \alpha _\ell  ^{1-s } 
    + (-) ^{ L+1 } \right ]
	\cos \frac{  s \pi }{2} 
, 
\end{align}
\end{subequations}
and $ \Gamma (\cdot ) $ is the gamma function. For Hahn echo, the control pulse parameters are 
$ L=1 $, $ \alpha _1 = \frac{ 1 }{2} $, and substituting the parameters into equation above lets us obtain the coefficients 
$ \mathcal{C} _{ \zeta, \mathrm{H} } 
= \frac{ 1 - 2 ^{ p +1 }  }{\pi}
	\Gamma  \left ( { p -1} \right )
	\sin \frac{  p  \pi }{2}   $ and 
$ \mathcal{C} _{\Phi, \mathrm{H} }
=  \frac{ 1- 2 ^{ s } }{\pi}
	\Gamma  \left ( {s-1} \right )
	\cos \frac{  s \pi }{2}  $ in the main text. 
Note that above equations are still well-defined if $ p,s $ are exact integers, where the gamma function in Eqs.~\eqref{appeq:coeff.asymp} alone might diverge: in this case, we could obtain the asymptotic coefficients by taking the continuous limit of Eqs.~\eqref{appeq:coeff.asymp} as the exponent approaches the corresponding integer value. The asymptotic limit of quench phase shift can be further simplified if the response function exponent take the value of $1$, as
\begin{align}
    \label{appeq:Phi.asymp.Ohmic}
s = 1 : & \,
\lim_{ t_f \to + \infty } 
\Phi _{\mathrm{q}} (t_f)
= A_0 /2  
. 
\end{align}

For exponents beyond the range of validity specified in  Eqs.~\eqref{appeq:qbdyn.pl}, the long-time behavior of the dephasing function (phase shift) may not have a well-defined asymptotic limit, or the asymptotic behavior would depend on details of the low- or high-frequency cutoff of the bath NSD (response function). To illustrate this, we discuss a concrete example where the long-time phase shift dynamics explicitly depends on details of the cutoff. We compare the Hahn echo phase shift dynamics for response function 
$ \text{Im}  
G ^R _{ \xi V } [ \omega ]
= -   
( A_0 /2 )
 \omega ^s 
 \mu _{G} \left (  
 \omega / \omega_\mathrm{c} \right ) $
with exponent $ s = 5/2 $, and 
two different UV cutoff functions: exponential cutoff with  
$ \mu _{G,\mathrm{exp} } \left ( x \right )
= e^{ -x }$, and step-function cutoff with  
$ \mu _{G,\mathrm{sp} } \left ( x \right )
= \Theta (1-x ) $, where $ \Theta (\cdot )
$ is the Heaviside step function. The quench phase shift is generally given by Eq.~\eqref{appeq:phi.pl.gen.expr}, which for Hahn echo can be computed analytically to yield  
\begin{align}
\Phi _{ \mathrm{q,exp} } (t_f) 
& = \frac{ 4 A_0 }{ \pi }
\int_{ 0 }^{ +\infty }  \!
    \omega ^{  \frac{1} { 2 } }
    e^{ - \frac{\omega} { \omega_\mathrm{c}} }
    \sin  \frac{ \omega  t_f }{ 2 } 
    \sin^2 
    \frac{ \omega  t_f }{ 4 } 
    d \omega 
    \\
& = \frac{ A_0 }{ 2 \sqrt{\pi} }
    t_f ^{ - \frac{3} { 2 } }
\left [ 2 ^{ \frac{3} { 2 } }
    e^{ i \frac{ \pi } { 4 } }
    \left ( 1  +  \frac{ 2i  }
    { \omega_\mathrm{c} t_f } 
    \right )^{ - \frac{3} { 2 } }
   - 2 ^{ -1 }
    e^{ i \frac{ \pi } { 4 } }
    \left ( 1  + \frac{ i  }{ \omega_\mathrm{c} t_f } 
    \right )^{ - \frac{3} { 2 } }
    + \text{c.c.}
    \right ]
    , \\
\Phi _{ \mathrm{q,sp} } (t_f) 
& = \frac{ 4 A_0 }{ \pi }
    \int_{ 0 }^{ \omega_\mathrm{c} }  \!
        \omega ^{  \frac{1} { 2 } }
    \sin  \frac{ \omega  t_f }{ 2 } 
        \sin^2 
        \frac{ \omega  t_f }{ 4 } 
        d \omega 
    \\
    \label{appeq:Phi.sp.osci}
& = \frac{ 8 A_0 }{ \pi   }
    \left [ 
    \frac{   \sqrt{\omega_\mathrm{c}} }{ t_f }
     \sin  ^4 
    \frac{ \omega _\mathrm{c}  t_f }{ 4 } 
   - \frac{  \sqrt{\omega_\mathrm{c}} }{ 2 t_f }
   \int_{ 0 }^{ 1 }  \!
    x ^{ - \frac{1} { 2 } }
    \sin  ^4 
    \frac{ x  \omega_\mathrm{c} t_f }{ 4 } 
    d x 
    \right ]
    .
\end{align} 
While the asymptotic 
$ t_f \gg \omega_\mathrm{c} ^{-1} $
limit of Hahn echo phase shift 
$ \Phi _{ \mathrm{q,exp} } (t_f) $ 
assuming exponential cutoff agrees with the universal result in Eq.~\eqref{appeq:phi.pl.asymp}, it is straightforward to see that the phase shift dynamics with step-function cutoff does not have a well-defined asymptotic long-time limit, and does not agree with Eq.~\eqref{appeq:phi.pl.asymp}. Although the oscillatory behavior of the first term in the square bracket in Eq.~\eqref{appeq:Phi.sp.osci} is typical when we have response functions with a step-function cutoff (e.g. see the dashed blue curve in Fig.~\ref{fig:QPS.Ohmic}(c) in the main text, depicting the QPS for Ohmic bath spectral function with a step-function UV cutoff), for exponents within the range of validity of Eq.~\eqref{appeq:Phi.sp.osci} such oscillations are negligible in the asymptotic long-time limit. However, as shown in Eq.~\eqref{appeq:Phi.sp.osci}, for exponents outside this range the oscillatory contribution is important even in the long-time limit. 
Generally, for bath NSD (response function) of the form given by Eqs.~\eqref{appeq:sum.spec.powerlaw} with exponent 
$ p \ge 1 $ ($ s \ge 2 $), the long-time behavior of the dephasing function 
$ \zeta  (t_f) $ (quench phase shift $ \Phi _{\mathrm{q}}  (t_f) $) depend on the detail of the UV cutoff of the spectrum. Similarly, for exponents 
$ p \le -3 $ ($ s \le - 2 $) below the regime of validity in Eq.~\eqref{appeq:qbdyn.pl}, the corresponding long-time behavior would depend on the low-frequency cutoff.

\section{Alternative derivation of Eq.~\eqref{eq:QPS.Ohmic.asymp} for quench phase shift in Ohmic environments}

\label{appsec:QPS.Ohmic.asymp}

As discussed in the main text, specifically for baths that exhibit Ohmic behavior (a flat NSD and a linear bath spectral function) in the asymptotic low-frequency limit, i.e., satisfying Eq.~\eqref{appeq:sum.spec.powerlaw} with 
$ p = 0 $, $ s = 1 $, the quench phase shift (QPS) under spin-echo or dynamical-decoupling control sequences tends to a constant in the long-time regime, as shown in Eq.~\eqref{eq:QPS.Ohmic.asymp} (see also Eq.~\eqref{appeq:Phi.asymp.Ohmic} in App.~\ref{appsec:QPS.asymp.deriv}). In this section, we provide an intuitive derivation of this result, which for a generic control sequence can be written as 
\begin{align}
\lim_{ t_f \to + \infty } 
\Phi _{\mathrm{q}} (t_f)
    = \frac{  F[ 0 ]  }{2} 
 	 \text{Re}  G ^R _{\xi \xi  } [ \omega = 0 ] 
	- \frac{ 1 }{2} 
	\left .\frac{ d \text{Im}  
	G ^R _{\xi \xi  } [ \omega ]  }
	{d \omega }  
	\right|_{ \omega =0 } 
	, 
\end{align}
where $ F [ 0 ]
\equiv	\int_0^{ t_f  } 
   \! F (t)  d t $. 
We start with the general linear response formula for QPS, assuming quench operator $ \hat{V}  = \hat{\xi}  /2 $, quench function $ \eta (t) 
=\Theta (t ) \Theta ( t_f - t ) $, and a generic filter function, which in the time domain is given by (see Eq.~\eqref{appeq:Phi.Gau.gen.temp}) 
\begin{align}
& \Phi _{\mathrm{q}} (t_f) 
    = \frac{1}{2} 
	\int_0^{ t_f  }  \!  d t_1 F(t_1) 
	\int _0^{ t_1  } \!\! d t_2  
	G ^R _{\xi \xi } (t_1 - t_2)
	.
\end{align}
We can rewrite the expression on the right hand side using integration by parts as 
\begin{align}
& \Phi _{\mathrm{q}} (t_f) 
     = \frac{F[ 0 ]}{2} 
	\int_0^{ t_f  } \!\! d t_1  
	G ^R _{\xi \xi } ( t_1 )
	-\frac{1}{2} 
	\int_0^{ t_f  } \!\! d t_1 G ^R _{\xi \xi } (t_1) 
	\int_0^{ t_1  } \!\! d t_2  
	F ( t_2)
	.
\end{align}
The first term on the RHS can be viewed as the net phase shift due to a constant qubit frequency shift 
$ \int_0^{ t_f  } \!\! d t  
G ^R _{\xi \xi } ( t ) $
accumulated during the time evolution, whereas the second term accounts for a residual phase correction due to the fact that the quench-induced frequency shift to the qubit is time dependent.   
For Ohmic baths whose response functions exhibit linear dependence in the asymptotic low-frequency regime, it is straightforward to show that the asymptotic long-time behaviors of these two terms are given by 
\begin{align}
\lim_{ t_f \to + \infty } 
\frac{F[ 0 ]}{2} 
    \int_0^{ t_f  } \!\! d t  
    G ^R _{\xi \xi } ( t )
&  =
    \frac{F[ 0 ] }{2} 
    \int_0^{ + \infty  } \!\! 
    G ^R _{\xi \xi } (t) dt 
    =  \frac{F[ 0 ]}{2} 
    \text{Re}  G ^R _{\xi \xi  } [ \omega =0 ] 
    , 
    \\
-\frac{1}{2} 
\lim_{ t_f \to + \infty } 
\int_0^{ t_f  } \!\! d t_1 G ^R _{\xi \xi } (t_1) 
	\int_0^{ t_1  } \!\! d t_2  
	F ( t_2)
	& = 
	-\frac{1}{2} \int_0^{ + \infty  } t 
G ^R _{\xi \xi  } (t) dt
= -\frac{1}{2} \left .\frac{ d \text{Im}  
	G ^R _{\xi \xi  } [ \omega ]  }
	{d \omega }  
	\right|_{ \omega =0 } 
	.
\end{align}
Thus, the asymptotic long-time behavior of QPS with Ohmic baths  can be viewed as the sum of phase shift due to a static frequency shift in the long-time limit, which is proportional to $  F[ 0 ] $, and a residual phase correction. Noting that the bath spectral function $ \mathcal{J}[\omega] $ is related to the response function via $ \mathcal{J}[\omega] = 
    - \frac{1 }{\pi} 
\text{Im } G^R_{\xi \xi}[\omega] $ (see also Eq.~\eqref{eq:spec.func.def} in the main text), we have 
\begin{align}
\lim_{ t_f \to + \infty } 
\Phi _{\mathrm{q}} (t_f)
    & = \frac{  F[ 0 ]  }{2} 
 	 \text{Re}  G ^R _{\xi \xi  } [ \omega = 0 ] 
	- \frac{ 1 }{2} 
	\left .\frac{ d \text{Im}  
	G ^R _{\xi \xi  } [ \omega ]  }
	{d \omega }  
	\right|_{ \omega =0 } 
	\nonumber \\
	& = \frac{  F[ 0 ]  }{2} 
 	 \text{Re}  G ^R _{\xi \xi  } [ \omega = 0 ] 
	+ \frac{ \pi }{2} 
	\left . \frac{ d \mathcal{J}[\omega] }{d \omega }  
	\right|_{ \omega =0 }
	.
\end{align}
Specifically for dynamical-decoupling-type control pulses with $ F [ 0 ] 
= 0 $, the first term would vanish, and we recover Eq.~\eqref{eq:QPS.Ohmic.asymp} in the main text. 
As a result, for approximately Ohmic baths with spectral function satisfying 
$ \mathcal{J}[\omega] \sim \omega $ at low frequencies, the asymptotic behavior of QPS under spin-echo control pulses in the long time 
$ t_f \to \infty  $ regime is universal (i.e., it only depends on the asymptotic linear dependence of the spectral function), and is independent of the specific UV cutoff of the response function and details of the qubit control sequence.

\section{Discussion on the use of Hahn echo versus Ramsey coherence times in quench-enhanced QNS for Ohmic bath thermometry}

\label{appsec:OhmThermo.NSD}

In the main text, we show that our quench-enhanced $T _2 $-style quantum noise spectroscopy offers a direct route (i.e. without any curve fitting) to estimating the temperature of any baths exhibiting Ohmic behavior in the asymptotic low-frequency limit, i.e. 
$  \bar{S}  [ \omega ]  \sim \text{const.} $ and 
$ \mathcal{J}[\omega] \sim \omega $ as $ \omega \to 0$. 
However, realistic systems may also experience a large amount of quasistatic noise, leading to deviations from perfect Ohmic behavior at infinitesimal frequencies. In this Appendix, we discuss how our thermometry protocol also works in the presence of such quasistatic noise.  

Recall that our protocol can be summarized in Eq.~(\ref{eq:thermo.Ohmic}) in the main text, where we can extract bath temperature $ T $ from the quench phase shift $ \Phi  _{\mathrm{q}} (t_f)  $ and low-frequency noise spectral density $ \bar{S}  [ 0 ] $, via the following relation 
\begin{align}
\label{appeq:thermo.Ohmic}
T =   
\frac{  \bar{S}  [ 0 ] }
{4 k_B }  
\left[ 
\lim_{ t_f \to + \infty } 
\Phi  _{\mathrm{q}} (t_f)  \right ] ^{-1} 
.
\end{align}
Note that this result assumes the NSD is flat in the low-frequency limit, in which case the qubit Ramsey and Hahn-echo coherence times are necessarily identical. 
Here we stress that even in the circumstances where the qubit Hahn-echo time 
$ T_2 $ differs from the Ramsey coherence time $ T _\mathrm{FID}  $, a modified version of Eq.~\eqref{appeq:thermo.Ohmic} is still applicable, as long as the slow noise disrupting Ohmic NSD behavior emerges at a much lower frequency scale compared to the Ohmic regime. More specifically, this means the NSD 
$  \bar{S}  [ \omega ] $ and the spectral function 
$ \mathcal{J}[\omega] $ has a low-frequency cutoff 
$ \omega_\mathrm{ir} $, below which the Ohmic behavior $  \bar{S}  [ \omega ]  \sim \text{const.} $ and 
$ \mathcal{J}[\omega] \sim \omega $ breaks down. As mentioned, this includes the common physical situations, where the environment also has a large amount of quasistatic noise, which can be described as an additional delta function peak in the NSD.

It then follows that, our thermometry protocol is applicable to baths with asymptotic low-frequency Ohmic behavior, which may exhibit a high- as well as a low-frequency cutoff. For this more general scenario, we should use asymptotic low-frequency NSD $  \lim_{ \omega \to 0  } 
\bar{S}  [ \omega  ] = 2 / T _2 $, instead of strictly zero-frequency noise 
$ \bar{S}  [ 0  ] = 2 / T _\mathrm{FID} $ in Eq.~(\ref{appeq:thermo.Ohmic}). This justifies the use of Hahn-echo coherence time $ T_2 $ in the main text.

\section{General strategy for reconstructing the environmental spectral function in a generic frequency range using time-dependent quench functions}

\label{appsec:QPS.recon.specfunc}

In the main text, we discussed using sensor qubits based on a single nitrogen vacancy center in diamond to engineer a time-dependent quench (c.f. Fig.~\ref{fig:quench.filter}), and we discussed its application in reconstructing the bath spectral function for a specific type of control pulses. In this Appendix, we discuss a general recipe to construct more general periodic control pulses, which lead to a powerful set of varying spectral filters that can be utilized to reconstruct the spectral function $ \mathcal{J}[\omega] $ in a broad range of frequencies.

As discussed in Sec~\ref{sec:gen.quench.spec} in the main text, to illustrate the idea we focus on case where the quench operator is directly related to noise, with $ \hat V =   \hat{\xi} /2 $. Without loss of generality, we also focus on periodic NV control pulses, which are suitable for reconstructing the spectral function at finite target frequencies.   Recall that the spin-$1$ structure of the NV lets us effectively realize a nontrivial quench function 
$ \eta (t) $, in addition to the standard noise filter function. More specifically, we can apply a periodic sequence of NV control pulses (period $ T $ with 
$2M $ repetitions, $M \in \mathbb{Z}$), switching between the qubit subspaces $\{ m_z = 0, m_z = +1 \} $ and 
$ \{ m_z = 0, m_z = -1 \}$ (see also Fig.~\ref{fig:quench.filter}), to realize a periodic quench function as 
\begin{align}
    \eta (t)  & = 
\sum _{ m = 0 } ^{ M-1 } 
     \eta _0 (t - 2mT ; 2T ) 
     , 
\end{align}
where $ \eta _0 ( t ; 2T )  $ denotes the base quench function, and satisfies the following relation
\begin{align} 
\eta _0 ( t ; 2T ) & =
    \begin{cases}
    +1 \textrm{ or } -1,  & 0 \le t \le 2T,
    \\
    0,  & t<0 \text{ or } t>2T.
    \end{cases}
\end{align}
The structure of switching pulses also ensure that 
$ \eta _0 ( t ; 2T ) = (-)^N \eta _0 ( t +T ; 2T ) $ for $ 0 < t < T $, where $ N$ is the total number of switching pulses per period $T $. For the example control pulse sequence depicted in Fig.~\ref{fig:quench.filter}, we have $ N=1$ and 
$ \eta _0 ( t ; 2T ) 
= - \Theta (t ) \Theta ( T - t ) 
+ \Theta (t - T ) 
\Theta ( 2T - t ) $, where $ \Theta ( \cdot ) $ denotes the Heaviside step function. Again introducing the total protocol time satisfying $ t_f = 2MT$, we can straightforwardly rewrite Fourier transform of the quench function as 
\begin{align} 
\label{appeq:eta.FT.DD}
&  \eta [\omega]  = 
e^{i (M-1) \frac{\omega t_f }{2M}  } 
    \frac{\sin \frac{\omega t_f }{2} }
    { \sin \frac{\omega t_f }{2M} }
    \eta _0 [\omega ; 2T ] 
    , \\ 
&  \eta _0 [\omega ; 2T ] 
   \equiv \int_0^{ 2T  } 
   \eta _0 ( t ; 2T ) 
    e^{ i \omega t }  d t . 
\end{align}

For reasons that will become clear, we also assume a periodic sequence of standard qubit control $\pi$-pulses, with a same period $ T $ and total evolution time $ t_f = 2MT$, so that we similarly have
\begin{align} 
F (t)  & = 
\sum _{ m = 0 } ^{ M-1 } 
     F _0 (t - 2mT ; 2T ) 
    , \\ 
    \label{appeq:F.FT.DD}
 F [\omega]  &  = 
e^{i (M-1) \frac{\omega t_f }{2M}  } 
    \frac{\sin \frac{\omega t_f }{2} }
    { \sin \frac{\omega t_f }{2M} }
    F  _0 [\omega ; 2T ]  . 
\end{align}
Substituting Eqs.~\eqref{appeq:F.FT.DD} and \eqref{appeq:eta.FT.DD} above into Eq.~\eqref{appeq:Phi.Gau.gen} (or Eq.~\eqref{eq:Phi.Gau.gen} in the main text), which described the general quench phase shift, and noting that $  \hat V =   \hat{\xi} /2 $, we obtain
\begin{align}
\Phi _{\mathrm{q}}  (t_f) 
 & = 
    \int _{-\infty }^{ + \infty } 
    \! \frac{d  \omega }{2 \pi}
	F^*[\omega]
	\eta [\omega]
	G ^R _{\xi V }[ \omega ]
	\nonumber \\ 
	& =  	 
	\int _{-\infty }^{ + \infty } \! 
	\frac{d  \omega }{4 \pi}
	F^*[\omega] \eta [\omega]
	G ^R _{\xi \xi }[ \omega ] 
	\nonumber \\
	\label{appeq:QPS.eta.comb.GR}
	& =  
	\int _{-\infty }^{ + \infty } \! 
    \frac{d  \omega }{4 \pi} 
	\frac{\sin ^2 \frac{\omega t_f }{2} }
    { \sin ^2 \frac{\omega t_f }{2M} }
    F  ^*_0 [\omega ; 2T ] 
	\eta _0 [\omega ; 2T ] 
	G ^R _{\xi \xi }[ \omega ] 
	.
\end{align}

We are now ready to present the recipe, or the necessary and sufficient conditions, to construct spectral filters that specifically probe the bath spectral function $ \mathcal{J}[\omega] $ (see also Eq.~\eqref{eq:spec.func.def} in the main text)
\begin{equation}
    \mathcal{J}[\omega] = 
    - \frac{1 }{\pi} 
    \text{Im} G^R_{\xi \xi}[\omega]
    .
\end{equation}
We essentially require that the base filter and quench functions exhibit the same periodicity, and satisfy the following conditions
\begin{itemize}
    \item {The base filter and quench functions must be mirror symmetric or anti-symmetric with respect to $t=T $, i.e., 
    $ F _0 ( t ; 2T ) = s_F F _0 ( 2T-t ; 2T )  $, and 
    $ \eta _0 ( t ; 2T ) = s_\eta \eta _0 ( 2T-t ; 2T )  $, where $s_F$, $ s_\eta = \pm 1$. }
    \item {The base filter and quench functions exhibit opposite mirror symmetries with respect to $t=T $, i.e.,
    $s_F  = - s_\eta = + 1 $ or $ -1 $.}
\end{itemize}
Above constraints ensure that the quench phase shift in Eq.~\eqref{appeq:QPS.eta.comb.GR} is only sensitive to the imaginary part of the response function 
$ \text{Im} G^R_{\xi \xi}[\omega]$, or equivalently the spectral function $ \mathcal{J}[\omega] $, so that we have 
\begin{align}
\label{appeq:QPS.filter.gen}
 & \Phi _{\mathrm{q}}  (t_f) 
=   \int _{-\infty }^{ + \infty } \! 
	\mathcal{F} _{ \mathcal{J} } [\omega ; t_f ] 
	\mathcal{J}[\omega] 
	d  \omega 
	= 2 \int _{ 0 }^{ + \infty } \! 
	\mathcal{F} _{ \mathcal{J} } [\omega ; t_f ] 
	\mathcal{J}[\omega] 
	d  \omega 
    , \\
 &  \mathcal{F} _{ \mathcal{J} } [\omega ; t_f ] 
    =  	\frac{\sin ^2 \frac{\omega t_f }{2} }
    { 4 \sin ^2 \frac{\omega t_f }{2M} }
    \text{Im} 
    ( F  ^*_0 [\omega ; 2T ] 
	\eta _0 [\omega ; 2T ] ) 
	.
\end{align}
The spectral filter $ \mathcal{F} _{ \mathcal{J} } [\omega ; t_f ] $ for $\mathcal{J}[\omega] $ forms a comb-like structure in frequency space, if we fix pulse periodicity 
$ T = t_f/2M $ and take the asymptotic large pulse number limit, i.e.  
\begin{subequations}
\label{appeq:eta.comb.sum}
\begin{align} 
\label{appeq:eta.comb.asymp}
\mathcal{F} _{ \mathcal{J} } [\omega ; t_f ] 
& \sim   
  \frac{ M \omega _0 }{ 4 } 
    \sum _{\ell =  -\infty } ^{ + \infty} 
     \text{Im} 
    ( F  ^*_0 [ \ell \omega _0 ; 2T ] 
	\eta _0 [ \ell \omega _0 ; 2T ] )  
    \, \delta ( \omega 
    - \ell \omega _0 )
   \quad  ( M \gg 1)
   , \\
\omega _0 & =  \pi / T 
= 2  M \pi / t_f  
    . 
\end{align}
\end{subequations}
Thus, given a finite target frequency range, we can construct a corresponding set of NV control pulses that specifically realize frequency comb filters for the spectral function at target frequencies. We can then measure the quench phase shifts under these control pulses in the comb limit (fix $ T = t_f/2M $ and choose $M \gg 1 $), which in turn enable reconstruction of the spectral function 
$ \mathcal{J}[\omega] $ via Eq.~\eqref{appeq:QPS.filter.gen}.

\twocolumngrid

\input{quenchQNS.bbl}


\begin{thebibliography}{54}%
\makeatletter
\providecommand \@ifxundefined [1]{%
 \@ifx{#1\undefined}
}%
\providecommand \@ifnum [1]{%
 \ifnum #1\expandafter \@firstoftwo
 \else \expandafter \@secondoftwo
 \fi
}%
\providecommand \@ifx [1]{%
 \ifx #1\expandafter \@firstoftwo
 \else \expandafter \@secondoftwo
 \fi
}%
\providecommand \natexlab [1]{#1}%
\providecommand \enquote  [1]{``#1''}%
\providecommand \bibnamefont  [1]{#1}%
\providecommand \bibfnamefont [1]{#1}%
\providecommand \citenamefont [1]{#1}%
\providecommand \href@noop [0]{\@secondoftwo}%
\providecommand \href [0]{\begingroup \@sanitize@url \@href}%
\providecommand \@href[1]{\@@startlink{#1}\@@href}%
\providecommand \@@href[1]{\endgroup#1\@@endlink}%
\providecommand \@sanitize@url [0]{\catcode `\\12\catcode `\$12\catcode
  `\&12\catcode `\#12\catcode `\^12\catcode `\_12\catcode `\%12\relax}%
\providecommand \@@startlink[1]{}%
\providecommand \@@endlink[0]{}%
\providecommand \url  [0]{\begingroup\@sanitize@url \@url }%
\providecommand \@url [1]{\endgroup\@href {#1}{\urlprefix }}%
\providecommand \urlprefix  [0]{URL }%
\providecommand \Eprint [0]{\href }%
\providecommand \doibase [0]{http://dx.doi.org/}%
\providecommand \selectlanguage [0]{\@gobble}%
\providecommand \bibinfo  [0]{\@secondoftwo}%
\providecommand \bibfield  [0]{\@secondoftwo}%
\providecommand \translation [1]{[#1]}%
\providecommand \BibitemOpen [0]{}%
\providecommand \bibitemStop [0]{}%
\providecommand \bibitemNoStop [0]{.\EOS\space}%
\providecommand \EOS [0]{\spacefactor3000\relax}%
\providecommand \BibitemShut  [1]{\csname bibitem#1\endcsname}%
\let\auto@bib@innerbib\@empty
\bibitem [{\citenamefont {Degen}\ \emph {et~al.}(2017)\citenamefont {Degen},
  \citenamefont {Reinhard},\ and\ \citenamefont {Cappellaro}}]{Cappellaro2017}%
  \BibitemOpen
  \bibfield  {author} {\bibinfo {author} {\bibfnamefont {C.~L.}\ \bibnamefont
  {Degen}}, \bibinfo {author} {\bibfnamefont {F.}~\bibnamefont {Reinhard}}, \
  and\ \bibinfo {author} {\bibfnamefont {P.}~\bibnamefont {Cappellaro}},\
  }\href {\doibase 10.1103/RevModPhys.89.035002} {\bibfield  {journal}
  {\bibinfo  {journal} {Rev. Mod. Phys.}\ }\textbf {\bibinfo {volume} {89}},\
  \bibinfo {pages} {035002} (\bibinfo {year} {2017})}\BibitemShut {NoStop}%
\bibitem [{\citenamefont {Kofman}\ and\ \citenamefont
  {Kurizki}(2004)}]{Kurizki2004}%
  \BibitemOpen
  \bibfield  {author} {\bibinfo {author} {\bibfnamefont {A.~G.}\ \bibnamefont
  {Kofman}}\ and\ \bibinfo {author} {\bibfnamefont {G.}~\bibnamefont
  {Kurizki}},\ }\href {\doibase 10.1103/PhysRevLett.93.130406} {\bibfield
  {journal} {\bibinfo  {journal} {Phys. Rev. Lett.}\ }\textbf {\bibinfo
  {volume} {93}},\ \bibinfo {pages} {130406} (\bibinfo {year}
  {2004})}\BibitemShut {NoStop}%
\bibitem [{\citenamefont {de~Sousa}(2009)}]{deSousa2009}%
  \BibitemOpen
  \bibfield  {author} {\bibinfo {author} {\bibfnamefont {R.}~\bibnamefont
  {de~Sousa}},\ }\enquote {\bibinfo {title} {Electron spin as a spectrometer of
  nuclear-spin noise and other fluctuations},}\ in\ \href {\doibase
  10.1007/978-3-540-79365-6_10} {\emph {\bibinfo {booktitle} {Electron Spin
  Resonance and Related Phenomena in Low-Dimensional Structures}}},\ \bibinfo
  {editor} {edited by\ \bibinfo {editor} {\bibfnamefont {M.}~\bibnamefont
  {Fanciulli}}}\ (\bibinfo  {publisher} {Springer Berlin Heidelberg},\ \bibinfo
  {address} {Berlin, Heidelberg},\ \bibinfo {year} {2009})\ pp.\ \bibinfo
  {pages} {183--220}\BibitemShut {NoStop}%
\bibitem [{\citenamefont {de~Lange}\ \emph {et~al.}(2010)\citenamefont
  {de~Lange}, \citenamefont {Wang}, \citenamefont {Rist{\`e}}, \citenamefont
  {Dobrovitski},\ and\ \citenamefont {Hanson}}]{Hanson2010}%
  \BibitemOpen
  \bibfield  {author} {\bibinfo {author} {\bibfnamefont {G.}~\bibnamefont
  {de~Lange}}, \bibinfo {author} {\bibfnamefont {Z.~H.}\ \bibnamefont {Wang}},
  \bibinfo {author} {\bibfnamefont {D.}~\bibnamefont {Rist{\`e}}}, \bibinfo
  {author} {\bibfnamefont {V.~V.}\ \bibnamefont {Dobrovitski}}, \ and\ \bibinfo
  {author} {\bibfnamefont {R.}~\bibnamefont {Hanson}},\ }\href {\doibase
  10.1126/science.1192739} {\bibfield  {journal} {\bibinfo  {journal}
  {Science}\ }\textbf {\bibinfo {volume} {330}},\ \bibinfo {pages} {60}
  (\bibinfo {year} {2010})}\BibitemShut {NoStop}%
\bibitem [{\citenamefont {{\'A}lvarez}\ and\ \citenamefont
  {Suter}(2011)}]{Suter2011}%
  \BibitemOpen
  \bibfield  {author} {\bibinfo {author} {\bibfnamefont {G.~A.}\ \bibnamefont
  {{\'A}lvarez}}\ and\ \bibinfo {author} {\bibfnamefont {D.}~\bibnamefont
  {Suter}},\ }\href {\doibase 10.1103/PhysRevLett.107.230501} {\bibfield
  {journal} {\bibinfo  {journal} {Phys. Rev. Lett.}\ }\textbf {\bibinfo
  {volume} {107}},\ \bibinfo {pages} {230501} (\bibinfo {year}
  {2011})}\BibitemShut {NoStop}%
\bibitem [{\citenamefont {Bylander}\ \emph {et~al.}(2011)\citenamefont
  {Bylander}, \citenamefont {Gustavsson}, \citenamefont {Yan}, \citenamefont
  {Yoshihara}, \citenamefont {Harrabi}, \citenamefont {Fitch}, \citenamefont
  {Cory}, \citenamefont {Nakamura}, \citenamefont {Tsai},\ and\ \citenamefont
  {Oliver}}]{Oliver2011}%
  \BibitemOpen
  \bibfield  {author} {\bibinfo {author} {\bibfnamefont {J.}~\bibnamefont
  {Bylander}}, \bibinfo {author} {\bibfnamefont {S.}~\bibnamefont
  {Gustavsson}}, \bibinfo {author} {\bibfnamefont {F.}~\bibnamefont {Yan}},
  \bibinfo {author} {\bibfnamefont {F.}~\bibnamefont {Yoshihara}}, \bibinfo
  {author} {\bibfnamefont {K.}~\bibnamefont {Harrabi}}, \bibinfo {author}
  {\bibfnamefont {G.}~\bibnamefont {Fitch}}, \bibinfo {author} {\bibfnamefont
  {D.~G.}\ \bibnamefont {Cory}}, \bibinfo {author} {\bibfnamefont
  {Y.}~\bibnamefont {Nakamura}}, \bibinfo {author} {\bibfnamefont {J.~S.}\
  \bibnamefont {Tsai}}, \ and\ \bibinfo {author} {\bibfnamefont {W.~D.}\
  \bibnamefont {Oliver}},\ }\href {\doibase 10.1038/nphys1994} {\bibfield
  {journal} {\bibinfo  {journal} {Nat. Phys.}\ }\textbf {\bibinfo {volume}
  {7}},\ \bibinfo {pages} {565} (\bibinfo {year} {2011})}\BibitemShut {NoStop}%
\bibitem [{\citenamefont {Dial}\ \emph {et~al.}(2013)\citenamefont {Dial},
  \citenamefont {Shulman}, \citenamefont {Harvey}, \citenamefont {Bluhm},
  \citenamefont {Umansky},\ and\ \citenamefont {Yacoby}}]{Yacoby2013}%
  \BibitemOpen
  \bibfield  {author} {\bibinfo {author} {\bibfnamefont {O.~E.}\ \bibnamefont
  {Dial}}, \bibinfo {author} {\bibfnamefont {M.~D.}\ \bibnamefont {Shulman}},
  \bibinfo {author} {\bibfnamefont {S.~P.}\ \bibnamefont {Harvey}}, \bibinfo
  {author} {\bibfnamefont {H.}~\bibnamefont {Bluhm}}, \bibinfo {author}
  {\bibfnamefont {V.}~\bibnamefont {Umansky}}, \ and\ \bibinfo {author}
  {\bibfnamefont {A.}~\bibnamefont {Yacoby}},\ }\href {\doibase
  10.1103/PhysRevLett.110.146804} {\bibfield  {journal} {\bibinfo  {journal}
  {Phys. Rev. Lett.}\ }\textbf {\bibinfo {volume} {110}},\ \bibinfo {pages}
  {146804} (\bibinfo {year} {2013})}\BibitemShut {NoStop}%
\bibitem [{\citenamefont {Romach}\ \emph {et~al.}(2015)\citenamefont {Romach},
  \citenamefont {M\"uller}, \citenamefont {Unden}, \citenamefont {Rogers},
  \citenamefont {Isoda}, \citenamefont {Itoh}, \citenamefont {Markham},
  \citenamefont {Stacey}, \citenamefont {Meijer}, \citenamefont {Pezzagna},
  \citenamefont {Naydenov}, \citenamefont {McGuinness}, \citenamefont
  {Bar-Gill},\ and\ \citenamefont {Jelezko}}]{Jelezko2015}%
  \BibitemOpen
  \bibfield  {author} {\bibinfo {author} {\bibfnamefont {Y.}~\bibnamefont
  {Romach}}, \bibinfo {author} {\bibfnamefont {C.}~\bibnamefont {M\"uller}},
  \bibinfo {author} {\bibfnamefont {T.}~\bibnamefont {Unden}}, \bibinfo
  {author} {\bibfnamefont {L.~J.}\ \bibnamefont {Rogers}}, \bibinfo {author}
  {\bibfnamefont {T.}~\bibnamefont {Isoda}}, \bibinfo {author} {\bibfnamefont
  {K.~M.}\ \bibnamefont {Itoh}}, \bibinfo {author} {\bibfnamefont
  {M.}~\bibnamefont {Markham}}, \bibinfo {author} {\bibfnamefont
  {A.}~\bibnamefont {Stacey}}, \bibinfo {author} {\bibfnamefont
  {J.}~\bibnamefont {Meijer}}, \bibinfo {author} {\bibfnamefont
  {S.}~\bibnamefont {Pezzagna}}, \bibinfo {author} {\bibfnamefont
  {B.}~\bibnamefont {Naydenov}}, \bibinfo {author} {\bibfnamefont {L.~P.}\
  \bibnamefont {McGuinness}}, \bibinfo {author} {\bibfnamefont
  {N.}~\bibnamefont {Bar-Gill}}, \ and\ \bibinfo {author} {\bibfnamefont
  {F.}~\bibnamefont {Jelezko}},\ }\href {\doibase
  10.1103/PhysRevLett.114.017601} {\bibfield  {journal} {\bibinfo  {journal}
  {Phys. Rev. Lett.}\ }\textbf {\bibinfo {volume} {114}},\ \bibinfo {pages}
  {017601} (\bibinfo {year} {2015})}\BibitemShut {NoStop}%
\bibitem [{\citenamefont {Frey}\ \emph {et~al.}(2017)\citenamefont {Frey},
  \citenamefont {Mavadia}, \citenamefont {Norris}, \citenamefont {de~Ferranti},
  \citenamefont {Lucarelli}, \citenamefont {Viola},\ and\ \citenamefont
  {Biercuk}}]{Biercuk2017}%
  \BibitemOpen
  \bibfield  {author} {\bibinfo {author} {\bibfnamefont {V.~M.}\ \bibnamefont
  {Frey}}, \bibinfo {author} {\bibfnamefont {S.}~\bibnamefont {Mavadia}},
  \bibinfo {author} {\bibfnamefont {L.~M.}\ \bibnamefont {Norris}}, \bibinfo
  {author} {\bibfnamefont {W.}~\bibnamefont {de~Ferranti}}, \bibinfo {author}
  {\bibfnamefont {D.}~\bibnamefont {Lucarelli}}, \bibinfo {author}
  {\bibfnamefont {L.}~\bibnamefont {Viola}}, \ and\ \bibinfo {author}
  {\bibfnamefont {M.~J.}\ \bibnamefont {Biercuk}},\ }\href
  {https://doi.org/10.1038/s41467-017-02298-2} {\bibfield  {journal} {\bibinfo
  {journal} {Nat. Commun.}\ }\textbf {\bibinfo {volume} {8}},\ \bibinfo {pages}
  {2189} (\bibinfo {year} {2017})}\BibitemShut {NoStop}%
\bibitem [{\citenamefont {Norris}\ \emph {et~al.}(2018)\citenamefont {Norris},
  \citenamefont {Lucarelli}, \citenamefont {Frey}, \citenamefont {Mavadia},
  \citenamefont {Biercuk},\ and\ \citenamefont {Viola}}]{Viola2018}%
  \BibitemOpen
  \bibfield  {author} {\bibinfo {author} {\bibfnamefont {L.~M.}\ \bibnamefont
  {Norris}}, \bibinfo {author} {\bibfnamefont {D.}~\bibnamefont {Lucarelli}},
  \bibinfo {author} {\bibfnamefont {V.~M.}\ \bibnamefont {Frey}}, \bibinfo
  {author} {\bibfnamefont {S.}~\bibnamefont {Mavadia}}, \bibinfo {author}
  {\bibfnamefont {M.~J.}\ \bibnamefont {Biercuk}}, \ and\ \bibinfo {author}
  {\bibfnamefont {L.}~\bibnamefont {Viola}},\ }\href {\doibase
  10.1103/PhysRevA.98.032315} {\bibfield  {journal} {\bibinfo  {journal} {Phys.
  Rev. A}\ }\textbf {\bibinfo {volume} {98}},\ \bibinfo {pages} {032315}
  (\bibinfo {year} {2018})}\BibitemShut {NoStop}%
\bibitem [{\citenamefont {Chan}\ \emph {et~al.}(2018)\citenamefont {Chan},
  \citenamefont {Huang}, \citenamefont {Yang}, \citenamefont {Hwang},
  \citenamefont {Hensen}, \citenamefont {Tanttu}, \citenamefont {Hudson},
  \citenamefont {Itoh}, \citenamefont {Laucht}, \citenamefont {Morello},\ and\
  \citenamefont {Dzurak}}]{Dzurak2018}%
  \BibitemOpen
  \bibfield  {author} {\bibinfo {author} {\bibfnamefont {K.~W.}\ \bibnamefont
  {Chan}}, \bibinfo {author} {\bibfnamefont {W.}~\bibnamefont {Huang}},
  \bibinfo {author} {\bibfnamefont {C.~H.}\ \bibnamefont {Yang}}, \bibinfo
  {author} {\bibfnamefont {J.~C.~C.}\ \bibnamefont {Hwang}}, \bibinfo {author}
  {\bibfnamefont {B.}~\bibnamefont {Hensen}}, \bibinfo {author} {\bibfnamefont
  {T.}~\bibnamefont {Tanttu}}, \bibinfo {author} {\bibfnamefont {F.~E.}\
  \bibnamefont {Hudson}}, \bibinfo {author} {\bibfnamefont {K.~M.}\
  \bibnamefont {Itoh}}, \bibinfo {author} {\bibfnamefont {A.}~\bibnamefont
  {Laucht}}, \bibinfo {author} {\bibfnamefont {A.}~\bibnamefont {Morello}}, \
  and\ \bibinfo {author} {\bibfnamefont {A.~S.}\ \bibnamefont {Dzurak}},\
  }\href {\doibase 10.1103/PhysRevApplied.10.044017} {\bibfield  {journal}
  {\bibinfo  {journal} {Phys. Rev. Applied}\ }\textbf {\bibinfo {volume}
  {10}},\ \bibinfo {pages} {044017} (\bibinfo {year} {2018})}\BibitemShut
  {NoStop}%
\bibitem [{\citenamefont {Ferrie}\ \emph {et~al.}(2018)\citenamefont {Ferrie},
  \citenamefont {Granade}, \citenamefont {Paz-Silva},\ and\ \citenamefont
  {Wiseman}}]{Wiseman2018}%
  \BibitemOpen
  \bibfield  {author} {\bibinfo {author} {\bibfnamefont {C.}~\bibnamefont
  {Ferrie}}, \bibinfo {author} {\bibfnamefont {C.}~\bibnamefont {Granade}},
  \bibinfo {author} {\bibfnamefont {G.}~\bibnamefont {Paz-Silva}}, \ and\
  \bibinfo {author} {\bibfnamefont {H.~M.}\ \bibnamefont {Wiseman}},\ }\href
  {\doibase 10.1088/1367-2630/aaf207} {\bibfield  {journal} {\bibinfo
  {journal} {New J. Phys.}\ }\textbf {\bibinfo {volume} {20}},\ \bibinfo
  {pages} {123005} (\bibinfo {year} {2018})}\BibitemShut {NoStop}%
\bibitem [{\citenamefont {Sinitsyn}\ and\ \citenamefont
  {Pershin}(2016)}]{Pershin2016}%
  \BibitemOpen
  \bibfield  {author} {\bibinfo {author} {\bibfnamefont {N.~A.}\ \bibnamefont
  {Sinitsyn}}\ and\ \bibinfo {author} {\bibfnamefont {Y.~V.}\ \bibnamefont
  {Pershin}},\ }\href {\doibase 10.1088/0034-4885/79/10/106501} {\bibfield
  {journal} {\bibinfo  {journal} {Rep. Prog. Phys.}\ }\textbf {\bibinfo
  {volume} {79}},\ \bibinfo {pages} {106501} (\bibinfo {year}
  {2016})}\BibitemShut {NoStop}%
\bibitem [{\citenamefont {Norris}\ \emph {et~al.}(2016)\citenamefont {Norris},
  \citenamefont {Paz-Silva},\ and\ \citenamefont {Viola}}]{Viola2016nG}%
  \BibitemOpen
  \bibfield  {author} {\bibinfo {author} {\bibfnamefont {L.~M.}\ \bibnamefont
  {Norris}}, \bibinfo {author} {\bibfnamefont {G.~A.}\ \bibnamefont
  {Paz-Silva}}, \ and\ \bibinfo {author} {\bibfnamefont {L.}~\bibnamefont
  {Viola}},\ }\href {\doibase 10.1103/PhysRevLett.116.150503} {\bibfield
  {journal} {\bibinfo  {journal} {Phys. Rev. Lett.}\ }\textbf {\bibinfo
  {volume} {116}},\ \bibinfo {pages} {150503} (\bibinfo {year}
  {2016})}\BibitemShut {NoStop}%
\bibitem [{\citenamefont {Sza{\'{n}}kowski}\ \emph {et~al.}(2017)\citenamefont
  {Sza{\'{n}}kowski}, \citenamefont {Ramon}, \citenamefont {Krzywda},
  \citenamefont {Kwiatkowski},\ and\ \citenamefont
  {Cywi{\'{n}}ski}}]{Cywinski2017}%
  \BibitemOpen
  \bibfield  {author} {\bibinfo {author} {\bibfnamefont {P.}~\bibnamefont
  {Sza{\'{n}}kowski}}, \bibinfo {author} {\bibfnamefont {G.}~\bibnamefont
  {Ramon}}, \bibinfo {author} {\bibfnamefont {J.}~\bibnamefont {Krzywda}},
  \bibinfo {author} {\bibfnamefont {D.}~\bibnamefont {Kwiatkowski}}, \ and\
  \bibinfo {author} {\bibfnamefont {{\L}.}~\bibnamefont {Cywi{\'{n}}ski}},\
  }\href {\doibase 10.1088/1361-648x/aa7648} {\bibfield  {journal} {\bibinfo
  {journal} {J. Phys.: Cond. Matter}\ }\textbf {\bibinfo {volume} {29}},\
  \bibinfo {pages} {333001} (\bibinfo {year} {2017})}\BibitemShut {NoStop}%
\bibitem [{\citenamefont {Yan}\ \emph {et~al.}(2018)\citenamefont {Yan},
  \citenamefont {Campbell}, \citenamefont {Krantz}, \citenamefont {Kjaergaard},
  \citenamefont {Kim}, \citenamefont {Yoder}, \citenamefont {Hover},
  \citenamefont {Sears}, \citenamefont {Kerman}, \citenamefont {Orlando},
  \citenamefont {Gustavsson},\ and\ \citenamefont {Oliver}}]{Oliver2018}%
  \BibitemOpen
  \bibfield  {author} {\bibinfo {author} {\bibfnamefont {F.}~\bibnamefont
  {Yan}}, \bibinfo {author} {\bibfnamefont {D.}~\bibnamefont {Campbell}},
  \bibinfo {author} {\bibfnamefont {P.}~\bibnamefont {Krantz}}, \bibinfo
  {author} {\bibfnamefont {M.}~\bibnamefont {Kjaergaard}}, \bibinfo {author}
  {\bibfnamefont {D.}~\bibnamefont {Kim}}, \bibinfo {author} {\bibfnamefont
  {J.~L.}\ \bibnamefont {Yoder}}, \bibinfo {author} {\bibfnamefont
  {D.}~\bibnamefont {Hover}}, \bibinfo {author} {\bibfnamefont
  {A.}~\bibnamefont {Sears}}, \bibinfo {author} {\bibfnamefont {A.~J.}\
  \bibnamefont {Kerman}}, \bibinfo {author} {\bibfnamefont {T.~P.}\
  \bibnamefont {Orlando}}, \bibinfo {author} {\bibfnamefont {S.}~\bibnamefont
  {Gustavsson}}, \ and\ \bibinfo {author} {\bibfnamefont {W.~D.}\ \bibnamefont
  {Oliver}},\ }\href {\doibase 10.1103/PhysRevLett.120.260504} {\bibfield
  {journal} {\bibinfo  {journal} {Phys. Rev. Lett.}\ }\textbf {\bibinfo
  {volume} {120}},\ \bibinfo {pages} {260504} (\bibinfo {year}
  {2018})}\BibitemShut {NoStop}%
\bibitem [{\citenamefont {Sung}\ \emph {et~al.}(2019)\citenamefont {Sung},
  \citenamefont {Beaudoin}, \citenamefont {Norris}, \citenamefont {Yan},
  \citenamefont {Kim}, \citenamefont {Qiu}, \citenamefont {von L{\"{u}}pke},
  \citenamefont {Yoder}, \citenamefont {Orlando}, \citenamefont {Gustavsson},
  \citenamefont {Viola},\ and\ \citenamefont {Oliver}}]{Oliver2019}%
  \BibitemOpen
  \bibfield  {author} {\bibinfo {author} {\bibfnamefont {Y.}~\bibnamefont
  {Sung}}, \bibinfo {author} {\bibfnamefont {F.}~\bibnamefont {Beaudoin}},
  \bibinfo {author} {\bibfnamefont {L.~M.}\ \bibnamefont {Norris}}, \bibinfo
  {author} {\bibfnamefont {F.}~\bibnamefont {Yan}}, \bibinfo {author}
  {\bibfnamefont {D.~K.}\ \bibnamefont {Kim}}, \bibinfo {author} {\bibfnamefont
  {J.~Y.}\ \bibnamefont {Qiu}}, \bibinfo {author} {\bibfnamefont
  {U.}~\bibnamefont {von L{\"{u}}pke}}, \bibinfo {author} {\bibfnamefont
  {J.~L.}\ \bibnamefont {Yoder}}, \bibinfo {author} {\bibfnamefont {T.~P.}\
  \bibnamefont {Orlando}}, \bibinfo {author} {\bibfnamefont {S.}~\bibnamefont
  {Gustavsson}}, \bibinfo {author} {\bibfnamefont {L.}~\bibnamefont {Viola}}, \
  and\ \bibinfo {author} {\bibfnamefont {W.~D.}\ \bibnamefont {Oliver}},\
  }\href {\doibase 10.1038/s41467-019-11699-4} {\bibfield  {journal} {\bibinfo
  {journal} {Nat. Commun.}\ }\textbf {\bibinfo {volume} {10}},\ \bibinfo
  {pages} {3715} (\bibinfo {year} {2019})}\BibitemShut {NoStop}%
\bibitem [{\citenamefont {von L\"upke}\ \emph {et~al.}(2020)\citenamefont {von
  L\"upke}, \citenamefont {Beaudoin}, \citenamefont {Norris}, \citenamefont
  {Sung}, \citenamefont {Winik}, \citenamefont {Qiu}, \citenamefont
  {Kjaergaard}, \citenamefont {Kim}, \citenamefont {Yoder}, \citenamefont
  {Gustavsson}, \citenamefont {Viola},\ and\ \citenamefont
  {Oliver}}]{Oliver2020}%
  \BibitemOpen
  \bibfield  {author} {\bibinfo {author} {\bibfnamefont {U.}~\bibnamefont {von
  L\"upke}}, \bibinfo {author} {\bibfnamefont {F.}~\bibnamefont {Beaudoin}},
  \bibinfo {author} {\bibfnamefont {L.~M.}\ \bibnamefont {Norris}}, \bibinfo
  {author} {\bibfnamefont {Y.}~\bibnamefont {Sung}}, \bibinfo {author}
  {\bibfnamefont {R.}~\bibnamefont {Winik}}, \bibinfo {author} {\bibfnamefont
  {J.~Y.}\ \bibnamefont {Qiu}}, \bibinfo {author} {\bibfnamefont
  {M.}~\bibnamefont {Kjaergaard}}, \bibinfo {author} {\bibfnamefont
  {D.}~\bibnamefont {Kim}}, \bibinfo {author} {\bibfnamefont {J.}~\bibnamefont
  {Yoder}}, \bibinfo {author} {\bibfnamefont {S.}~\bibnamefont {Gustavsson}},
  \bibinfo {author} {\bibfnamefont {L.}~\bibnamefont {Viola}}, \ and\ \bibinfo
  {author} {\bibfnamefont {W.~D.}\ \bibnamefont {Oliver}},\ }\href {\doibase
  10.1103/PRXQuantum.1.010305} {\bibfield  {journal} {\bibinfo  {journal} {PRX
  Quantum}\ }\textbf {\bibinfo {volume} {1}},\ \bibinfo {pages} {010305}
  (\bibinfo {year} {2020})}\BibitemShut {NoStop}%
\bibitem [{\citenamefont {Sung}\ \emph {et~al.}(2021)\citenamefont {Sung},
  \citenamefont {Veps{\"a}l{\"a}inen}, \citenamefont {Braum{\"u}ller},
  \citenamefont {Yan}, \citenamefont {Wang}, \citenamefont {Kjaergaard},
  \citenamefont {Winik}, \citenamefont {Krantz}, \citenamefont {Bengtsson},
  \citenamefont {Melville}, \citenamefont {Niedzielski}, \citenamefont
  {Schwartz}, \citenamefont {Kim}, \citenamefont {Yoder}, \citenamefont
  {Orlando}, \citenamefont {Gustavsson},\ and\ \citenamefont
  {Oliver}}]{Oliver2021}%
  \BibitemOpen
  \bibfield  {author} {\bibinfo {author} {\bibfnamefont {Y.}~\bibnamefont
  {Sung}}, \bibinfo {author} {\bibfnamefont {A.}~\bibnamefont
  {Veps{\"a}l{\"a}inen}}, \bibinfo {author} {\bibfnamefont {J.}~\bibnamefont
  {Braum{\"u}ller}}, \bibinfo {author} {\bibfnamefont {F.}~\bibnamefont {Yan}},
  \bibinfo {author} {\bibfnamefont {J.~I.-J.}\ \bibnamefont {Wang}}, \bibinfo
  {author} {\bibfnamefont {M.}~\bibnamefont {Kjaergaard}}, \bibinfo {author}
  {\bibfnamefont {R.}~\bibnamefont {Winik}}, \bibinfo {author} {\bibfnamefont
  {P.}~\bibnamefont {Krantz}}, \bibinfo {author} {\bibfnamefont
  {A.}~\bibnamefont {Bengtsson}}, \bibinfo {author} {\bibfnamefont {A.~J.}\
  \bibnamefont {Melville}}, \bibinfo {author} {\bibfnamefont {B.~M.}\
  \bibnamefont {Niedzielski}}, \bibinfo {author} {\bibfnamefont {M.~E.}\
  \bibnamefont {Schwartz}}, \bibinfo {author} {\bibfnamefont {D.~K.}\
  \bibnamefont {Kim}}, \bibinfo {author} {\bibfnamefont {J.~L.}\ \bibnamefont
  {Yoder}}, \bibinfo {author} {\bibfnamefont {T.~P.}\ \bibnamefont {Orlando}},
  \bibinfo {author} {\bibfnamefont {S.}~\bibnamefont {Gustavsson}}, \ and\
  \bibinfo {author} {\bibfnamefont {W.~D.}\ \bibnamefont {Oliver}},\ }\href
  {https://doi.org/10.1038/s41467-021-21098-3} {\bibfield  {journal} {\bibinfo
  {journal} {Nat. Commun.}\ }\textbf {\bibinfo {volume} {12}},\ \bibinfo
  {pages} {967} (\bibinfo {year} {2021})}\BibitemShut {NoStop}%
\bibitem [{\citenamefont {Bar-Gill}\ \emph {et~al.}(2012)\citenamefont
  {Bar-Gill}, \citenamefont {Pham}, \citenamefont {Belthangady}, \citenamefont
  {Sage}, \citenamefont {Cappellaro}, \citenamefont {Maze}, \citenamefont
  {Lukin}, \citenamefont {Yacoby},\ and\ \citenamefont
  {Walsworth}}]{Walsworth2012}%
  \BibitemOpen
  \bibfield  {author} {\bibinfo {author} {\bibfnamefont {N.}~\bibnamefont
  {Bar-Gill}}, \bibinfo {author} {\bibfnamefont {L.}~\bibnamefont {Pham}},
  \bibinfo {author} {\bibfnamefont {C.}~\bibnamefont {Belthangady}}, \bibinfo
  {author} {\bibfnamefont {D.~L.}\ \bibnamefont {Sage}}, \bibinfo {author}
  {\bibfnamefont {P.}~\bibnamefont {Cappellaro}}, \bibinfo {author}
  {\bibfnamefont {J.}~\bibnamefont {Maze}}, \bibinfo {author} {\bibfnamefont
  {M.}~\bibnamefont {Lukin}}, \bibinfo {author} {\bibfnamefont
  {A.}~\bibnamefont {Yacoby}}, \ and\ \bibinfo {author} {\bibfnamefont
  {R.}~\bibnamefont {Walsworth}},\ }\href {https://doi.org/10.1038/ncomms1856}
  {\bibfield  {journal} {\bibinfo  {journal} {Nat. Commun.}\ }\textbf {\bibinfo
  {volume} {3}},\ \bibinfo {pages} {858} (\bibinfo {year} {2012})}\BibitemShut
  {NoStop}%
\bibitem [{\citenamefont {Peng}\ \emph {et~al.}(2015)\citenamefont {Peng},
  \citenamefont {Zhou}, \citenamefont {Wei}, \citenamefont {Cui}, \citenamefont
  {Du},\ and\ \citenamefont {Liu}}]{Liu2015}%
  \BibitemOpen
  \bibfield  {author} {\bibinfo {author} {\bibfnamefont {X.}~\bibnamefont
  {Peng}}, \bibinfo {author} {\bibfnamefont {H.}~\bibnamefont {Zhou}}, \bibinfo
  {author} {\bibfnamefont {B.-B.}\ \bibnamefont {Wei}}, \bibinfo {author}
  {\bibfnamefont {J.}~\bibnamefont {Cui}}, \bibinfo {author} {\bibfnamefont
  {J.}~\bibnamefont {Du}}, \ and\ \bibinfo {author} {\bibfnamefont {R.-B.}\
  \bibnamefont {Liu}},\ }\href {\doibase 10.1103/PhysRevLett.114.010601}
  {\bibfield  {journal} {\bibinfo  {journal} {Phys. Rev. Lett.}\ }\textbf
  {\bibinfo {volume} {114}},\ \bibinfo {pages} {010601} (\bibinfo {year}
  {2015})}\BibitemShut {NoStop}%
\bibitem [{\citenamefont {Casola}\ \emph {et~al.}(2018)\citenamefont {Casola},
  \citenamefont {van~der Sar},\ and\ \citenamefont {Yacoby}}]{Yacoby2018}%
  \BibitemOpen
  \bibfield  {author} {\bibinfo {author} {\bibfnamefont {F.}~\bibnamefont
  {Casola}}, \bibinfo {author} {\bibfnamefont {T.}~\bibnamefont {van~der Sar}},
  \ and\ \bibinfo {author} {\bibfnamefont {A.}~\bibnamefont {Yacoby}},\ }\href
  {https://doi.org/10.1038/natrevmats.2017.88} {\bibfield  {journal} {\bibinfo
  {journal} {Nat. Rev. Mater.}\ }\textbf {\bibinfo {volume} {3}},\ \bibinfo
  {pages} {17088} (\bibinfo {year} {2018})}\BibitemShut {NoStop}%
\bibitem [{\citenamefont {Ye}\ \emph {et~al.}(2020)\citenamefont {Ye},
  \citenamefont {Machado}, \citenamefont {White}, \citenamefont {Mong},\ and\
  \citenamefont {Yao}}]{Yao2020}%
  \BibitemOpen
  \bibfield  {author} {\bibinfo {author} {\bibfnamefont {B.}~\bibnamefont
  {Ye}}, \bibinfo {author} {\bibfnamefont {F.}~\bibnamefont {Machado}},
  \bibinfo {author} {\bibfnamefont {C.~D.}\ \bibnamefont {White}}, \bibinfo
  {author} {\bibfnamefont {R.~S.~K.}\ \bibnamefont {Mong}}, \ and\ \bibinfo
  {author} {\bibfnamefont {N.~Y.}\ \bibnamefont {Yao}},\ }\href {\doibase
  10.1103/PhysRevLett.125.030601} {\bibfield  {journal} {\bibinfo  {journal}
  {Phys. Rev. Lett.}\ }\textbf {\bibinfo {volume} {125}},\ \bibinfo {pages}
  {030601} (\bibinfo {year} {2020})}\BibitemShut {NoStop}%
\bibitem [{\citenamefont {Davis}\ \emph {et~al.}(2021)\citenamefont {Davis},
  \citenamefont {Ye}, \citenamefont {Machado}, \citenamefont {Meynell},
  \citenamefont {Mittiga}, \citenamefont {Schenken}, \citenamefont {Joos},
  \citenamefont {Kobrin}, \citenamefont {Lyu}, \citenamefont {Bluvstein},
  \citenamefont {Choi}, \citenamefont {Zu}, \citenamefont {Jayich},\ and\
  \citenamefont {Yao}}]{Yao2021}%
  \BibitemOpen
  \bibfield  {author} {\bibinfo {author} {\bibfnamefont {E.~J.}\ \bibnamefont
  {Davis}}, \bibinfo {author} {\bibfnamefont {B.}~\bibnamefont {Ye}}, \bibinfo
  {author} {\bibfnamefont {F.}~\bibnamefont {Machado}}, \bibinfo {author}
  {\bibfnamefont {S.~A.}\ \bibnamefont {Meynell}}, \bibinfo {author}
  {\bibfnamefont {T.}~\bibnamefont {Mittiga}}, \bibinfo {author} {\bibfnamefont
  {W.}~\bibnamefont {Schenken}}, \bibinfo {author} {\bibfnamefont
  {M.}~\bibnamefont {Joos}}, \bibinfo {author} {\bibfnamefont {B.}~\bibnamefont
  {Kobrin}}, \bibinfo {author} {\bibfnamefont {Y.}~\bibnamefont {Lyu}},
  \bibinfo {author} {\bibfnamefont {D.}~\bibnamefont {Bluvstein}}, \bibinfo
  {author} {\bibfnamefont {S.}~\bibnamefont {Choi}}, \bibinfo {author}
  {\bibfnamefont {C.}~\bibnamefont {Zu}}, \bibinfo {author} {\bibfnamefont
  {A.~C.~B.}\ \bibnamefont {Jayich}}, \ and\ \bibinfo {author} {\bibfnamefont
  {N.~Y.}\ \bibnamefont {Yao}},\ }\href {https://arxiv.org/abs/2103.12742}
  {\bibfield  {journal} {\bibinfo  {journal} {arXiv preprint arXiv:2103.12742}\
  } (\bibinfo {year} {2021})}\BibitemShut {NoStop}%
\bibitem [{\citenamefont {Zhao}\ \emph {et~al.}(2011)\citenamefont {Zhao},
  \citenamefont {Wang},\ and\ \citenamefont {Liu}}]{RBLiu2011}%
  \BibitemOpen
  \bibfield  {author} {\bibinfo {author} {\bibfnamefont {N.}~\bibnamefont
  {Zhao}}, \bibinfo {author} {\bibfnamefont {Z.-Y.}\ \bibnamefont {Wang}}, \
  and\ \bibinfo {author} {\bibfnamefont {R.-B.}\ \bibnamefont {Liu}},\ }\href
  {\doibase 10.1103/PhysRevLett.106.217205} {\bibfield  {journal} {\bibinfo
  {journal} {Phys. Rev. Lett.}\ }\textbf {\bibinfo {volume} {106}},\ \bibinfo
  {pages} {217205} (\bibinfo {year} {2011})}\BibitemShut {NoStop}%
\bibitem [{\citenamefont {Huang}\ \emph {et~al.}(2011)\citenamefont {Huang},
  \citenamefont {Kong}, \citenamefont {Zhao}, \citenamefont {Shi},
  \citenamefont {Wang}, \citenamefont {Rong}, \citenamefont {Liu},\ and\
  \citenamefont {Du}}]{Du2011}%
  \BibitemOpen
  \bibfield  {author} {\bibinfo {author} {\bibfnamefont {P.}~\bibnamefont
  {Huang}}, \bibinfo {author} {\bibfnamefont {X.}~\bibnamefont {Kong}},
  \bibinfo {author} {\bibfnamefont {N.}~\bibnamefont {Zhao}}, \bibinfo {author}
  {\bibfnamefont {F.}~\bibnamefont {Shi}}, \bibinfo {author} {\bibfnamefont
  {P.}~\bibnamefont {Wang}}, \bibinfo {author} {\bibfnamefont {X.}~\bibnamefont
  {Rong}}, \bibinfo {author} {\bibfnamefont {R.-B.}\ \bibnamefont {Liu}}, \
  and\ \bibinfo {author} {\bibfnamefont {J.}~\bibnamefont {Du}},\ }\href
  {\doibase 10.1038/ncomms1579} {\bibfield  {journal} {\bibinfo  {journal}
  {Nat. Commun.}\ }\textbf {\bibinfo {volume} {2}},\ \bibinfo {pages} {570}
  (\bibinfo {year} {2011})}\BibitemShut {NoStop}%
\bibitem [{\citenamefont {Yang}\ \emph {et~al.}(2016)\citenamefont {Yang},
  \citenamefont {Ma},\ and\ \citenamefont {Liu}}]{Yang2016}%
  \BibitemOpen
  \bibfield  {author} {\bibinfo {author} {\bibfnamefont {W.}~\bibnamefont
  {Yang}}, \bibinfo {author} {\bibfnamefont {W.-L.}\ \bibnamefont {Ma}}, \ and\
  \bibinfo {author} {\bibfnamefont {R.-B.}\ \bibnamefont {Liu}},\ }\href
  {\doibase 10.1088/0034-4885/80/1/016001} {\bibfield  {journal} {\bibinfo
  {journal} {Rep. Prog. Phys.}\ }\textbf {\bibinfo {volume} {80}},\ \bibinfo
  {pages} {016001} (\bibinfo {year} {2016})}\BibitemShut {NoStop}%
\bibitem [{\citenamefont {Wang}\ \emph {et~al.}(2019)\citenamefont {Wang},
  \citenamefont {Chen}, \citenamefont {Peng}, \citenamefont {Wrachtrup},\ and\
  \citenamefont {Liu}}]{RBLiu2019}%
  \BibitemOpen
  \bibfield  {author} {\bibinfo {author} {\bibfnamefont {P.}~\bibnamefont
  {Wang}}, \bibinfo {author} {\bibfnamefont {C.}~\bibnamefont {Chen}}, \bibinfo
  {author} {\bibfnamefont {X.}~\bibnamefont {Peng}}, \bibinfo {author}
  {\bibfnamefont {J.}~\bibnamefont {Wrachtrup}}, \ and\ \bibinfo {author}
  {\bibfnamefont {R.-B.}\ \bibnamefont {Liu}},\ }\href {\doibase
  10.1103/PhysRevLett.123.050603} {\bibfield  {journal} {\bibinfo  {journal}
  {Phys. Rev. Lett.}\ }\textbf {\bibinfo {volume} {123}},\ \bibinfo {pages}
  {050603} (\bibinfo {year} {2019})}\BibitemShut {NoStop}%
\bibitem [{\citenamefont {Wang}\ and\ \citenamefont {Clerk}(2020)}]{Clerk2020}%
  \BibitemOpen
  \bibfield  {author} {\bibinfo {author} {\bibfnamefont {Y.-X.}\ \bibnamefont
  {Wang}}\ and\ \bibinfo {author} {\bibfnamefont {A.~A.}\ \bibnamefont
  {Clerk}},\ }\href {\doibase 10.1103/PhysRevResearch.2.033196} {\bibfield
  {journal} {\bibinfo  {journal} {Phys. Rev. Research}\ }\textbf {\bibinfo
  {volume} {2}},\ \bibinfo {pages} {033196} (\bibinfo {year}
  {2020})}\BibitemShut {NoStop}%
\bibitem [{\citenamefont {Sza{\'{n}}kowski}\ and\ \citenamefont
  {Cywi{\'{n}}ski}(2020)}]{Cywinski2020noise}%
  \BibitemOpen
  \bibfield  {author} {\bibinfo {author} {\bibfnamefont {P.}~\bibnamefont
  {Sza{\'{n}}kowski}}\ and\ \bibinfo {author} {\bibfnamefont
  {{\L}.}~\bibnamefont {Cywi{\'{n}}ski}},\ }\href
  {https://doi.org/10.1038/s41598-020-78079-7} {\bibfield  {journal} {\bibinfo
  {journal} {Sci. Rep.}\ }\textbf {\bibinfo {volume} {10}},\ \bibinfo {pages}
  {22189} (\bibinfo {year} {2020})}\BibitemShut {NoStop}%
\bibitem [{\citenamefont {Schoelkopf}\ \emph {et~al.}(2003)\citenamefont
  {Schoelkopf}, \citenamefont {Clerk}, \citenamefont {Girvin}, \citenamefont
  {Lehnert},\ and\ \citenamefont {Devoret}}]{Devoret2003}%
  \BibitemOpen
  \bibfield  {author} {\bibinfo {author} {\bibfnamefont {R.~J.}\ \bibnamefont
  {Schoelkopf}}, \bibinfo {author} {\bibfnamefont {A.~A.}\ \bibnamefont
  {Clerk}}, \bibinfo {author} {\bibfnamefont {S.~M.}\ \bibnamefont {Girvin}},
  \bibinfo {author} {\bibfnamefont {K.~W.}\ \bibnamefont {Lehnert}}, \ and\
  \bibinfo {author} {\bibfnamefont {M.~H.}\ \bibnamefont {Devoret}},\ }\enquote
  {\bibinfo {title} {Qubits as spectrometers of quantum noise},}\ in\ \href
  {\doibase 10.1007/978-94-010-0089-5_9} {\emph {\bibinfo {booktitle} {Quantum
  Noise in Mesoscopic Physics}}},\ \bibinfo {editor} {edited by\ \bibinfo
  {editor} {\bibfnamefont {Y.~V.}\ \bibnamefont {Nazarov}}}\ (\bibinfo
  {publisher} {Springer},\ \bibinfo {address} {Dordrecht},\ \bibinfo {year}
  {2003})\ pp.\ \bibinfo {pages} {175--203}\BibitemShut {NoStop}%
\bibitem [{\citenamefont {Clerk}\ \emph {et~al.}(2010)\citenamefont {Clerk},
  \citenamefont {Devoret}, \citenamefont {Girvin}, \citenamefont {Marquardt},\
  and\ \citenamefont {Schoelkopf}}]{QN.RMP2010}%
  \BibitemOpen
  \bibfield  {author} {\bibinfo {author} {\bibfnamefont {A.~A.}\ \bibnamefont
  {Clerk}}, \bibinfo {author} {\bibfnamefont {M.~H.}\ \bibnamefont {Devoret}},
  \bibinfo {author} {\bibfnamefont {S.~M.}\ \bibnamefont {Girvin}}, \bibinfo
  {author} {\bibfnamefont {F.}~\bibnamefont {Marquardt}}, \ and\ \bibinfo
  {author} {\bibfnamefont {R.~J.}\ \bibnamefont {Schoelkopf}},\ }\href
  {\doibase 10.1103/RevModPhys.82.1155} {\bibfield  {journal} {\bibinfo
  {journal} {Rev. Mod. Phys.}\ }\textbf {\bibinfo {volume} {82}},\ \bibinfo
  {pages} {1155} (\bibinfo {year} {2010})}\BibitemShut {NoStop}%
\bibitem [{\citenamefont {Paz-Silva}\ \emph {et~al.}(2016)\citenamefont
  {Paz-Silva}, \citenamefont {Lee}, \citenamefont {Green},\ and\ \citenamefont
  {Viola}}]{Viola2016}%
  \BibitemOpen
  \bibfield  {author} {\bibinfo {author} {\bibfnamefont {G.~A.}\ \bibnamefont
  {Paz-Silva}}, \bibinfo {author} {\bibfnamefont {S.-W.}\ \bibnamefont {Lee}},
  \bibinfo {author} {\bibfnamefont {T.~J.}\ \bibnamefont {Green}}, \ and\
  \bibinfo {author} {\bibfnamefont {L.}~\bibnamefont {Viola}},\ }\href
  {\doibase 10.1088/1367-2630/18/7/073020} {\bibfield  {journal} {\bibinfo
  {journal} {New J. Phys.}\ }\textbf {\bibinfo {volume} {18}},\ \bibinfo
  {pages} {073020} (\bibinfo {year} {2016})}\BibitemShut {NoStop}%
\bibitem [{\citenamefont {Paz-Silva}\ \emph {et~al.}(2017)\citenamefont
  {Paz-Silva}, \citenamefont {Norris},\ and\ \citenamefont
  {Viola}}]{Viola2017}%
  \BibitemOpen
  \bibfield  {author} {\bibinfo {author} {\bibfnamefont {G.~A.}\ \bibnamefont
  {Paz-Silva}}, \bibinfo {author} {\bibfnamefont {L.~M.}\ \bibnamefont
  {Norris}}, \ and\ \bibinfo {author} {\bibfnamefont {L.}~\bibnamefont
  {Viola}},\ }\href {\doibase 10.1103/PhysRevA.95.022121} {\bibfield  {journal}
  {\bibinfo  {journal} {Phys. Rev. A}\ }\textbf {\bibinfo {volume} {95}},\
  \bibinfo {pages} {022121} (\bibinfo {year} {2017})}\BibitemShut {NoStop}%
\bibitem [{\citenamefont {Kwiatkowski}\ \emph {et~al.}(2020)\citenamefont
  {Kwiatkowski}, \citenamefont {Sza\ifmmode~\acute{n}\else \'{n}\fi{}kowski},\
  and\ \citenamefont {Cywi\ifmmode~\acute{n}\else
  \'{n}\fi{}ski}}]{Cywinski2020}%
  \BibitemOpen
  \bibfield  {author} {\bibinfo {author} {\bibfnamefont {D.}~\bibnamefont
  {Kwiatkowski}}, \bibinfo {author} {\bibfnamefont {P.}~\bibnamefont
  {Sza\ifmmode~\acute{n}\else \'{n}\fi{}kowski}}, \ and\ \bibinfo {author}
  {\bibfnamefont {L.}~\bibnamefont {Cywi\ifmmode~\acute{n}\else
  \'{n}\fi{}ski}},\ }\href {\doibase 10.1103/PhysRevB.101.155412} {\bibfield
  {journal} {\bibinfo  {journal} {Phys. Rev. B}\ }\textbf {\bibinfo {volume}
  {101}},\ \bibinfo {pages} {155412} (\bibinfo {year} {2020})}\BibitemShut
  {NoStop}%
\bibitem [{\citenamefont {Polkovnikov}\ \emph {et~al.}(2011)\citenamefont
  {Polkovnikov}, \citenamefont {Sengupta}, \citenamefont {Silva},\ and\
  \citenamefont {Vengalattore}}]{Vengalattore2011}%
  \BibitemOpen
  \bibfield  {author} {\bibinfo {author} {\bibfnamefont {A.}~\bibnamefont
  {Polkovnikov}}, \bibinfo {author} {\bibfnamefont {K.}~\bibnamefont
  {Sengupta}}, \bibinfo {author} {\bibfnamefont {A.}~\bibnamefont {Silva}}, \
  and\ \bibinfo {author} {\bibfnamefont {M.}~\bibnamefont {Vengalattore}},\
  }\href {\doibase 10.1103/RevModPhys.83.863} {\bibfield  {journal} {\bibinfo
  {journal} {Rev. Mod. Phys.}\ }\textbf {\bibinfo {volume} {83}},\ \bibinfo
  {pages} {863} (\bibinfo {year} {2011})}\BibitemShut {NoStop}%
\bibitem [{\citenamefont {Weiss}(2012)}]{Weiss2012book}%
  \BibitemOpen
  \bibfield  {author} {\bibinfo {author} {\bibfnamefont {U.}~\bibnamefont
  {Weiss}},\ }\href {\doibase 10.1142/8334} {\emph {\bibinfo {title} {Quantum
  Dissipative Systems}}},\ \bibinfo {edition} {4th}\ ed.\ (\bibinfo
  {publisher} {World Scientific},\ \bibinfo {address} {Singapore},\ \bibinfo
  {year} {2012})\BibitemShut {NoStop}%
\bibitem [{\citenamefont {Callen}\ and\ \citenamefont
  {Welton}(1951)}]{Welton1951}%
  \BibitemOpen
  \bibfield  {author} {\bibinfo {author} {\bibfnamefont {H.~B.}\ \bibnamefont
  {Callen}}\ and\ \bibinfo {author} {\bibfnamefont {T.~A.}\ \bibnamefont
  {Welton}},\ }\href {\doibase 10.1103/PhysRev.83.34} {\bibfield  {journal}
  {\bibinfo  {journal} {Phys. Rev.}\ }\textbf {\bibinfo {volume} {83}},\
  \bibinfo {pages} {34} (\bibinfo {year} {1951})}\BibitemShut {NoStop}%
\bibitem [{\citenamefont {Kubo}(1966)}]{Kubo1966}%
  \BibitemOpen
  \bibfield  {author} {\bibinfo {author} {\bibfnamefont {R.}~\bibnamefont
  {Kubo}},\ }\href {\doibase 10.1088/0034-4885/29/1/306} {\bibfield  {journal}
  {\bibinfo  {journal} {Rep. Prog. Phys.}\ }\textbf {\bibinfo {volume} {29}},\
  \bibinfo {pages} {255} (\bibinfo {year} {1966})}\BibitemShut {NoStop}%
\bibitem [{\citenamefont {Zamponi}\ \emph {et~al.}(2005)\citenamefont
  {Zamponi}, \citenamefont {Bonetto}, \citenamefont {Cugliandolo},\ and\
  \citenamefont {Kurchan}}]{Kurchan2005}%
  \BibitemOpen
  \bibfield  {author} {\bibinfo {author} {\bibfnamefont {F.}~\bibnamefont
  {Zamponi}}, \bibinfo {author} {\bibfnamefont {F.}~\bibnamefont {Bonetto}},
  \bibinfo {author} {\bibfnamefont {L.~F.}\ \bibnamefont {Cugliandolo}}, \ and\
  \bibinfo {author} {\bibfnamefont {J.}~\bibnamefont {Kurchan}},\ }\href
  {\doibase 10.1088/1742-5468/2005/09/p09013} {\bibfield  {journal} {\bibinfo
  {journal} {J. Stat. Mech.}\ }\textbf {\bibinfo {volume} {2005}},\ \bibinfo
  {pages} {P09013} (\bibinfo {year} {2005})}\BibitemShut {NoStop}%
\bibitem [{\citenamefont {Cugliandolo}(2011)}]{Cugliandolo2011}%
  \BibitemOpen
  \bibfield  {author} {\bibinfo {author} {\bibfnamefont {L.~F.}\ \bibnamefont
  {Cugliandolo}},\ }\href {\doibase 10.1088/1751-8113/44/48/483001} {\bibfield
  {journal} {\bibinfo  {journal} {J. Phys. A: Math. Theor.}\ }\textbf {\bibinfo
  {volume} {44}},\ \bibinfo {pages} {483001} (\bibinfo {year}
  {2011})}\BibitemShut {NoStop}%
\bibitem [{\citenamefont {Schriefl}\ \emph {et~al.}(2006)\citenamefont
  {Schriefl}, \citenamefont {Makhlin}, \citenamefont {Shnirman},\ and\
  \citenamefont {Sch\"{o}n}}]{Schoen2006}%
  \BibitemOpen
  \bibfield  {author} {\bibinfo {author} {\bibfnamefont {J.}~\bibnamefont
  {Schriefl}}, \bibinfo {author} {\bibfnamefont {Y.}~\bibnamefont {Makhlin}},
  \bibinfo {author} {\bibfnamefont {A.}~\bibnamefont {Shnirman}}, \ and\
  \bibinfo {author} {\bibfnamefont {G.}~\bibnamefont {Sch\"{o}n}},\ }\href
  {\doibase 10.1088/1367-2630/8/1/001} {\bibfield  {journal} {\bibinfo
  {journal} {New J. Phys.}\ }\textbf {\bibinfo {volume} {8}},\ \bibinfo {pages}
  {1} (\bibinfo {year} {2006})}\BibitemShut {NoStop}%
\bibitem [{\citenamefont {Cywi\ifmmode~\acute{n}\else \'{n}\fi{}ski}\ \emph
  {et~al.}(2008)\citenamefont {Cywi\ifmmode~\acute{n}\else \'{n}\fi{}ski},
  \citenamefont {Lutchyn}, \citenamefont {Nave},\ and\ \citenamefont
  {Das~Sarma}}]{DasSarma2008b}%
  \BibitemOpen
  \bibfield  {author} {\bibinfo {author} {\bibfnamefont {L.}~\bibnamefont
  {Cywi\ifmmode~\acute{n}\else \'{n}\fi{}ski}}, \bibinfo {author}
  {\bibfnamefont {R.~M.}\ \bibnamefont {Lutchyn}}, \bibinfo {author}
  {\bibfnamefont {C.~P.}\ \bibnamefont {Nave}}, \ and\ \bibinfo {author}
  {\bibfnamefont {S.}~\bibnamefont {Das~Sarma}},\ }\href {\doibase
  10.1103/PhysRevB.77.174509} {\bibfield  {journal} {\bibinfo  {journal} {Phys.
  Rev. B}\ }\textbf {\bibinfo {volume} {77}},\ \bibinfo {pages} {174509}
  (\bibinfo {year} {2008})}\BibitemShut {NoStop}%
\bibitem [{\citenamefont {Bauch}\ \emph {et~al.}(2018)\citenamefont {Bauch},
  \citenamefont {Hart}, \citenamefont {Schloss}, \citenamefont {Turner},
  \citenamefont {Barry}, \citenamefont {Kehayias}, \citenamefont {Singh},\ and\
  \citenamefont {Walsworth}}]{Walsworth2018}%
  \BibitemOpen
  \bibfield  {author} {\bibinfo {author} {\bibfnamefont {E.}~\bibnamefont
  {Bauch}}, \bibinfo {author} {\bibfnamefont {C.~A.}\ \bibnamefont {Hart}},
  \bibinfo {author} {\bibfnamefont {J.~M.}\ \bibnamefont {Schloss}}, \bibinfo
  {author} {\bibfnamefont {M.~J.}\ \bibnamefont {Turner}}, \bibinfo {author}
  {\bibfnamefont {J.~F.}\ \bibnamefont {Barry}}, \bibinfo {author}
  {\bibfnamefont {P.}~\bibnamefont {Kehayias}}, \bibinfo {author}
  {\bibfnamefont {S.}~\bibnamefont {Singh}}, \ and\ \bibinfo {author}
  {\bibfnamefont {R.~L.}\ \bibnamefont {Walsworth}},\ }\href {\doibase
  10.1103/PhysRevX.8.031025} {\bibfield  {journal} {\bibinfo  {journal} {Phys.
  Rev. X}\ }\textbf {\bibinfo {volume} {8}},\ \bibinfo {pages} {031025}
  (\bibinfo {year} {2018})}\BibitemShut {NoStop}%
\bibitem [{\citenamefont {Bluvstein}\ \emph {et~al.}(2019)\citenamefont
  {Bluvstein}, \citenamefont {Zhang}, \citenamefont {McLellan}, \citenamefont
  {Williams},\ and\ \citenamefont {Jayich}}]{Jayich2019}%
  \BibitemOpen
  \bibfield  {author} {\bibinfo {author} {\bibfnamefont {D.}~\bibnamefont
  {Bluvstein}}, \bibinfo {author} {\bibfnamefont {Z.}~\bibnamefont {Zhang}},
  \bibinfo {author} {\bibfnamefont {C.~A.}\ \bibnamefont {McLellan}}, \bibinfo
  {author} {\bibfnamefont {N.~R.}\ \bibnamefont {Williams}}, \ and\ \bibinfo
  {author} {\bibfnamefont {A.~C.~B.}\ \bibnamefont {Jayich}},\ }\href {\doibase
  10.1103/PhysRevLett.123.146804} {\bibfield  {journal} {\bibinfo  {journal}
  {Phys. Rev. Lett.}\ }\textbf {\bibinfo {volume} {123}},\ \bibinfo {pages}
  {146804} (\bibinfo {year} {2019})}\BibitemShut {NoStop}%
\bibitem [{\citenamefont {Sakurai}\ and\ \citenamefont
  {Napolitano}(2017)}]{Sakurai2011book}%
  \BibitemOpen
  \bibfield  {author} {\bibinfo {author} {\bibfnamefont {J.~J.}\ \bibnamefont
  {Sakurai}}\ and\ \bibinfo {author} {\bibfnamefont {J.}~\bibnamefont
  {Napolitano}},\ }\href {\doibase 10.1017/9781108499996} {\emph {\bibinfo
  {title} {Modern Quantum Mechanics}}},\ \bibinfo {edition} {2nd}\ ed.\
  (\bibinfo  {publisher} {Cambridge University Press},\ \bibinfo {address}
  {Cambridge},\ \bibinfo {year} {2017})\BibitemShut {NoStop}%
\bibitem [{\citenamefont {Gardiner}\ and\ \citenamefont
  {Zoller}(2004)}]{Gardiner2004book}%
  \BibitemOpen
  \bibfield  {author} {\bibinfo {author} {\bibfnamefont {C.}~\bibnamefont
  {Gardiner}}\ and\ \bibinfo {author} {\bibfnamefont {P.}~\bibnamefont
  {Zoller}},\ }\href@noop {} {\emph {\bibinfo {title} {Quantum Noise: A
  Handbook of Markovian and Non-Markovian Quantum Stochastic Methods with
  Applications to Quantum Optics}}},\ Springer Series in Synergetics\ (\bibinfo
   {publisher} {Springer},\ \bibinfo {address} {Berlin},\ \bibinfo {year}
  {2004})\BibitemShut {NoStop}%
\bibitem [{\citenamefont {Barry}\ \emph {et~al.}(2020)\citenamefont {Barry},
  \citenamefont {Schloss}, \citenamefont {Bauch}, \citenamefont {Turner},
  \citenamefont {Hart}, \citenamefont {Pham},\ and\ \citenamefont
  {Walsworth}}]{Walsworth2020}%
  \BibitemOpen
  \bibfield  {author} {\bibinfo {author} {\bibfnamefont {J.~F.}\ \bibnamefont
  {Barry}}, \bibinfo {author} {\bibfnamefont {J.~M.}\ \bibnamefont {Schloss}},
  \bibinfo {author} {\bibfnamefont {E.}~\bibnamefont {Bauch}}, \bibinfo
  {author} {\bibfnamefont {M.~J.}\ \bibnamefont {Turner}}, \bibinfo {author}
  {\bibfnamefont {C.~A.}\ \bibnamefont {Hart}}, \bibinfo {author}
  {\bibfnamefont {L.~M.}\ \bibnamefont {Pham}}, \ and\ \bibinfo {author}
  {\bibfnamefont {R.~L.}\ \bibnamefont {Walsworth}},\ }\href {\doibase
  10.1103/RevModPhys.92.015004} {\bibfield  {journal} {\bibinfo  {journal}
  {Rev. Mod. Phys.}\ }\textbf {\bibinfo {volume} {92}},\ \bibinfo {pages}
  {015004} (\bibinfo {year} {2020})}\BibitemShut {NoStop}%
\bibitem [{\citenamefont {Kubo}(1962)}]{Kubo1962}%
  \BibitemOpen
  \bibfield  {author} {\bibinfo {author} {\bibfnamefont {R.}~\bibnamefont
  {Kubo}},\ }\href {\doibase 10.1143/JPSJ.17.1100} {\bibfield  {journal}
  {\bibinfo  {journal} {J. Phys. Soc. Jpn.}\ }\textbf {\bibinfo {volume}
  {17}},\ \bibinfo {pages} {1100} (\bibinfo {year} {1962})}\BibitemShut
  {NoStop}%
\bibitem [{\citenamefont {Sieberer}\ \emph {et~al.}(2016)\citenamefont
  {Sieberer}, \citenamefont {Buchhold},\ and\ \citenamefont
  {Diehl}}]{Diehl2016}%
  \BibitemOpen
  \bibfield  {author} {\bibinfo {author} {\bibfnamefont {L.~M.}\ \bibnamefont
  {Sieberer}}, \bibinfo {author} {\bibfnamefont {M.}~\bibnamefont {Buchhold}},
  \ and\ \bibinfo {author} {\bibfnamefont {S.}~\bibnamefont {Diehl}},\ }\href
  {\doibase 10.1088/0034-4885/79/9/096001} {\bibfield  {journal} {\bibinfo
  {journal} {Rep. Prog. Phys.}\ }\textbf {\bibinfo {volume} {79}},\ \bibinfo
  {pages} {096001} (\bibinfo {year} {2016})}\BibitemShut {NoStop}%
\bibitem [{\citenamefont {Gabelli}\ and\ \citenamefont
  {Reulet}(2008)}]{Reulet2008}%
  \BibitemOpen
  \bibfield  {author} {\bibinfo {author} {\bibfnamefont {J.}~\bibnamefont
  {Gabelli}}\ and\ \bibinfo {author} {\bibfnamefont {B.}~\bibnamefont
  {Reulet}},\ }\href {\doibase 10.1103/PhysRevLett.100.026601} {\bibfield
  {journal} {\bibinfo  {journal} {Phys. Rev. Lett.}\ }\textbf {\bibinfo
  {volume} {100}},\ \bibinfo {pages} {026601} (\bibinfo {year}
  {2008})}\BibitemShut {NoStop}%
\bibitem [{\citenamefont {Stratonovich}(1994)}]{Stratonovich1992book}%
  \BibitemOpen
  \bibfield  {author} {\bibinfo {author} {\bibfnamefont {R.~L.}\ \bibnamefont
  {Stratonovich}},\ }\href {\doibase 10.1007/978-3-642-77343-3} {\emph
  {\bibinfo {title} {Nonlinear nonequilibrium thermodynamics I}}}\ (\bibinfo
  {publisher} {Springer-Verlag},\ \bibinfo {address} {Berlin},\ \bibinfo {year}
  {1994})\BibitemShut {NoStop}%
\bibitem [{\citenamefont {Abragam}(1961)}]{Abragam1961book}%
  \BibitemOpen
  \bibfield  {author} {\bibinfo {author} {\bibfnamefont {A.}~\bibnamefont
  {Abragam}},\ }\href@noop {} {\emph {\bibinfo {title} {The Principles of
  Nuclear Magnetism}}},\ Comparative Pathobiology - Studies in the Postmodern
  Theory of Education\ (\bibinfo  {publisher} {Clarendon Press},\ \bibinfo
  {address} {Oxford},\ \bibinfo {year} {1961})\BibitemShut {NoStop}%
\bibitem [{\citenamefont {Petit}\ \emph {et~al.}(2018)\citenamefont {Petit},
  \citenamefont {Boter}, \citenamefont {Eenink}, \citenamefont {Droulers},
  \citenamefont {Tagliaferri}, \citenamefont {Li}, \citenamefont {Franke},
  \citenamefont {Singh}, \citenamefont {Clarke}, \citenamefont {Schouten},
  \citenamefont {Dobrovitski}, \citenamefont {Vandersypen},\ and\ \citenamefont
  {Veldhorst}}]{Veldhorst2018}%
  \BibitemOpen
  \bibfield  {author} {\bibinfo {author} {\bibfnamefont {L.}~\bibnamefont
  {Petit}}, \bibinfo {author} {\bibfnamefont {J.~M.}\ \bibnamefont {Boter}},
  \bibinfo {author} {\bibfnamefont {H.~G.~J.}\ \bibnamefont {Eenink}}, \bibinfo
  {author} {\bibfnamefont {G.}~\bibnamefont {Droulers}}, \bibinfo {author}
  {\bibfnamefont {M.~L.~V.}\ \bibnamefont {Tagliaferri}}, \bibinfo {author}
  {\bibfnamefont {R.}~\bibnamefont {Li}}, \bibinfo {author} {\bibfnamefont
  {D.~P.}\ \bibnamefont {Franke}}, \bibinfo {author} {\bibfnamefont {K.~J.}\
  \bibnamefont {Singh}}, \bibinfo {author} {\bibfnamefont {J.~S.}\ \bibnamefont
  {Clarke}}, \bibinfo {author} {\bibfnamefont {R.~N.}\ \bibnamefont
  {Schouten}}, \bibinfo {author} {\bibfnamefont {V.~V.}\ \bibnamefont
  {Dobrovitski}}, \bibinfo {author} {\bibfnamefont {L.~M.~K.}\ \bibnamefont
  {Vandersypen}}, \ and\ \bibinfo {author} {\bibfnamefont {M.}~\bibnamefont
  {Veldhorst}},\ }\href {\doibase 10.1103/PhysRevLett.121.076801} {\bibfield
  {journal} {\bibinfo  {journal} {Phys. Rev. Lett.}\ }\textbf {\bibinfo
  {volume} {121}},\ \bibinfo {pages} {076801} (\bibinfo {year}
  {2018})}\BibitemShut {NoStop}%
\end{thebibliography}%
\end{document}